\documentclass[lettersize,journal]{IEEEtran}
\usepackage{amsmath,amsfonts}
\usepackage{algorithmic}
\usepackage{array}
\usepackage{textcomp}
\usepackage{stfloats}
\usepackage{url}
\usepackage{xcolor}
\usepackage{verbatim}
\usepackage{graphicx}
\usepackage{subfigure}
\usepackage{booktabs}
\usepackage{cite}
\usepackage{multirow}
\def\BibTeX{{\rm B\kern-.05em{\sc i\kern-.025em b}\kern-.08em
    T\kern-.1667em\lower.7ex\hbox{E}\kern-.125emX}}
\usepackage{balance}

\begin{document}
\title{Hybrid Ray-Tracing and Physics-Embedded Neural Modeling for High-Fidelity Channel Reconstruction in Wireless Digital Twins}
\author{Huiwen Zhang,~\IEEEmembership{Graduate Student Member,~IEEE,} Chu Ma,~\IEEEmembership{Member,~IEEE,} Feng Ye,~\IEEEmembership{Senior Member,~IEEE}
\thanks{This project is partially supported by the U.S. National Science Foundation under Grants IIS-2542859 and CCSS-2237619.}
\thanks{Huiwen Zhang, Chu Ma and Feng Ye (corresponding) are with the Department of Electrical and Computer Engineering, University of Wisconsin-Madison, Wisconsin, WI, USA. Emails: \{hzhang2279, chu.ma, feng.ye\}@wisc.edu. 
}
}

\markboth{Journal of \LaTeX\ Class Files,~Vol.~18, No.~9, September~2020}%
{How to Use the IEEEtran \LaTeX \ Templates}

\maketitle

\begin{abstract}
Many existing wireless digital twins (DTs) rely heavily on ray‑tracing (RT)-based channel simulators to achieve scalability in large‑scale environments. However, the geometric-optics (GO) approximations underlying RT become inaccurate in propagation regimes dominated by diffraction and refraction, which are prevalent at sub-6 GHz carrier frequencies. This paper presents a hybrid RT and physics‑embedded physics‑informed neural network (PE‑PINN) framework that enhances site‑specific wireless DTs with physics‑consistent full‑wave accuracy. Building on prior PE‑PINN work, we introduce targeted architectural extensions, including a finite‑length edge diffraction kernel and a lossy dielectric formulation based on complex wavenumbers, to accurately model electromagnetic wave interactions with realistic objects and materials. To balance accuracy and efficiency, we further propose a localized tunnel strategy that deploys PE‑PINN models only along dominant propagation paths, selectively replacing or augmenting RT predictions in regions where GO assumptions break down. Comprehensive evaluations against full‑wave COMSOL simulations demonstrate that the proposed framework closely matches reference solutions and significantly outperforms RT‑only approaches in diffraction‑ and refraction‑dominated scenarios. We further demonstrate seamless integration of the proposed approach into a realistic indoor wireless DT, highlighting its practicality and scalability for next‑generation wireless network design.
\end{abstract}


\section{Introduction}

\IEEEPARstart{D}{igital} twins (DTs) of wireless environments are emerging as a foundational tool for next-generation network design, deployment, and operation. By providing a virtual replica of the radio propagation environment, wireless DTs enable key tasks such as coverage prediction, beam management, network planning, and joint communication and sensing without requiring repeated, costly field measurements~\cite{9854866, thelen2022comprehensivereviewdigitaltwin}. As wireless systems evolve toward increasingly complex and heterogeneous environments, the accuracy of electromagnetic (EM) wave propagation modeling has become a critical bottleneck that directly impacts DT reliability. 
Ray-tracing (RT)-based simulators have emerged as the backbone of large-scale wireless channel modeling and are commonly treated as ground truth in DT development. Among these tools, the open-source library Sionna~\cite{sionna} has gained widespread adoption, while commercial and research oriented tools such as MATLAB's RT toolbox~\cite{matlab_comm_toolbox} continue to be used across industry and academia. RT engines compute channel impulse responses by tracing discrete propagation paths using geometric optics (GO) and the uniform theory of diffraction (UTD). These asymptotic methods perform well at millimeter-wave and sub-THz frequencies, where the wavelength is much smaller than typical scatterers. At lower carrier frequencies, such as 2.4\,GHz ($\lambda \approx 12.5$\,cm), however, the wavelength becomes comparable to common indoor structures, rendering diffraction, refraction, and near-field coupling effects increasingly significant and poorly captured by ray-based approximations~\cite{901903,7771852}. 

Fully physics-based numerical solvers, such as the finite element method (FEM), address this limitation by rigorously solving Maxwell’s equations. However, because mesh resolution must remain a small fraction of the wavelength to ensure numerical convergence, applying FEM at room or building scales incurs prohibitive computational and memory costs, limiting its practicality for DT construction. Recently, Physics-Informed Neural Networks (PINNs)~\cite{RAISSI2019686,rasht2022physics,baldan2023physics,song2022versatile} have emerged as a promising mesh-free alternative for solving partial differential equations (PDEs). Despite their flexibility, standard PINNs suffer from spectral bias and optimization instability when applied to high-frequency wave problems~\cite{rahaman2019spectral}. To address these challenges, we previously proposed the Physics-Embedded PINN (PE-PINN) framework~\cite{zhang2026physicsinformedneuralnetworksarchitectural}, which embeds physics directly into the network architecture to improve convergence and stability. Although PE-PINN has demonstrated high-fidelity 2D and 3D wave-field reconstruction with substantial memory savings compared to FEM, its initial formulation was limited to relatively controlled settings, leaving a gap for practical deployment in realistic, highly heterogeneous wireless environments.

To bridge this gap, this paper proposes a hybrid RT and PE‑PINN framework tailored for site‑specific wireless DTs. Rather than introducing a new DT construction paradigm, we leverage PE‑PINN and develop targeted architectural enhancements to supply high‑fidelity wireless channel reconstruction that can be readily integrated into existing RT‑based DT systems. Specifically, we introduce a finite‑length edge diffraction kernel to accurately model wave diffraction from physical edges, and extend the material‑aware domain decomposition of PE‑PINN to support wave propagation through lossy dielectrics by embedding complex wavenumbers into the network. Building on these enhancements, we further propose a localized \emph{tunnel} strategy, in which PE‑PINN models are constructed along dominant propagation paths to selectively replace or augment RT predictions in regions where GO approximations break down, achieving a favorable balance between accuracy and computational efficiency.
The main contributions of this paper are summarized as follows:
\begin{itemize}
\item \textbf{Physics‑Constrained High‑Fidelity Channel Reconstruction:}
We enhance PE‑PINN with a finite‑length edge diffraction kernel that discretizes physical edges into coherent secondary point sources, enabling accurate learning of diffraction effects. In addition, we introduce a lossy dielectric formulation by incorporating complex wavenumbers, allowing PE‑PINN to jointly model phase accumulation and attenuation within real‑world materials.
\item \textbf{Localized Digital Twin Enhancement Strategy:}  
We propose a scalable \emph{tunnel} strategy that confines full‑wave PE‑PINN computations to critical propagation paths, enabling seamless integration of neural PDE solvers with existing RT‑based wireless DT architectures.

\item \textbf{Rigorous Validation and Practical Deployment:}  
We validate the proposed framework against full‑wave COMSOL simulations in controlled diffraction and refraction scenarios, and further demonstrate its practical deployment in a realistic three‑dimensional laboratory DT, highlighting physics‑consistent accuracy improvements in regions where standard RT engines fall short.
\end{itemize}
The rest of this paper is organized as follows. Sec.~\ref{sec: related work} reviews related work; Sec.~\ref{sec: prelim} presents a preliminary analysis of RT-based simulators; Sec.~\ref{sec: method} describes the proposed PE-PINN-based DT enhancement; Sec.~\ref{sec: evaluation} reports the evaluation results; and Sec.~\ref{sec:conclusion} concludes the paper.

\section{Related Work}\label{sec: related work}

Wireless DTs aim to construct a virtual replica of the radio environment that accurately mirrors real-world propagation characteristics, enabling applications such as network planning, beamforming optimization, and spectrum management~\cite{9854866,9429703}. A central challenge lies in the channel modeling component: how to efficiently and accurately represent the electromagnetic field across the entire scene.

Recent work has explored neural scene representations, originally developed in the computer vision community, for RF field reconstruction~\cite{11440325}. NeRF$^2$~\cite{Zhao_2023} adapts Neural Radiance Fields to learn a continuous volumetric RF function that maps spatial coordinates to signal attenuation, trained on sparse in-situ signal measurements collected at physical receiver locations. RF-3DGS~\cite{11355734} represents the propagation environment as a set of anisotropic 3D Gaussian ellipsoids, each encoding the scattering and reflection characteristics of a local region; trained on measured spatial 
spectrum data and calibrated with real-world measurements. GSRF~\cite{yang2025gsrfcomplexvalued3dgaussian} further extends this representation to the complex-valued domain to capture both amplitude and phase, enabling the modeling of constructive and destructive interference. Jiang et al.~\cite{10886938} propose a deep learning framework for learnable wireless DTs, where per-object neural representations are combined with explicit RT to reconstruct the EM field from crowd-sourced signal measurements. Across these methods, the RF ground truth used for training is obtained either from dense real-world measurements or from RT simulations. Other studies have explored physics-informed neural networks for wireless channel modeling. Zhu et al.~\cite{zhu2024physicsinformedgeneralizablewirelesschannel} provide a comprehensive overview of PINN-based channel modeling and demonstrate indoor path loss prediction using semantic segmentation. Jia et al.~\cite{jia2025physicsinformedrepresentationalignmentsparse} incorporate a simplified 2D Helmholtz equation as a regularization term within a diffusion model to reconstruct radio maps from sparse samples. In these approaches, physical laws serve as auxiliary constraints to improve data efficiency and generalization, while the primary supervision still comes from RT or measurement datasets. 


In contrast, the PE-PINN in this work operates as a direct PDE solver: the Helmholtz equation and boundary conditions alone determine the predicted field, without requiring any ground-truth field data for supervision. Rather than replacing RT entirely, we adopt a hybrid strategy that integrates the PE-PINN solver into the DT. RT provides the global field map across the full environment, while the PE-PINN solves the full-wave equation within localized tunnel domains where phenomena such as diffraction, multi-material refraction, and near-field interference are physically significant, replacing the RT predictions in those regions.

\section{Inaccuracy of RT-based Channel Simulation}\label{sec: prelim}

Most existing wireless DT frameworks implicitly treat simulation outputs as ground truth, commonly relying on RT-based tools such as Sionna~\cite{sionna}. Although its GPU‑accelerated design enables efficient simulation of large environments, the channel reconstruction capability of Sionna and other RT-based approaches is fundamentally limited by the reliance on high‑frequency asymptotic approximations. GO and UTD model EM waves as rays, an assumption that degrades when the characteristic dimensions of objects become comparable to the wavelength, as is often the case in sub-6 GHz indoor environments. In such regimes, full-wave phenomena, including complex diffraction and refraction, cannot be faithfully captured by a finite number of traced rays. Moreover, these RT engines usually evaluate the EM field only at discrete receiver locations by summing ray contributions, rather than solving for the field throughout the spatial domain. This point‑wise formulation prevents the generation of continuous, spatially resolved field maps, limiting the suitability of RT simulations as ground truth for DT applications that require dense field information or accurate wave–object interaction modeling.

To explicitly expose these limitations, this section focuses on scenarios dominated by diffraction and refraction, particularly when object dimensions are comparable to the operating wavelength. These cases highlight inaccuracies that are intrinsic to RT‑based wireless simulators, rather than implementation specific artifacts, and motivate the use of full‑wave references. Ground truth is generated using COMSOL Multiphysics~\cite{comsol}, a commercial FEM solver for Maxwell’s equations. While being physically faithful to wave propagation, however, FEM is fundamentally constrained by mesh resolution requirements, leading to rapidly increasing computational and memory costs as the domain size grows. For example, generating a full wavefield at 2.4 GHz over a $5 \times 5 \times 5$ m$^3$ region requires more than 12.5 TB of memory. As a result, FEM‑based solvers are impractical for large‑scale or system‑level DT simulations. Consequently, our analysis focuses on small‑scale scenarios where full‑wave reference solutions remain tractable.

\subsection{Diffraction Case Demonstration}




To evaluate the accuracy of Sionna’s diffraction modeling, we design a controlled diffraction scenario within a rectangular computational domain of dimensions $2 \times 0.5 \times 0.5~\text{m}$. A perfect electric conductor (PEC) cube with a side length of $0.25~\text{m}$ is placed at the origin $(0,0,0)$, serving as the diffracting obstacle. The operating frequency is set to $2.4~\text{GHz}$, corresponding to a free-space wavelength of approximately $0.125~\text{m}$. The excitation source is located at $(-0.75, 0, 0)~\text{m}$. In Sionna, the source is modeled as an ideal point source. In COMSOL, however, a true point source introduces a singularity in the finite-element formulation. To avoid this issue, the source is modeled as a spherical radiator with a total radiated power of $1~\text{W}$, defined on a sphere whose radius equals one wavelength. This radius is chosen such that, beyond the source sphere, the phase of the radiated field matches that of an ideal point source, thereby ensuring a consistent phase reference between the two solvers. Due to the memory limitations of our laboratory workstation, the simulation domain cannot be further expanded while maintaining the mesh density required for numerical convergence of the FEM solver. For the Sionna simulation, receivers are uniformly distributed throughout the domain with a spacing of $0.01~\text{m}$ in each dimension. The electric field values recorded at all receiver locations are aggregated to reconstruct the full three-dimensional spatial field distribution for comparison with the FEM results.

\begin{figure*}[ht]
    \centering
    \includegraphics[width=.9\linewidth]{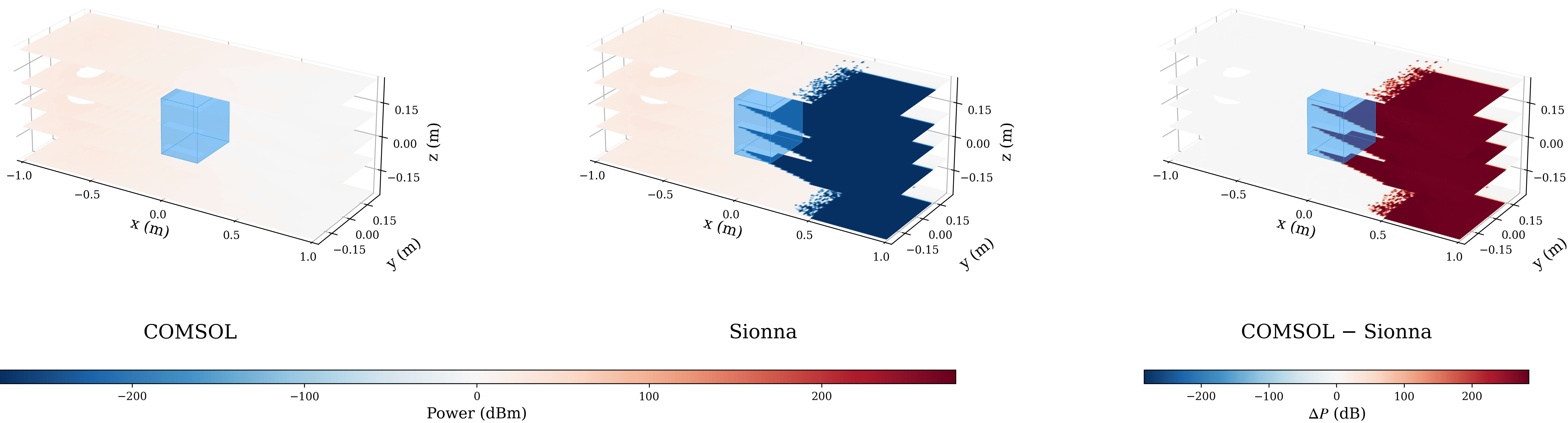}
    \caption{Received power (dBm) and difference heatmap (dB) in diffraction case from COMSOL and Sionna. }
    \label{fig:PrelimDiffra_mag}
    \vspace{-3mm}
\end{figure*}

\begin{figure*}[ht]
    \centering
    \includegraphics[width=.9\linewidth]{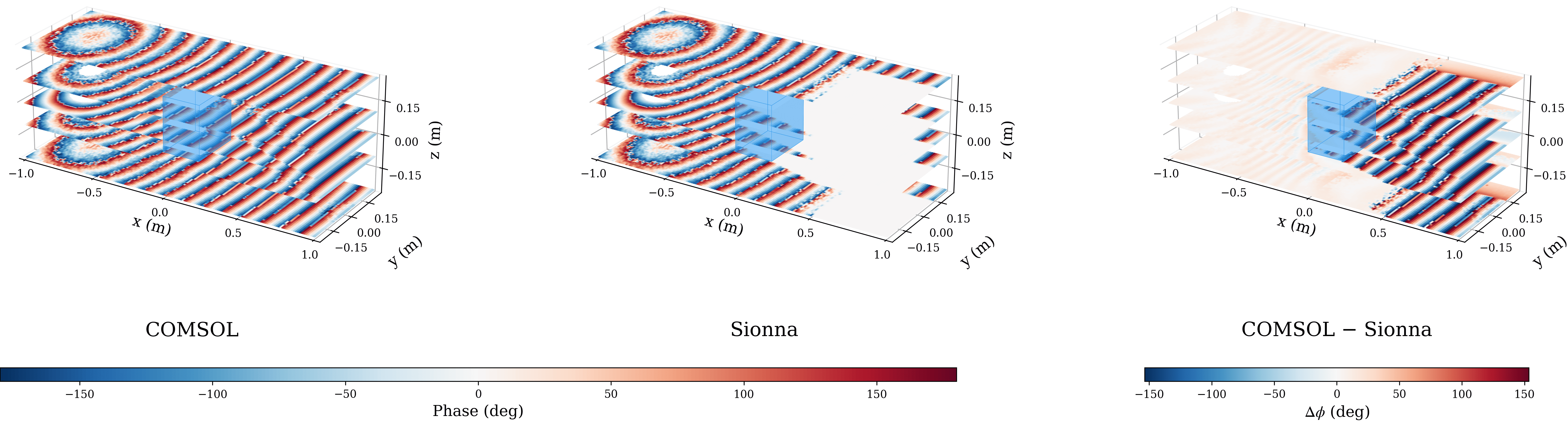}
    \caption{Received phase ($^\circ$) and difference heatmap in diffraction case between COMSOL and Sionna.}
    \label{fig:PrelimDiffra_phase}
    \vspace{-3mm}
\end{figure*}


\begin{table}[t]

\centering
\caption{Results comparison in diffraction case.} 
\begin{tabular}{ll|ccc}
\toprule
\textbf{Region} &  & \textbf{MAE} & \textbf{RMSE} & \textbf{Median} \\
\midrule
\multirow{2}{*}{Whole domain}
&Magnitude (dB)     & 88.45 & 156.36 & 1.92 \\
&Phase ($^\circ$)   & 35.14 & 58.81  & 11.45 \\
\midrule
\multirow{2}{*}{$x > -0.25$}
  & Magnitude (dB)       & 141.11  & 197.87  & 22.68 \\
  & Phase ($^\circ$)     & 51.64 & 74.01  & 27.41  \\
\bottomrule
\end{tabular}
\label{tab:PrelimDiffra}
 \vspace{-4mm}
\end{table}

For consistency, the electric field is converted into received magnitude and phase, matching the channel impulse response representation produced by Sionna. Figs.~\ref{fig:PrelimDiffra_mag} and \ref{fig:PrelimDiffra_phase} show the difference heatmaps (COMSOL$-$Sionna) for field magnitude and phase, each visualized through five planar slices along the $z$-axis at $z = -0.25$, $-0.1$, $0.0$, $0.1$, and $0.25~\text{m}$, stacked vertically to reveal the spatial variation across the domain. The heatmaps show that Sionna produces a nearly empty shadow region behind the obstacle, whereas COMSOL captures the expected edge diffraction into the geometrical shadow zone. Notably, despite the symmetric placement of the obstacle, the diffraction errors are visibly asymmetric. As summarized in Table~\ref{tab:PrelimDiffra}, the full-domain magnitude errors (MAE $88.45$\,dB, RMSE $156.36$\,dB, median only $1.92$\,dB) and phase errors ($35.14^\circ$, $58.81^\circ$, $11.45^\circ$) confirm that RT resolves line-of-sight regions well but fails in the shadow zone, where extreme outliers dominate the aggregate statistics. Restricting the evaluation to $x > -0.25$\,m amplifies this: the magnitude median jumps to $22.68$\,dB and the MAE to $141.11$\,dB; phase errors degrade comparably (MAE $51.64^\circ$, RMSE $74.01^\circ$). These results demonstrate the limitations of RT-based simulation in the vicinity of diffracting obstacles compared with full-wave modeling.

\subsection{Refraction Case Demonstration}\label{sec: prelimrefra}

\begin{figure*}[t]
    \centering
    \includegraphics[width=.9\linewidth]{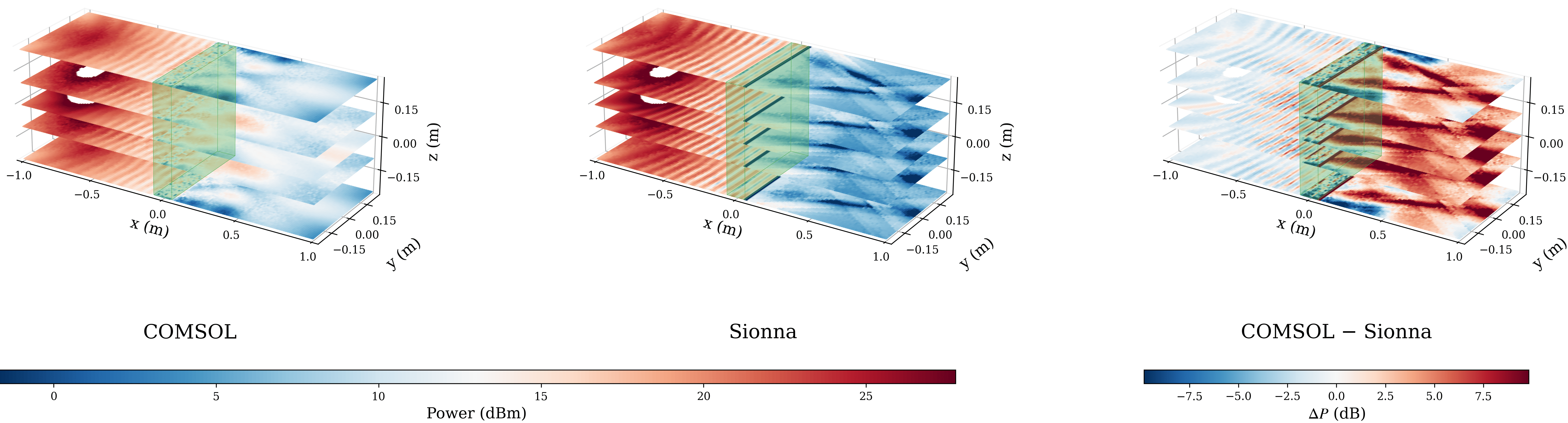}
    \caption{Received power (dBm) and difference heatmap (dB) in refraction case from COMSOL and Sionna. }
    \label{fig:PrelimRefra_mag}
    \vspace{-3mm}
\end{figure*}

\begin{figure*}[ht]
    \centering
    \includegraphics[width=.9\linewidth]{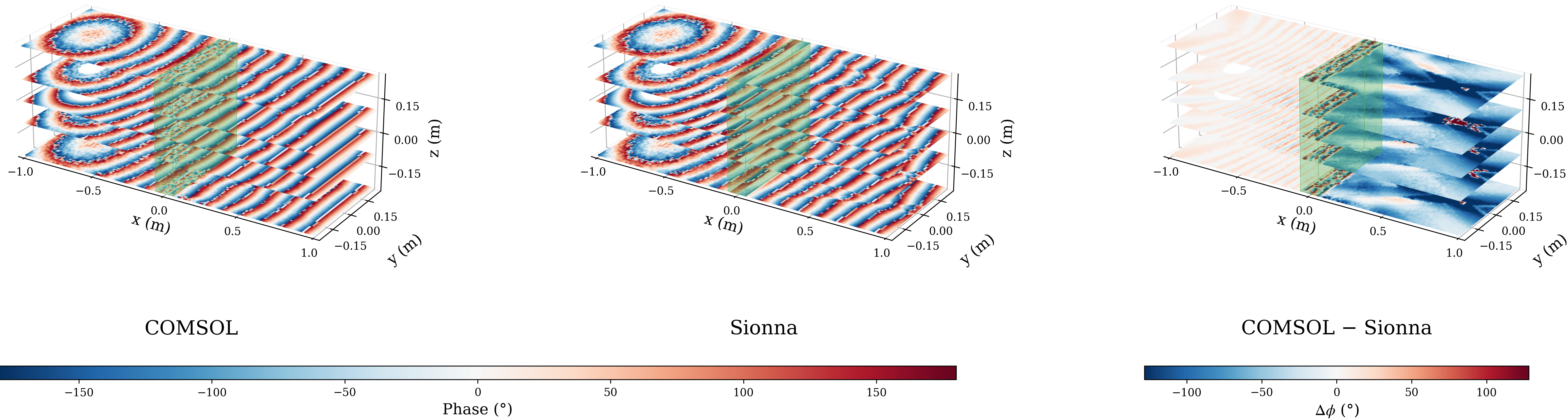}
    \caption{Received phase ($^\circ$) and difference heatmap in refraction case between COMSOL and Sionna.}
    \label{fig:PrelimRefra_phase}
    \vspace{-3mm}
\end{figure*}

We further assess Sionna’s accuracy in modeling refraction by introducing a dielectric strip with a relative permittivity of $\varepsilon_r = 9$ into the same $2 \times 0.5 \times 0.5~\text{m}$ computational domain. The dielectric strip occupies the region $x \in (-0.0625,\, 0.0625)~\text{m}$ and extends across the full cross section of the domain. The excitation source configuration remains identical to that used in the diffraction experiment. The corresponding field magnitude, phase, and difference results are shown in Figs.~\ref{fig:PrelimRefra_mag} and \ref{fig:PrelimRefra_phase}.

\begin{table}[b]
\vspace{-4mm}

\centering
\caption{Results comparison in refraction case.}

\begin{tabular}{ll|ccc}
\toprule
\textbf{Region} & & \textbf{MAE} & \textbf{RMSE} & \textbf{Median} \\
\midrule
\multirow{2}{*}{Whole domain}
  & Magnitude (dB)       & 6.43  & 40.42 & 2.47  \\
  & Phase ($^\circ$)     & 42.74 & 60.74 & 25.73 \\
\midrule
\multirow{2}{*}{$x > -0.125$}
  & Magnitude (dB)       & 10.19  & 53.57  & 4.64  \\
  & Phase ($^\circ$)     & 67.81  & 79.75  & 61.80  \\
\bottomrule
\end{tabular}
\label{tab:PrelimRefra}
\vspace{-4mm}
\end{table}

As illustrated by the difference heatmap, the modeling errors exhibit pronounced spatial dependence. On the source side of the dielectric strip, where propagation remains line-of-sight, the two solvers largely agree; the residual differences form an alternating $+/-$ pattern with a spatial period of approximately half a wavelength, consistent with a standing wave produced by incident\textendash reflected interference at the dielectric interface. On the transmission side, a systematic mismatch emerges, indicating that the dominant source of error lies in Sionna's ray-based refraction model rather than in free-space propagation. The quantitative results in Table~\ref{tab:PrelimRefra} corroborate this interpretation. The full-domain magnitude errors (MAE of $6.43$\,dB, median of $2.47$\,dB, RMSE of $40.42$\,dB) show reasonable overall amplitude fidelity, though the elevated RMSE reveals localized large deviations. Phase errors expose a more fundamental limitation: a median of $25.73^\circ$ (MAE $42.74^\circ$, RMSE $60.74^\circ$) points to a systematic bias distributed broadly beyond the strip rather than isolated outliers. When the evaluation is restricted to $x > -0.125$\,m, which encompasses the vicinity of the dielectric strip and the entire transmitted-wave domain, both trends are amplified: the magnitude MAE rises to $10.19$\,dB, the phase MAE to $67.81^\circ$, and the phase median surges to $61.80^\circ$, confirming that the dominant errors are concentrated near and beyond the dielectric interface.

\begin{figure*}[t]
\centering
\setlength{\tabcolsep}{1pt}
\begin{tabular}{cccccc}
    & $P=2$ & $P=5$ & $P=10$  & $P=25$ & $P=100$\\
    \rotatebox{90}{\small\; Magnitude} &
    \includegraphics[width=0.18\textwidth]{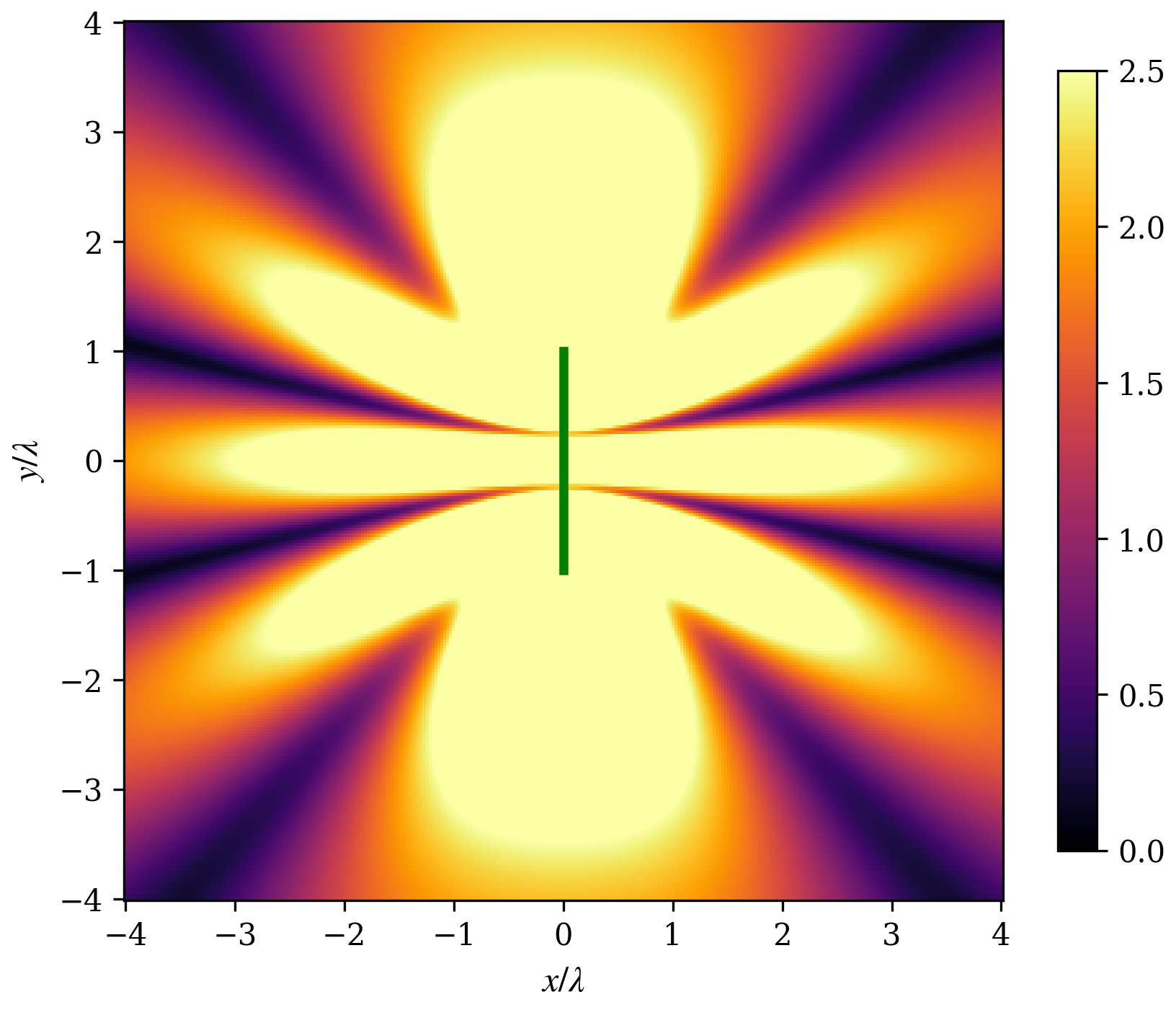} &
    \includegraphics[width=0.18\textwidth]{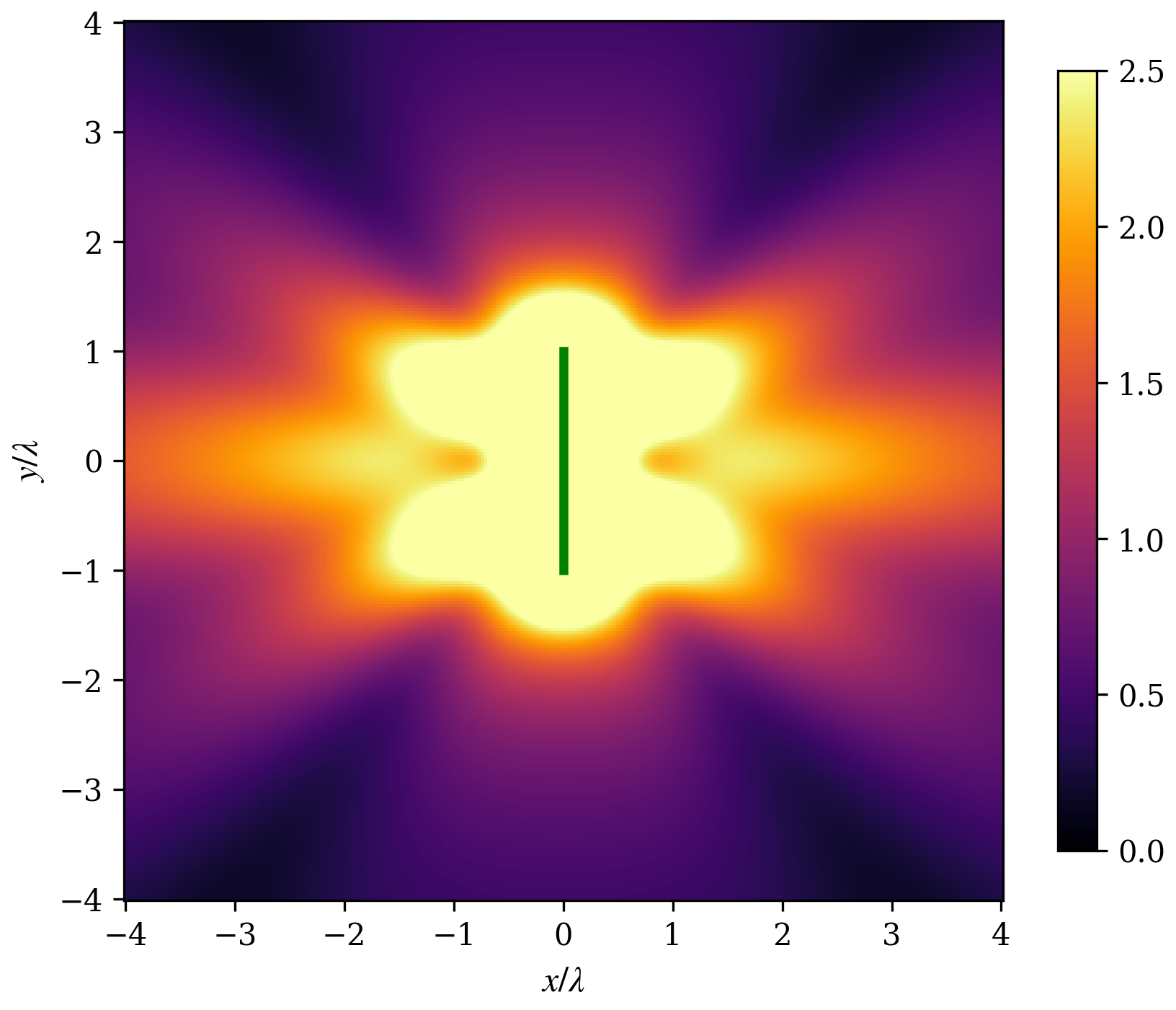} &
    \includegraphics[width=0.18\textwidth]{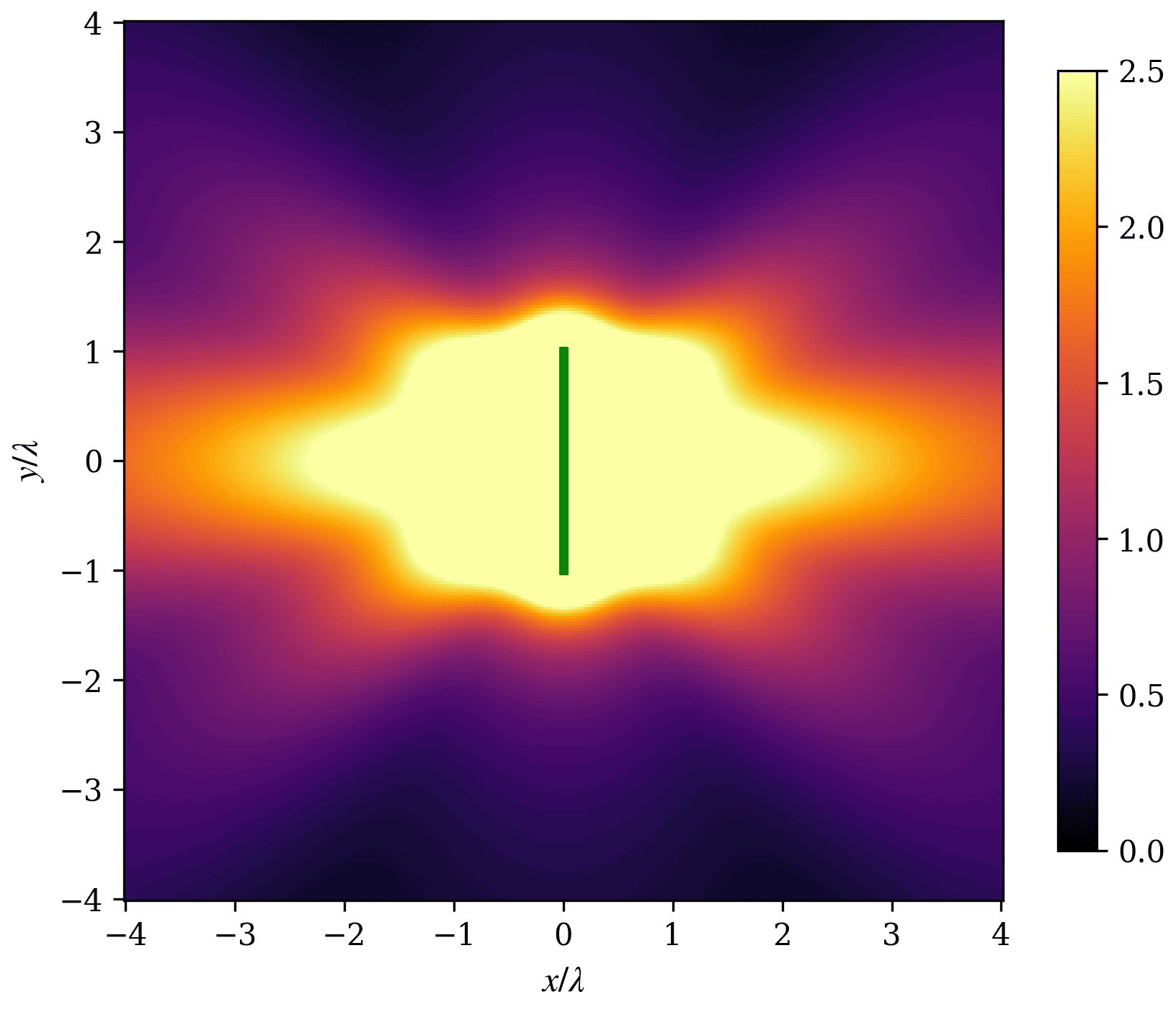} &
    \includegraphics[width=0.18\textwidth]{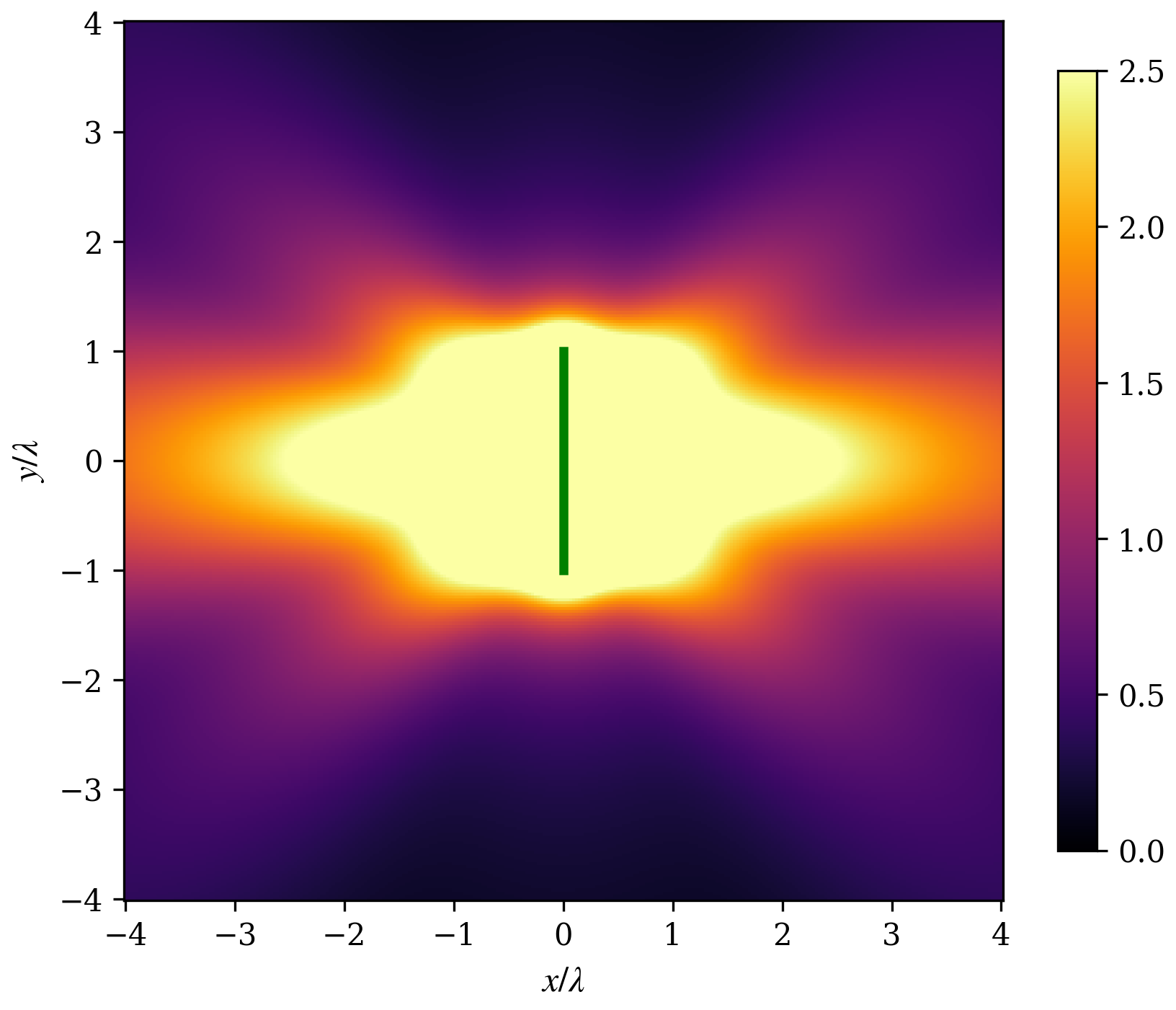} &
    \includegraphics[width=0.18\textwidth]{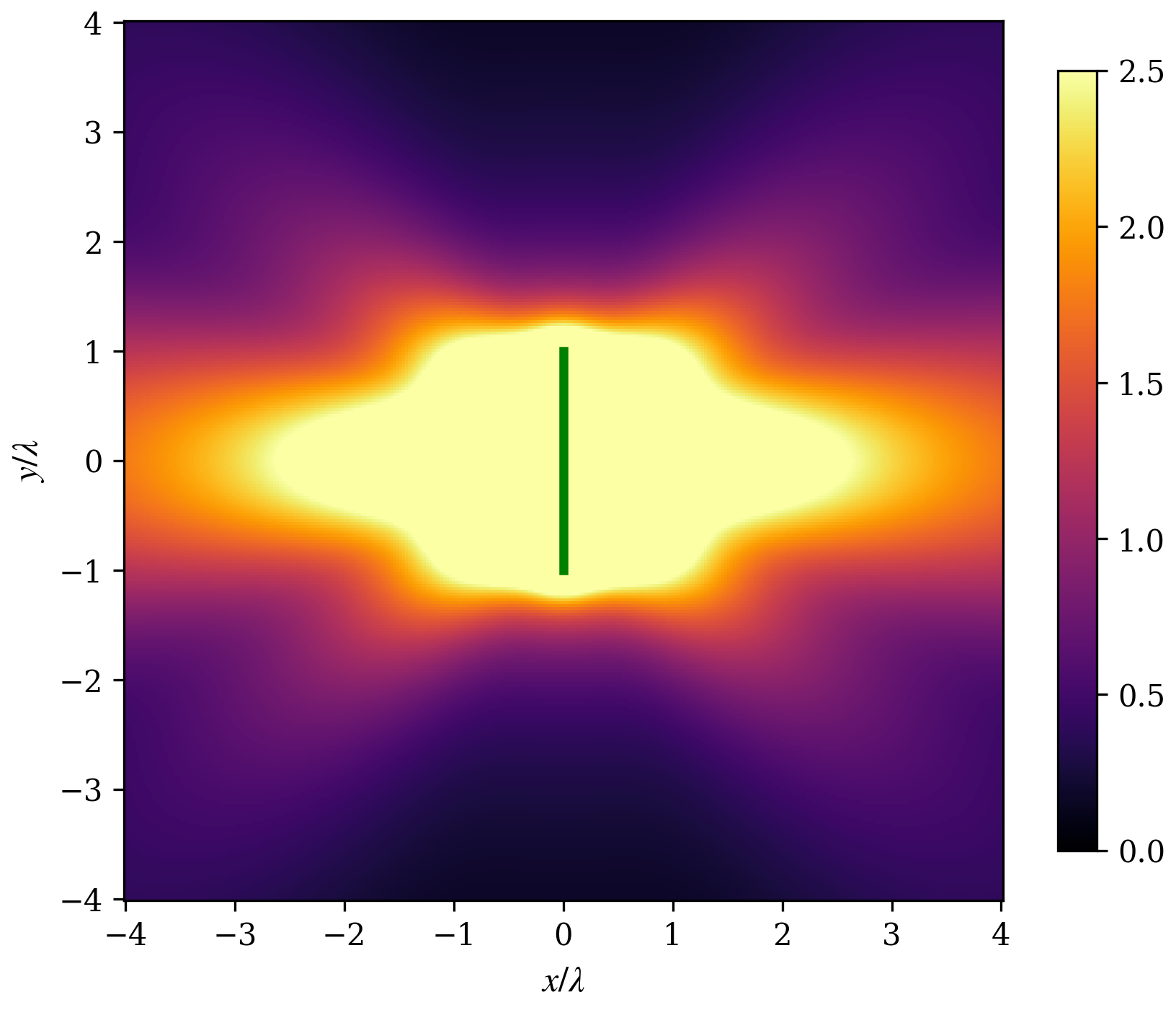} \\
    \rotatebox{90}{\small\; Phase} &
    \includegraphics[width=0.18\textwidth]{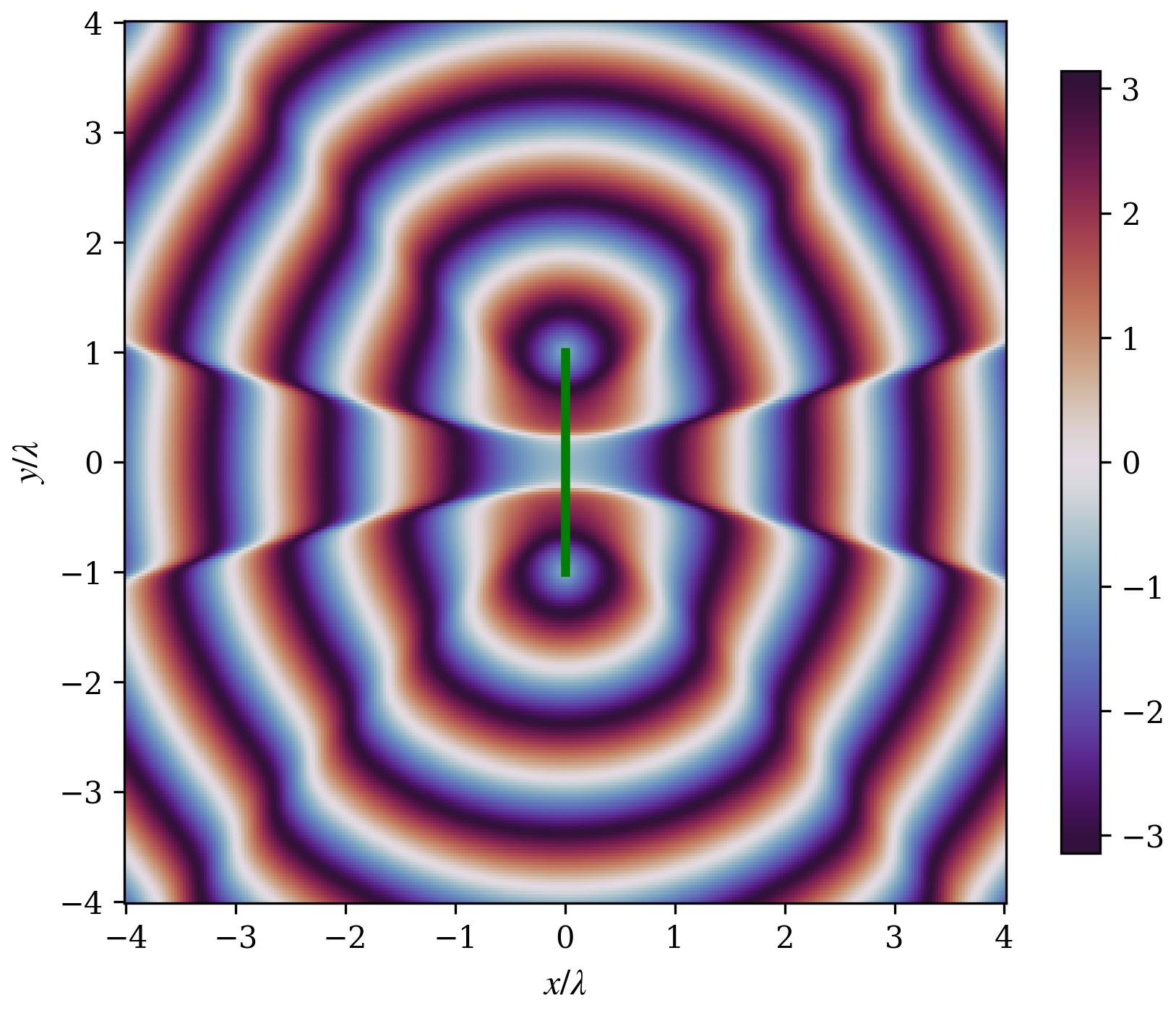} &
    \includegraphics[width=0.18\textwidth]{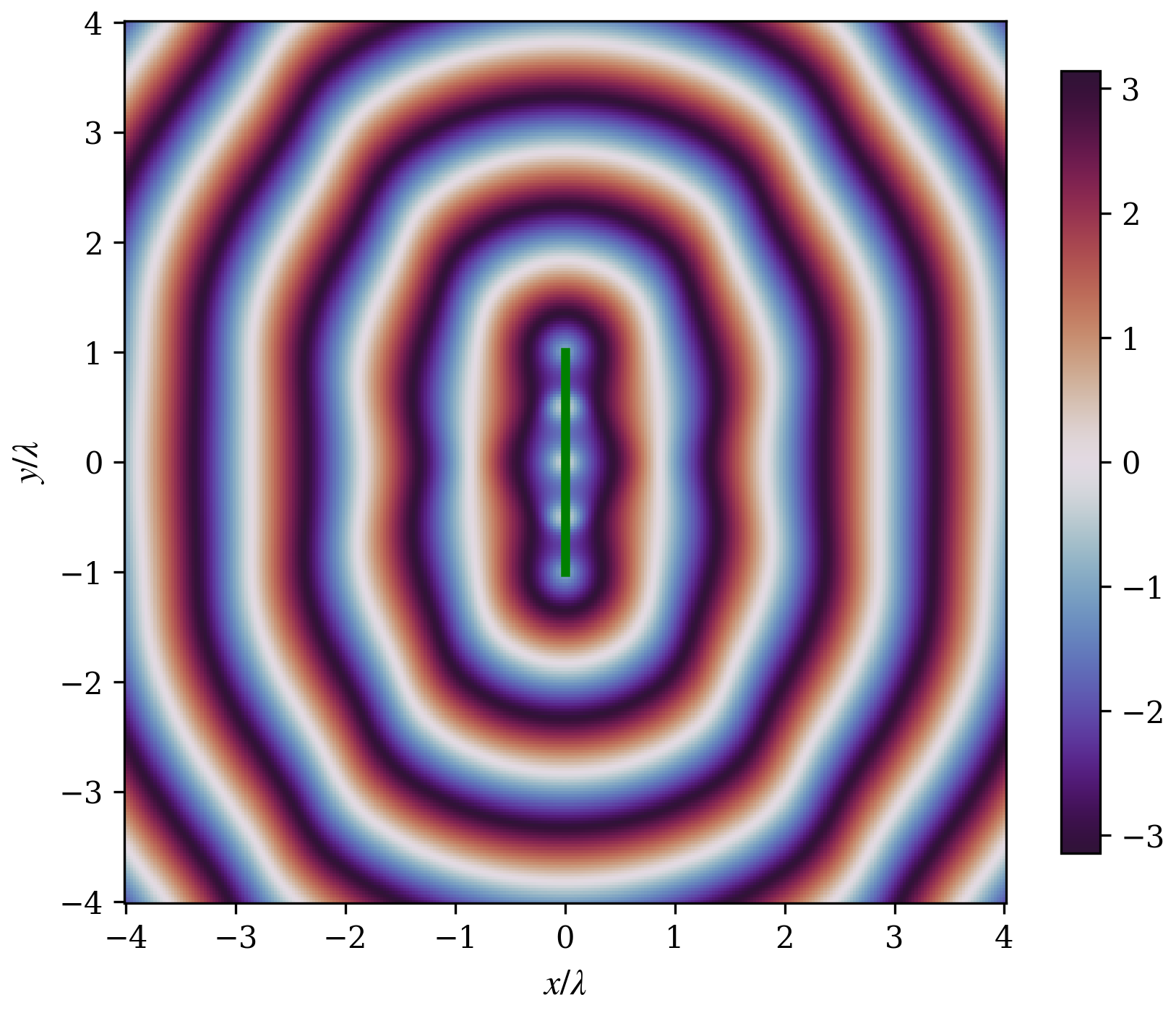} &
    \includegraphics[width=0.18\textwidth]{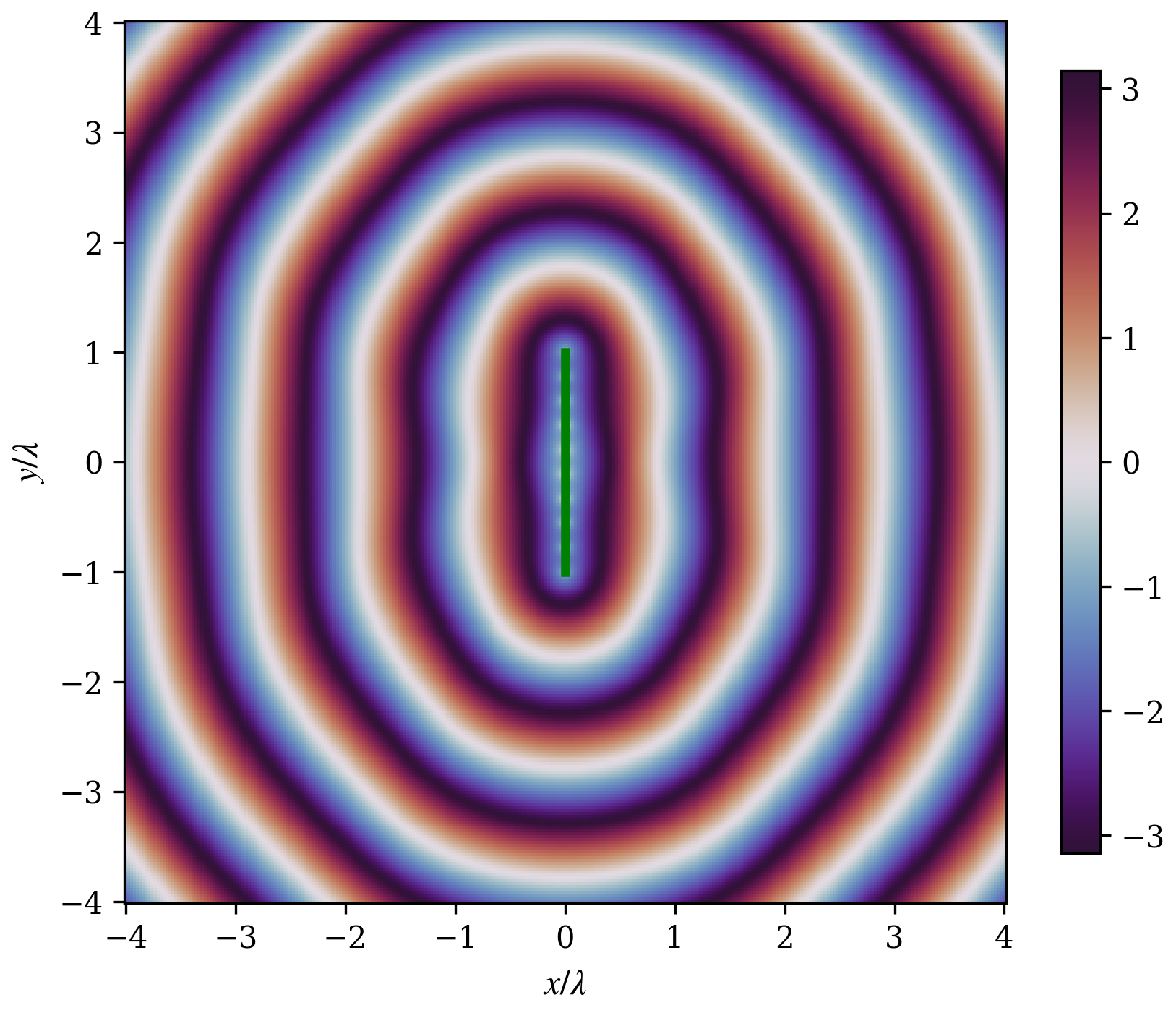} &
    \includegraphics[width=0.18\textwidth]{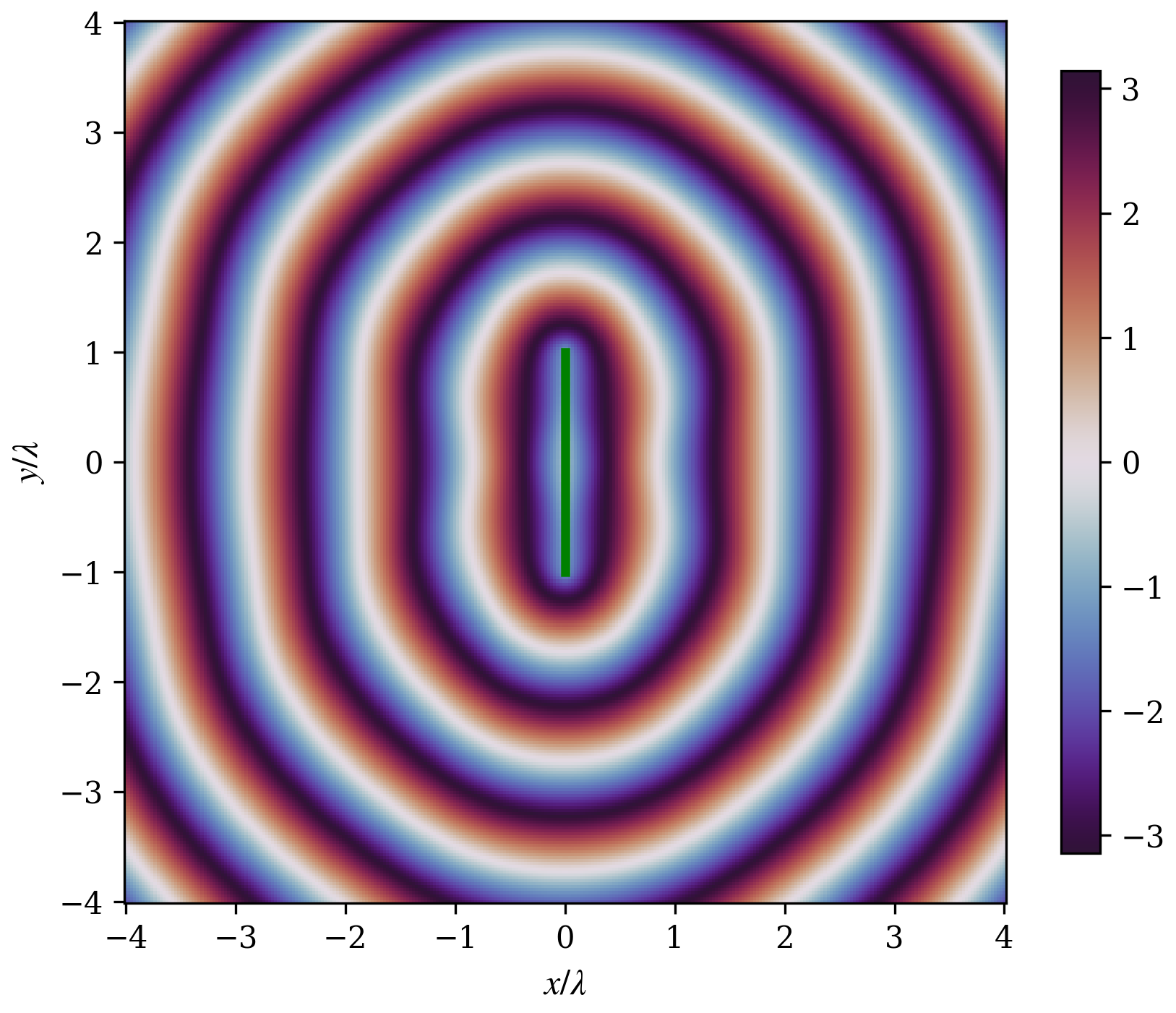} &
    \includegraphics[width=0.18\textwidth]{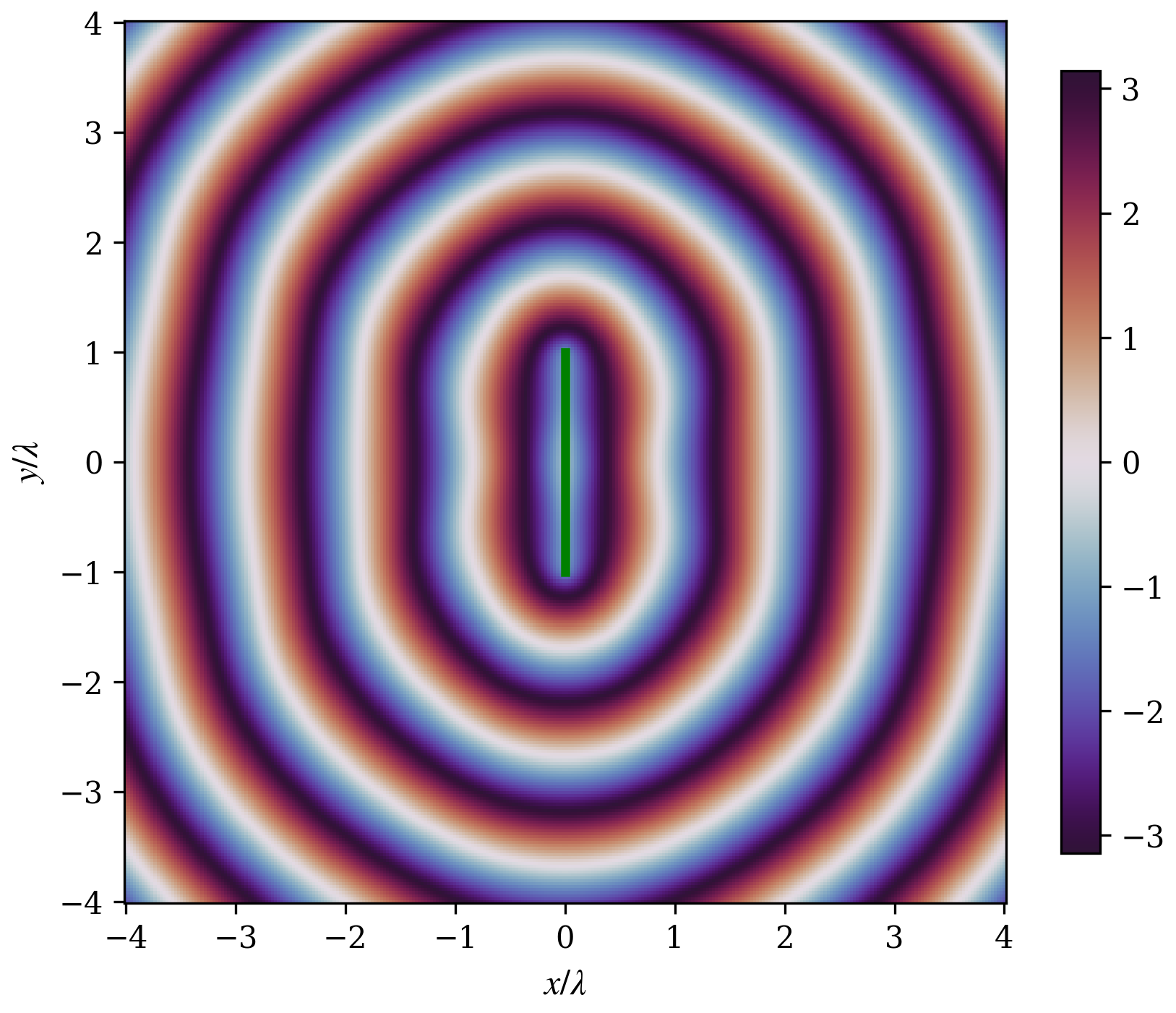} \\
\end{tabular}
\caption{Magnitude and phase of the edge diffraction kernel $\Psi_{\text{edge}}$ on the XY plane for varying numbers of secondary sources $P \in \{2, 5, 10, 25, 100\}$.}
\label{fig:edge_kernel_slices}
 \vspace{-4mm}

\end{figure*}

\section{PE-PINN based High-Fidelity Channel Reconstruction and Wireless DT Enhancement}\label{sec: method}

\subsection{Preliminary on PE-PINN}

This work builds upon our prior study on PE‑PINN~\cite{zhang2026physicsinformedneuralnetworksarchitectural}, which was proposed to overcome fundamental limitations of standard physics‑informed neural networks (PINNs) in large‑scale wave‑field reconstruction. Conventional PINNs encode physical laws only through residuals of the Helmholtz equation and boundary conditions within the loss function. While conceptually simple, this formulation suffers from spectral bias, slow convergence, and optimization instability when applied to high‑frequency wave phenomena that are central to wireless communication environments.
PE‑PINN addresses these challenges by embedding physics directly into the network architecture. Its key innovation is an envelope transformation layer that decomposes the complex electric field into a superposition of structured components,
\begin{equation}
E^{(m)}(\mathbf{x}) = w_m(\mathbf{x}) \cdot A_m(\mathbf{x}) \cdot \Psi_m(\mathbf{x}),
\end{equation}
where $\mathbf{x}$ denotes the spatial coordinate, $\Psi_m$ is a physics‑parameterized kernel (e.g., spherical or plane‑wave basis functions whose wavenumbers and centers are derived from source and material properties), $A_m$ is a smoothly varying envelope learned by the network, and $w_m$ is a spatial gating function. This factorization shifts the learning target from highly oscillatory fields to slowly varying envelopes, effectively mitigating spectral bias and accelerating convergence by more than an order of magnitude.
Beyond this architectural reformulation, PE‑PINN incorporates several physics‑aware mechanisms, including explicit separation of incident and scattered fields via dedicated subnetworks, material‑aware domain decomposition with enforced interface continuity, and Residual Adaptive Refinement (RAR) for efficient collocation sampling. These principles were validated on 2D and 3D EM wave reconstruction problems involving reflection, refraction, and diffraction, achieving high fidelity while reducing memory consumption by orders of magnitude compared with COMSOL.
To better align PE‑PINN with the requirements of communication digital twins, this work introduces targeted architectural refinements for diffraction‑ (particularly edge diffraction) and refraction‑dominated propagation scenarios, where RT approximations inherently break down.

\subsection{Edge Diffraction Modeling}\label{sec:edge_kernel}

In the PE-PINN framework, scattered field components are represented using spherical wave kernels centered at discrete point locations such as obstacle corners, which effectively capture corner diffraction. However, for 3D scattering from objects with finite-length edges, such as the cube obstacle, diffraction is not well described by isolated point sources alone. When a wave strikes a finite-length edge, each point along the edge can be treated as a secondary spherical wave source, consistent with the Huygens-Fresnel principle~\cite{peatross2015physics, Born_Wolf_Bhatia_Clemmow_Gabor_Stokes_Taylor_Wayman_Wilcock_1999}, and the total diffracted field is the coherent superposition of all these contributions.
To incorporate this physics into PE-PINN, we discretize each diffracting edge into $P$ equally spaced secondary point sources. For an edge segment along direction $\hat{\mathbf{a}}$ (e.g., $\hat{\mathbf{x}}$, $\hat{\mathbf{y}}$ or $\hat{\mathbf{z}}$) between coordinates $s_{\min}$ and $s_{\max}$, the edge diffraction kernel is defined as the coherent average over all $P$ sources:
\begin{equation}\label{eq:edge_kernel}
    \Psi_{\text{edge}}(\mathbf{x}) = \frac{1}{P} \sum_{p=1}^{P} e^{-jk_0 \, r_p(\mathbf{x})}, \quad 
    r_p(\mathbf{x}) = \|\mathbf{x} - \mathbf{x}_p^{\text{edge}}\|,
\end{equation}
where $\mathbf{x}_p^{\text{edge}}$ is the $p$-th source location uniformly sampled along the edge. Following the envelope-kernel factorization of PE-PINN, the edge diffraction component of the scattered field is:
\begin{equation}\label{eq:edge_field}
    E^{(\text{edge})}(\mathbf{x}) = w_{\text{edge}}(\mathbf{x}) \cdot A_{\text{edge}}(\mathbf{x}) \cdot \Psi_{\text{edge}}(\mathbf{x}).
\end{equation}
The kernel $\Psi_{\text{edge}}$ analytically encodes the oscillatory interference pattern produced by the coherent wavelet array, so the neural network only needs to learn the slowly varying envelope $A_{\text{edge}}(\mathbf{x})$ and the spatial gating weight $w_{\text{edge}}(\mathbf{x})$.

To illustrate the behavior of the edge diffraction kernel $\Psi_{\text{edge}}$, we visualize the diffracted field produced by a single finite-length edge illuminated by a point source. The edge is oriented along $\hat{\mathbf{x}}$ with length $2\lambda$, and the source is placed at a distance of $4\lambda$ along $\hat{\mathbf{z}}$. As shown in Fig.~\ref{fig:edge_kernel_slices} when $P=2$, the kernel exhibits a clear two-source interference pattern. As $P$ increases, the interference among the line-array sources shapes the diffraction pattern, transitioning from discrete interference fringes to continuous edge diffraction. At $P = 10$, the kernel already closely reproduces the physical edge diffraction pattern, and further increasing $P$ to 25 or 100 yields no visible difference.

\subsection{Lossy Dielectric Modeling for Refraction}

Accurate modeling of electromagnetic wave propagation in material media requires accounting for both permittivity and conductivity. The refraction scenario considered in \cite{zhang2026physicsinformedneuralnetworksarchitectural} focused exclusively on lossless dielectrics characterized by a real‑valued relative permittivity $\varepsilon_r$. In realistic communication environments, however, common materials such as concrete, glass, and wood exhibit non‑negligible electrical conductivity $\sigma$. This conductivity introduces attenuation as the electromagnetic wave propagates through the medium and therefore plays a critical role in accurately characterizing signal penetration and loss. To capture this effect, we extend the material‑aware domain decomposition framework of PE‑PINN to support lossy dielectric media.
In a medium characterized by both relative permittivity $\varepsilon_r$ and conductivity $\sigma$, the governing Helmholtz equation is given by
\begin{equation}\label{eq:helmholtz_lossy}
\nabla^2 E + k_c^2 E = 0,
\end{equation}
where $k_c$ denotes the complex wavenumber that accounts for both phase propagation and amplitude attenuation within the lossy material. Specifically,
\begin{equation}\label{eq:complex_k}
k_c^2 = \omega^2 \mu_0 \left( \varepsilon_r \varepsilon_0 - j \frac{\sigma}{\omega} \right)
= k_0^2 \varepsilon_r - j \omega \mu_0 \sigma,
\end{equation}
with $k_0$ denoting the free‑space wavenumber. The real component of $k_c^2$ governs the phase velocity of the wave inside the medium, while the imaginary component accounts for exponential attenuation of the field magnitude due to conductive loss.
By incorporating this complex‑valued wavenumber formulation into PE‑PINN, the proposed method is able to accurately reconstruct the pointwise magnitude and phase of the electric field in indoor environments containing lossy dielectric objects. This capability is essential for achieving high‑fidelity wireless DTs, particularly in practical wireless communication scenarios where material absorption significantly impacts signal coverage, penetration loss, and spatial channel characteristics.


\subsection{PE-PINN Integration in Wireless DTs}

To apply PE‑PINN to realistic indoor environments, we extract a localized computational domain from a full three‑dimensional scene model using the following procedure:
\begin{enumerate}
    \item \textbf{Hotspot Identification:} Obstacles $\mathcal{O}$ are identified where strong diffraction or transmission effects are expected, such that the high-frequency approximations underlying GO/UTD become unreliable. Two criteria are considered: (i) a \emph{diffraction criterion}, which flags  $\mathcal{O}$ with sharp edges that induce significant diffracted fields not accurately captured by GO/UTD; and (ii) a \emph{transmission criterion}, which marks penetrable obstacles (e.g., walls, partitions, furniture) whose finite thickness and material properties give rise to non-negligible transmitted fields with amplitude and phase distortions that ray-based models tend to mispredict.
    \item \textbf{Path Identification:} Starting from the transmitter location $\mathbf{r}_{\mathrm{tx}}$ and the target obstacle $\mathcal{O}$, RT is performed to identify the dominant propagation path $\mathcal{P}$ (i.e., typically the direct path) and the primary interaction point on the obstacle surface. 

    \item \textbf{Domain Construction:} A hexahedral sub-volume $\Omega$ is constructed along $\mathcal{P}$. The boundary closest to the transmitter encloses the source location, while the boundary on the far side extends beyond $\mathcal{O}$ to capture reflected, diffracted, or transmitted wave components induced by the obstacle.

    \item \textbf{Material and Boundary Assignment: } The obstacle within $\Omega$ is classified as either a dielectric body, for which the relative permittivity $\varepsilon_r$ and conductivity $\sigma$ are assigned and continuity of the electric field and its normal derivative is enforced at the interface, or a metallic object, for which the perfect electric conductor (PEC) condition $\hat{\mathbf{n}} \times \mathbf{E} = 0$ is imposed on its surface.
\end{enumerate}
Once the domain $\Omega$ and its associated material and boundary conditions are defined, PE‑PINN training is performed within this localized region. Upon convergence, the trained PE‑PINN provides a physics‑consistent solution for the electromagnetic field throughout $\Omega$. This localized high‑fidelity solution is then used to enhance the global DT generated by a RT‑based solver such as Sionna. Specifically, for receiver locations within $\Omega$, the RT prediction is augmented by replacing the dominant path contributions with the corresponding PE‑PINN results, while non‑dominant path components and predictions outside $\Omega$ are preserved unchanged. Through this selective replacement, field estimates are corrected precisely in regions where ray‑based solvers are least accurate, without incurring the computational cost of full‑wave simulation over the entire scene. The final wireless channel response is obtained by coherently combining the corrected dominant path with all remaining reflected path components.

\begin{table*}[t]
    \centering    
    \caption{Free-space comparison among analytical spherical wave solution, Sionna, and PE-PINN at selected receiver locations. $\Delta|E|$ and $\Delta\phi$ denote the PE-PINN normalized magnitude and phase errors relative to the analytical solution.}
    \label{tab:Freespace_comparison}
    \begin{tabular}{cccccccccc}
    \toprule
    \multirow{2}{*}{\textbf{Rx position}} &\multirow{2}{*}{\textbf{Distance to Source}} & \multicolumn{3}{c}{\textbf{Phase} ($^\circ$)} & \multicolumn{3}{c}{\textbf{Normalized Magnitude}} & \multicolumn{2}{c}{\textbf{PE-PINN Error}} \\
     \cmidrule(lr){3-5}  \cmidrule(lr){6-8}  \cmidrule(lr){9-10}
     && \textbf{Analytical} & \textbf{Sionna} & \textbf{PE-PINN} & \textbf{Analytical} & \textbf{Sionna} & \textbf{PE-PINN} & $\Delta|E|$ & $\Delta\phi$ ($^\circ$) \\
    \midrule
    (-0.9, 0, 0)        & 1.10 & 69.81  & 69.81  & 69.11  & 0.2273 & 0.2273 & 0.2314 & 0.0041 & -0.70 \\
    (-0.5, 0, 0)        & 1.50 & -2.99  & -2.99  & -4.06  & 0.1667 & 0.1667 & 0.1771 & 0.0104 & -1.07 \\
    (0.5, 0, 0)         & 2.50 & -4.98  & -4.98  & -6.97  & 0.1000 & 0.1000 & 0.1066 & 0.0066 & -1.99 \\
    (-1, -1, -1)        & 1.73 & 48.24  & 48.24  & 48.82  & 0.1443 & 0.1443 & 0.1411 & -0.0032 & 0.58 \\
    (-1, -2.5, -2.5)    & 3.67 & -149.12 & -149.12 & -146.48 & 0.0680 & 0.0680 & 0.0599 & -0.0081 & 2.64 \\
    (-1.5, -2.5, -2.5)  & 3.57 & 149.22 & 149.22 & 151.80 & 0.0700 & 0.0700 & 0.0623 & -0.0077 & 2.58 \\
    (-1.5, -1, -1)      & 1.50 & -2.99  & -2.99  & -2.17  & 0.1667 & 0.1667 & 0.1584 & -0.0083 & 0.82 \\
    (-1.75, 0, 0)       & 0.25 & -0.50  & -0.50  & -0.40  & 1.0000 & 1.0000 & 1.0000 & 0.0000 & 0.10 \\
    (0, 0, 0)           & 2.00 & -3.99  & -3.99  & -5.60  & 0.1250 & 0.1250 & 0.1318 & 0.0068 & -1.61 \\
    (1, 1, 1)           & 3.32 & 161.51 & 161.51 & 161.77 & 0.0754 & 0.0754 & 0.0784 & 0.0030 & 0.26 \\
    (1.5, 0, 0)         & 3.50 & -6.98  & -6.98  & -8.58  & 0.0714 & 0.0714 & 0.0789 & 0.0075 & -1.60 \\
    (2.5, 2.5, 0)       & 5.15 & -75.97 & -75.97 & -73.43 & 0.0486 & 0.0486 & 0.0522 & 0.0036 & 2.54 \\
    (2.5, 2.5, 2.5)     & 5.72 & 67.04  & 67.04  & 66.92  & 0.0437 & 0.0437 & 0.0495 & 0.0058 & -0.12 \\
    \bottomrule
    \end{tabular}
\end{table*}

\section{Evaluation Results}\label{sec: evaluation}

\subsection{Evaluation Settings}

All simulations are conducted at an operating frequency of 2.4 GHz. The evaluation is organized into two parts.

The first part revisits the controlled scenarios introduced in Sec.~\ref{sec: prelim}, namely diffraction and refraction within the same $2 \times 0.5 \times 0.5$ m computational domain and simulation configuration. In these scenarios, results from COMSOL, Sionna, and PE‑PINN are compared side by side. In addition, a free-space baseline is evaluated in a larger $5 \times 5 \times 5$ m domain, with the source located at $(-2,0,0)$ m. In this case, PE‑PINN and Sionna are compared against the analytical spherical‑wave solution, which serves as the ground truth, as it provides the exact solution in the absence of obstacles. For consistency with COMSOL, the PE‑PINN source is configured such that the electric field amplitude is normalized to unity on a sphere of radius equal to one wavelength centered at the source location. To ensure a fair three‑way comparison, the field distributions generated by COMSOL, Sionna, and PE‑PINN are all normalized by their respective maximum values at the source location prior to computing the error metrics.

The second part of the evaluation applies the PE‑PINN framework to the University of Wisconsin-Madison (UW–Madison) wireless laboratory DT, where diffraction and refraction tunnels are constructed around real objects in the environment. Due to memory limitations in COMSOL, full‑wave ground‑truth simulations are not feasible for these large‑scale scenarios. Consequently, the comparison in this setting is limited to PE‑PINN and Sionna, and the reported metrics characterize the discrepancy between the two approaches.

All simulations were conducted on two separate workstations. The COMSOL Multiphysics (v6.3) simulations were executed on a workstation equipped with a dual‑socket Intel Xeon E5‑2630 v3 processor (2.40 GHz) and 256 GB of RAM. The Sionna (v1.2.1) RT simulations, as well as PE‑PINN training and inference, were performed on a second workstation featuring an Intel Core i9‑13900K processor, an NVIDIA GeForce RTX 4090 GPU with 24 GB of VRAM, and 128 GB of RAM. The PE‑PINN model was implemented in PyTorch and trained on the RTX 4090 GPU.

\subsection{Evaluation on PE-PINN based Channel Simulation}

We first evaluate the performance of PE-PINN on channel simulation, with a focus on free space, diffraction, and refraction, respectively.

\textbf{Free Space Case: } The PE-PINN and Sionna based channel simulation results are evaluated at a set of locations distributed throughout the domain, and compared against the analytical spherical wave solution. Since the reference is the analytical spherical wave solution, the magnitude is computed directly from the real and imaginary parts of the electric field as $|E| = \sqrt{\text{Re}(E)^2 + \text{Im}(E)^2}$, rather than converting to received power in dBm. The magnitudes from all three are independently normalized by their respective maximum value among the selected evaluation points. Table~\ref{tab:Freespace_comparison} presents the comparison at thirteen locations spanning various distances and directions from the source. Sionna produces results that are identical to the analytical solution in both phase and normalized magnitude at all evaluation points, which shows its ray-based engine is exact in free-space propagation where no scattering or transmission is involved. The PE-PINN predictions exhibit small deviations from the analytical solution. These minor discrepancies are expected due to the stochastic nature of neural network training and are well within acceptable bounds in practice.

\begin{figure*}[t]
    \centering
    \subfigure{\includegraphics[width=.9\linewidth]{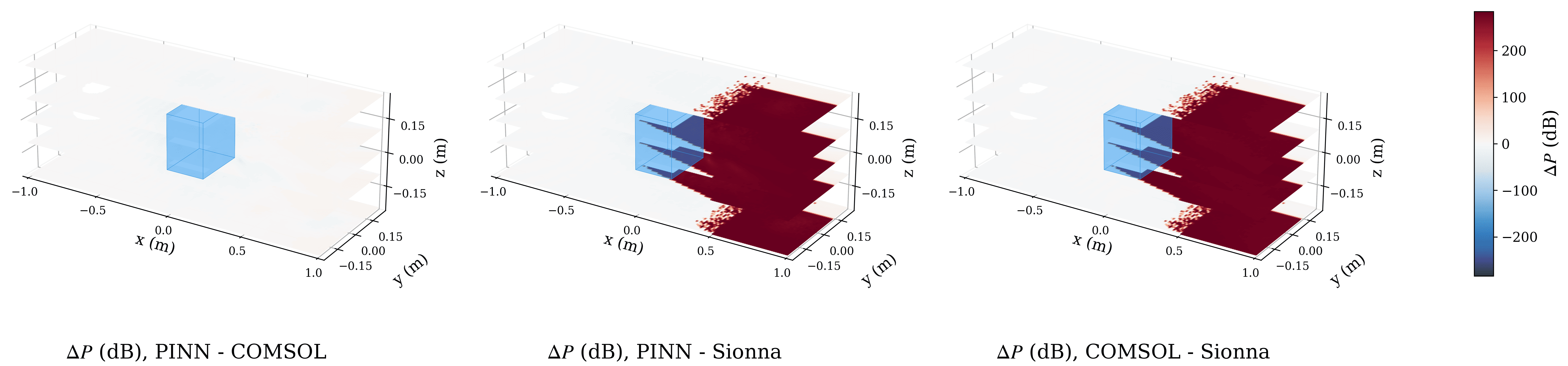}}\\
    \subfigure{\includegraphics[width=.9\linewidth]{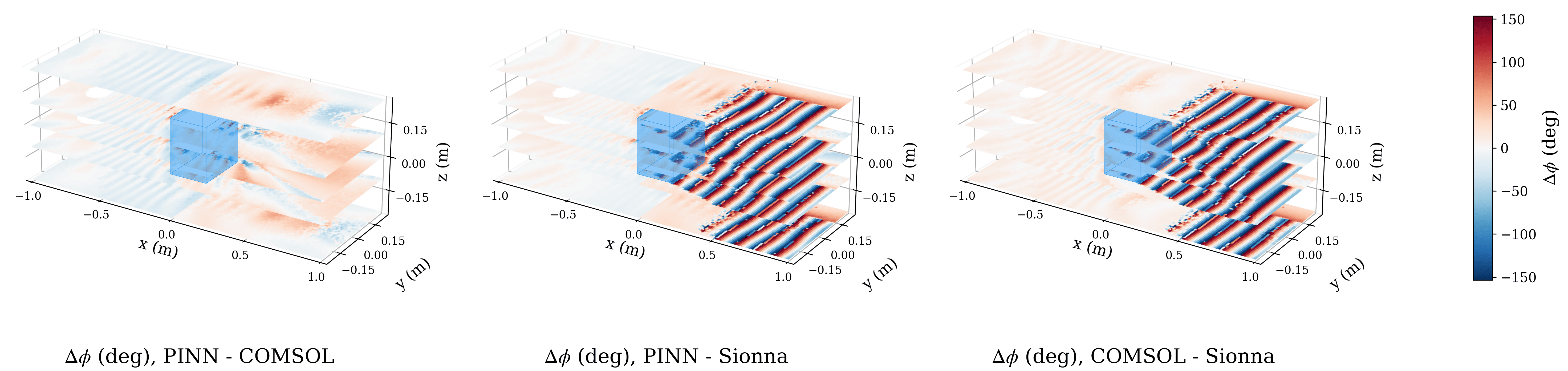}}
    \caption{Difference between different solvers on diffraction scenario given a two-wavelength cube object.}
    \label{fig:eval_diffra_diff}
     \vspace{-4mm}

\end{figure*}

\begin{figure*}[ht]
    \centering
    \subfigure{\includegraphics[width=.9\linewidth]{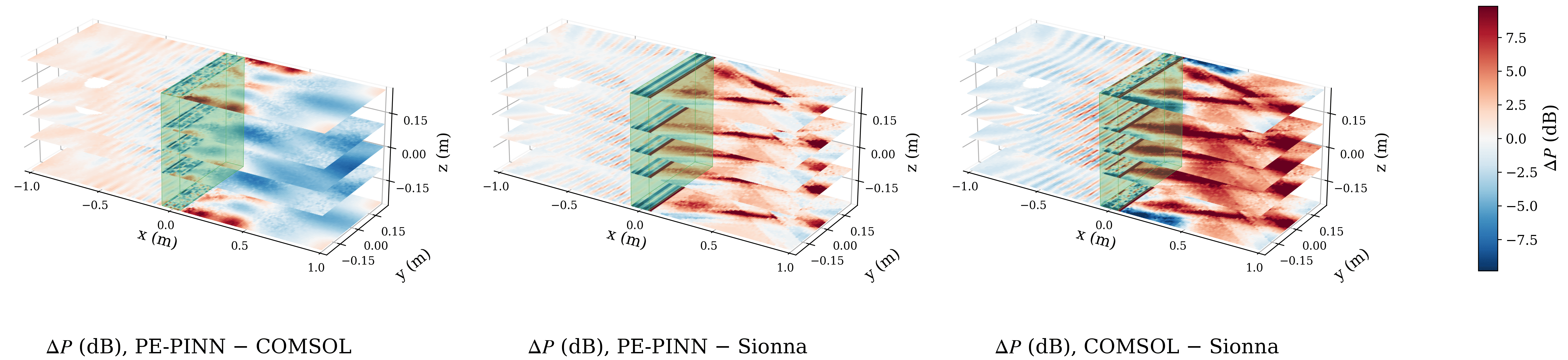}}\\
    \vspace{-3mm}
    \subfigure{\includegraphics[width=.9\linewidth]{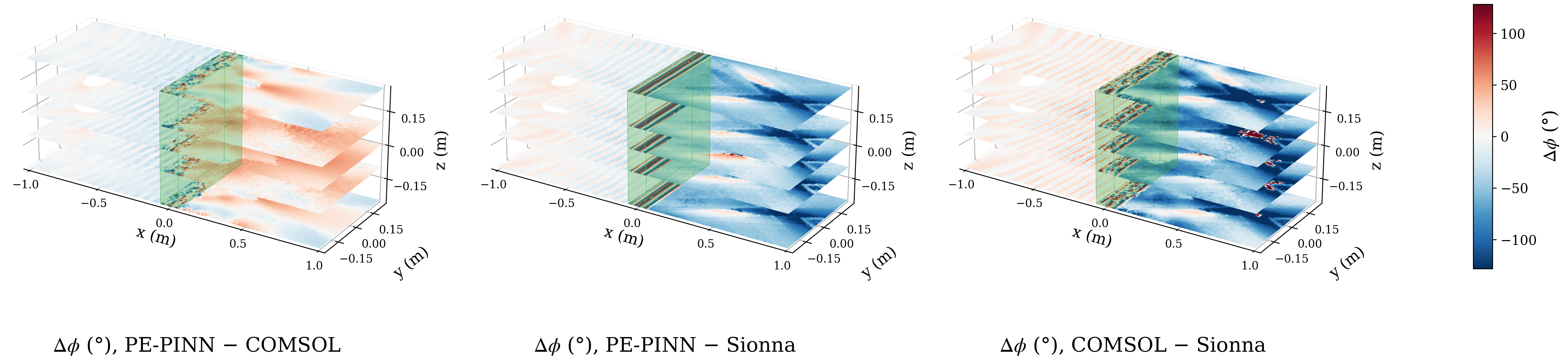}}
    \caption{Difference between different solvers on refraction scenario given a one-wavelength strip.}
    \label{fig:eval_refra_diff}
     \vspace{-4mm}
\end{figure*}

\textbf{Diffraction Scenes: }Table~\ref{tab:eval_comparison} summarizes the three-way comparison among PE-PINN, COMSOL, and Sionna for both the diffraction and refraction scenarios. Figs.~\ref{fig:eval_diffra_diff} and ~\ref{fig:eval_refra_diff} present the corresponding difference heatmaps in magnitude (top row) and phase (bottom row) for each pair of solvers. For the controlled diffraction scenario, the PE-PINN is configured as follows. The source network $\mathcal{N}_{\text{src}}$ models the incident field using the spherical wave kernel centered at $\mathbf{x}_{\text{src}} = (-0.75,\, 0,\, 0)$. The scatter network $\mathcal{N}_{\text{sct}}$ employs a multi-kernel structure tailored to the cube geometry. A reflection kernel is placed at the image source location $\mathbf{x}_{\text{ref}} = (0.5,\, 0,\, 0)$, mirrored across the illuminated front face ($x = -0.125$~m). A source kernel $\Psi_{\mathbf{x}_{\text{src}}}$ is retained in $\mathcal{N}_{\text{sct}}$ to account for the shadowing effect behind the cube. Four edge diffraction kernels are constructed along the four edges of the front face, each discretized with $P = 10$ secondary sources: $\mathbf{e}_{1}$--$\mathbf{e}_{2}$ along the two vertical edges (parallel to $z$) at $(x,y) = (-0.125,\, \pm 0.125)$, and $\mathbf{e}_{3}$--$\mathbf{e}_{4}$ along the two horizontal edges (parallel to $y$) at $(x,z) = (-0.125,\, \pm 0.125)$. On all six faces and four front edges of the cube, the PEC boundary condition $E^{\text{src}} + E^{\text{sct}} = 0$ is enforced. The total field is then expressed as:
\begin{equation}
    E(\mathbf{x}) \approx \mathcal{N}_{\text{src}}(\mathbf{x}) 
    + \mathcal{N}_{\text{sct}}\!\left(\mathbf{x};\, 
    \Psi_{\mathbf{x}_{\text{src}}},\, 
    \Psi_{\mathbf{x}_{\text{ref}}},\, 
    \{\Psi_{\mathbf{e}_{j}}\}_{j=1}^{4}\right).
\end{equation}
As shown in Fig.~\ref{fig:eval_diffra_diff} and Table~\ref{tab:eval_comparison}, the PE-PINN solution closely matches the COMSOL reference, with a magnitude MAE of 3.00~dB (median 1.75~dB) and a phase MAE of $14.84^\circ$ (median $10.04^\circ$). The difference heatmap in the left column of Fig.~\ref{fig:eval_diffra_diff} confirms that the PE-PINN$-$COMSOL discrepancy remains small throughout the domain. In contrast, both PE-PINN$-$Sionna and COMSOL$-$Sionna (middle and right columns) exhibit nearly identical error patterns, with large discrepancies concentrated in the shadow region behind the cube. The visual similarity between the middle and right columns confirms that PE-PINN aligns with COMSOL.

\begin{table}[t]
\centering
\caption{Performance comparison (difference) for diffraction and refraction scenarios.}
\label{tab:eval_comparison}
\scriptsize
\setlength{\tabcolsep}{3.5pt}
\renewcommand{\arraystretch}{1.1}
\begin{tabular}{ll|ccc|ccc}
\toprule
\textbf{} & \textbf{Comparison} 
& \multicolumn{3}{c|}{\textbf{Mag. (dB)}} 
& \multicolumn{3}{c}{\textbf{Phase ($^\circ$)}} \\
\cline{3-8}
 &  & MAE & RMSE & Med. & MAE & RMSE & Med. \\
\midrule
\multirow{3}{*}{Diff.}
& PE-PINN--COMSOL & 3.00 & 4.51 & 1.75 & 14.84 & 22.17 & 10.04 \\
& PE-PINN--Sionna & 89.47 & 156.64 & 3.82 & 36.29 & 59.70 & 12.39 \\
& COMSOL--Sionna  & 88.45 & 156.36 & 1.92 & 35.14 & 58.81 & 11.45 \\
\midrule
\multirow{3}{*}{Refr.}
& PE-PINN--COMSOL & 2.19 & 2.92 & 1.50 & 19.14 & 27.86 & 12.73 \\
& PE-PINN--Sionna & 4.91 & 40.10 & 1.30 & 33.39 & 47.76 & 23.00 \\
& COMSOL--Sionna  & 6.43 & 40.42 & 2.47 & 42.74 & 60.74 & 25.73 \\
\bottomrule
\end{tabular}
\vspace{-2mm}
\end{table}

\textbf{3D Refraction Scenes: }For the controlled refraction scenario, the PE-PINN is configured as follows. The domain is partitioned into three regions along the $x$-axis based on the dielectric strip boundaries at $x_L = -0.0625$~m and $x_R = 0.0625$~m. The source network $\mathcal{N}_{\text{src}}$ models the incident field using the spherical wave kernel centered at $\mathbf{x}_{\text{src}} = (-0.75,\, 0,\, 0)$. A separate scatter network is assigned to each region. On the source side (Region~1, $x < x_L$), two reflection kernels at $\mathbf{x}_{\text{ref},1} = (0.625,\, 0,\, 0)$ and $\mathbf{x}_{\text{ref},2} = (0.8125,\, 0,\, 0)$ model the reflected field from each dielectric interface. Inside the strip (Region~2, $x_L \le x \le x_R$), two spherical wave kernels at $\mathbf{x}_{\text{m},1} = (-2.125,\, 0,\, 0)$ and $\mathbf{x}_{\text{m},2} = (2.25,\, 0,\, 0)$ operate at the dielectric wavenumber $k_{\text{in}} = 3k_0$, together with two planewave kernels propagating in the $\pm x$ directions with learnable weights, capturing both forward and backward waves within the strip. On the transmission side (Region~3, $x > x_R$), a single kernel at $\mathbf{x}_{\text{t},1} = (-0.67,\, 0,\, 0)$ represents the transmitted field. Continuity of $E$ and its normal derivative $\partial E / \partial x$ is enforced at both interfaces. The total field is:
\begin{equation}
    E(\mathbf{x}) \approx 
    \begin{cases} 
    \mathcal{N}_{\text{src}} + \mathcal{N}_{\text{sct}}^{(1)}(\Psi_{\mathbf{x}_{\text{ref},1}},\, 
    \Psi_{\mathbf{x}_{\text{ref},2}}), 
    & x < x_L \\[6pt]
    \mathcal{N}_{\text{sct}}^{(2)}(\Psi_{\mathbf{x}_{\text{m},1}},\, 
    \Psi_{\mathbf{x}_{\text{m},2}},\, 
    \Psi_{\text{pw}}^{\pm}), 
    & x_L \le x \le x_R \\[6pt]
    \mathcal{N}_{\text{sct}}^{(3)}(\Psi_{\mathbf{x}_{\text{t},1}}), 
    & x > x_R
    \end{cases}
\end{equation}
where $\Psi_{\mathbf{x}}$ denotes the spherical wave kernel centered at $\mathbf{x}$, and $\Psi_{\text{pw}}^{\pm}$ are plane-wave kernels propagating in the $\pm x$ direction.
As shown in Fig.~\ref{fig:eval_refra_diff} and Table~\ref{tab:eval_comparison}, the PE-PINN$-$COMSOL difference (left column) exhibits moderate deviations primarily near the dielectric interfaces, with a magnitude MAE of 2.19~dB (median 1.50~dB) and a phase MAE of $19.14^\circ$ (median $12.73^\circ$).  The PE-PINN$-$Sionna and COMSOL$-$Sionna heatmaps (middle and right columns) again share a consistent spatial pattern, both exhibiting the alternating +/- signature on the strip, arising from the standing-wave interference that ray-based solvers misinterpreted, along with systematic deviations on the transmission side. The consistently low PE-PINN$-$COMSOL errors across both scenarios validate that results from PE-PINN are faithful to full wave physics.

\subsection{Wireless DT Enhancement}

\begin{figure}[ht!]
    \centering
    \subfigure[Photo of the lab.]{\includegraphics[width=.48\linewidth]{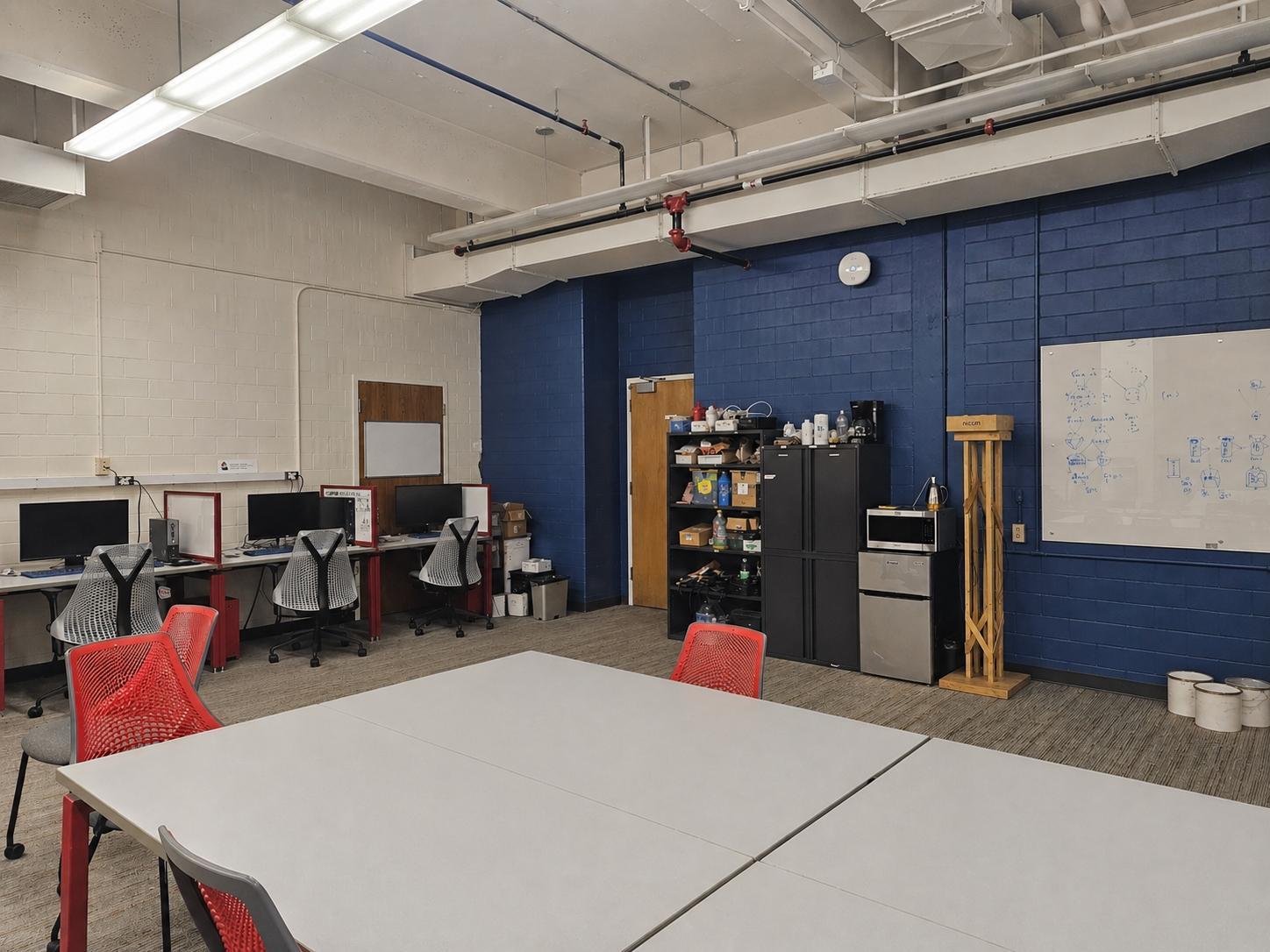}}%
    \hfill
       \subfigure[3D modeling of the lab.] {\includegraphics[width=.48\linewidth]{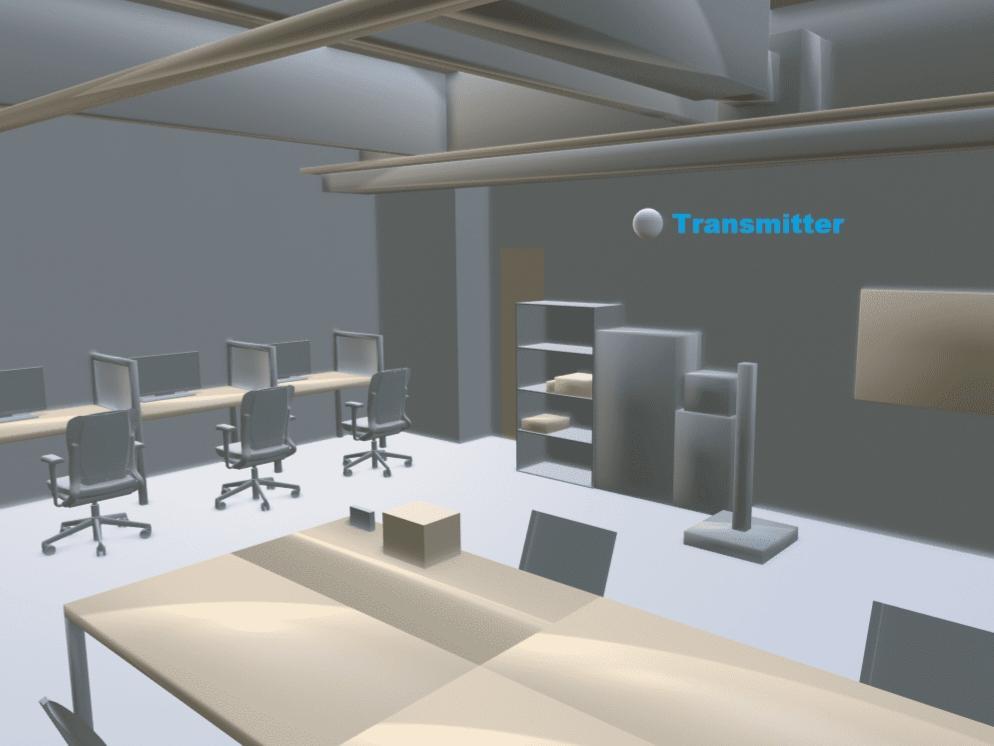}}
    \caption{Testing environment at UW-Madison.}
    \label{fig:lab}
    \vspace{-4mm}
\end{figure}

The wireless laboratory at UW-Madison (Fig.~\ref{fig:lab}) is used as the target environment for demonstrating wireless DT construction. A detailed 3D model of the laboratory is generated using the tool in~\cite{xu2026llmsupported3dmodelingtool}. The simulation domain spans
$x \in [-5.78, 5.75]$~m,
$y \in [-3.88, 3.93]$~m, and
$z \in [0.00, 3.80]$~m,
with the transmitter located at $(2.0,-3.2,2.3)$~m.
To enhance digital‑twin fidelity, Sionna’s predictions are selectively augmented using PE‑PINN solutions in regions dominated by diffraction or refraction. Localized computational tunnels are first identified along dominant propagation paths where wave–object interactions are significant. A PE‑PINN model is then constructed within each tunnel to compute a physics‑consistent full‑wave solution.
The PE‑PINN output can either replace only the dominant path contribution while preserving Sionna’s remaining multipath components, or substitute the entire field prediction within the tunnel. In practice, correcting only the dominant path is often sufficient, as it typically carries the majority of the received power.
To demonstrate this approach, one diffraction scenario and one refraction scenario are selected from the laboratory environment. For fair comparison, the corresponding Sionna simulations are performed with only the isolated scatterer present in an otherwise empty tunnel, matching the single‑object geometry assumed by PE‑PINN. For the diffraction case, PE‑PINN is further evaluated across tunnels of varying sizes to assess robustness and scalability. Finally, we demonstrate seamless integration of PE‑PINN into the complete RT‑based DT.

\begin{figure*}[hb]
    \centering
    \subfigure[small tunnel]{\includegraphics[width=0.31\linewidth]{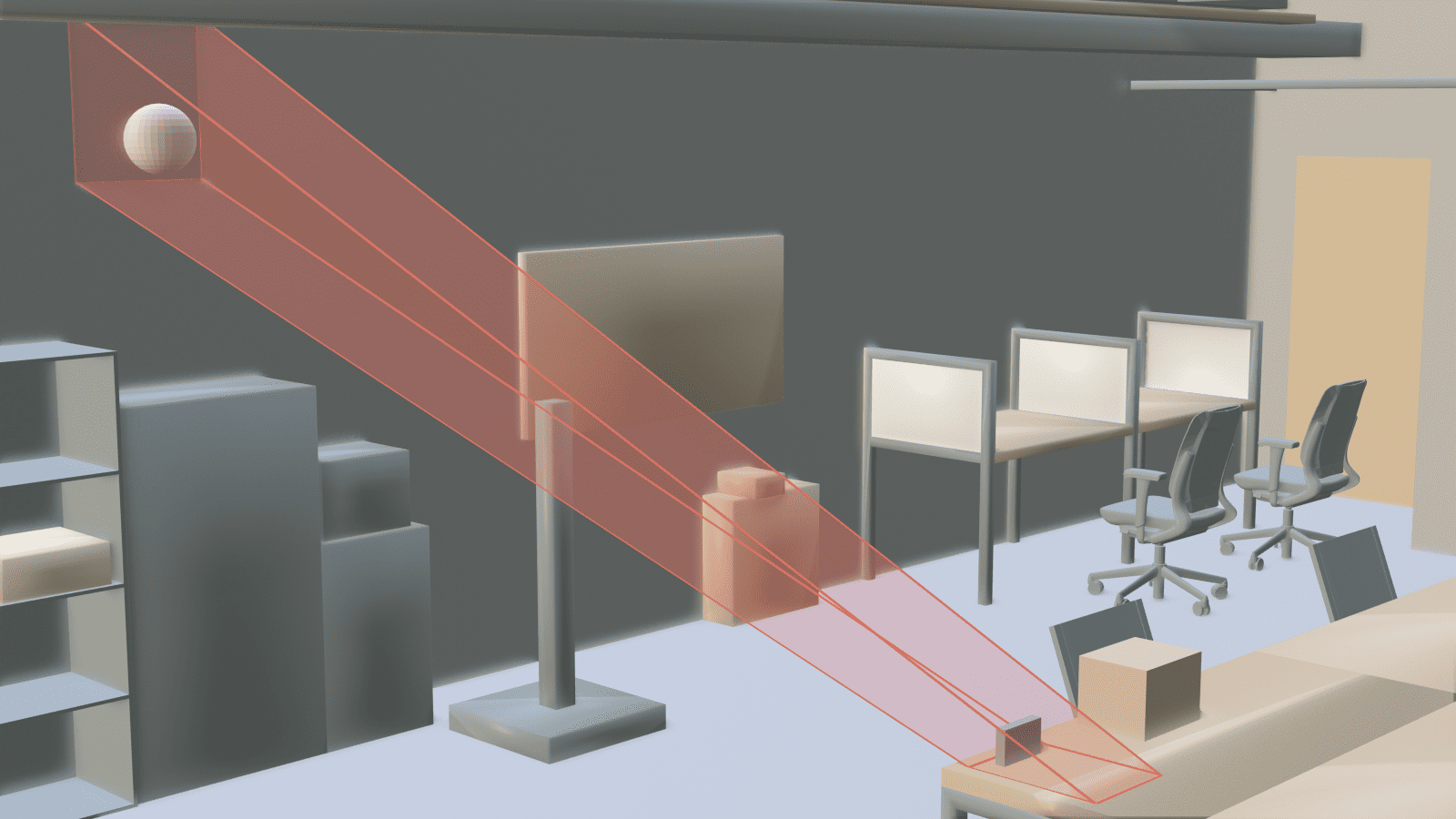}}
    \subfigure[wide tunnel]{\includegraphics[width=0.31\linewidth]{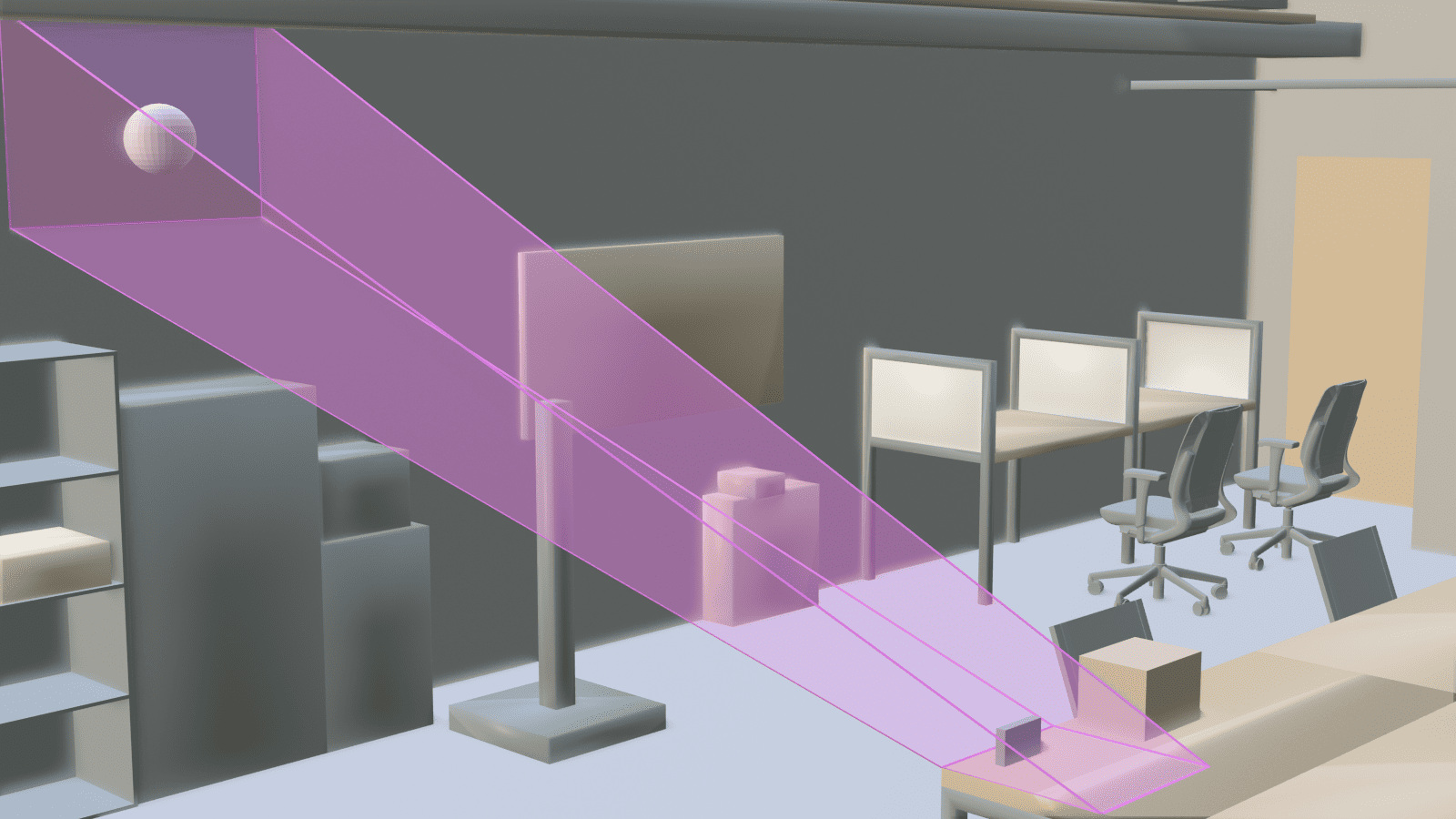}}
    \subfigure[tunnels]{\includegraphics[width=0.31\linewidth]{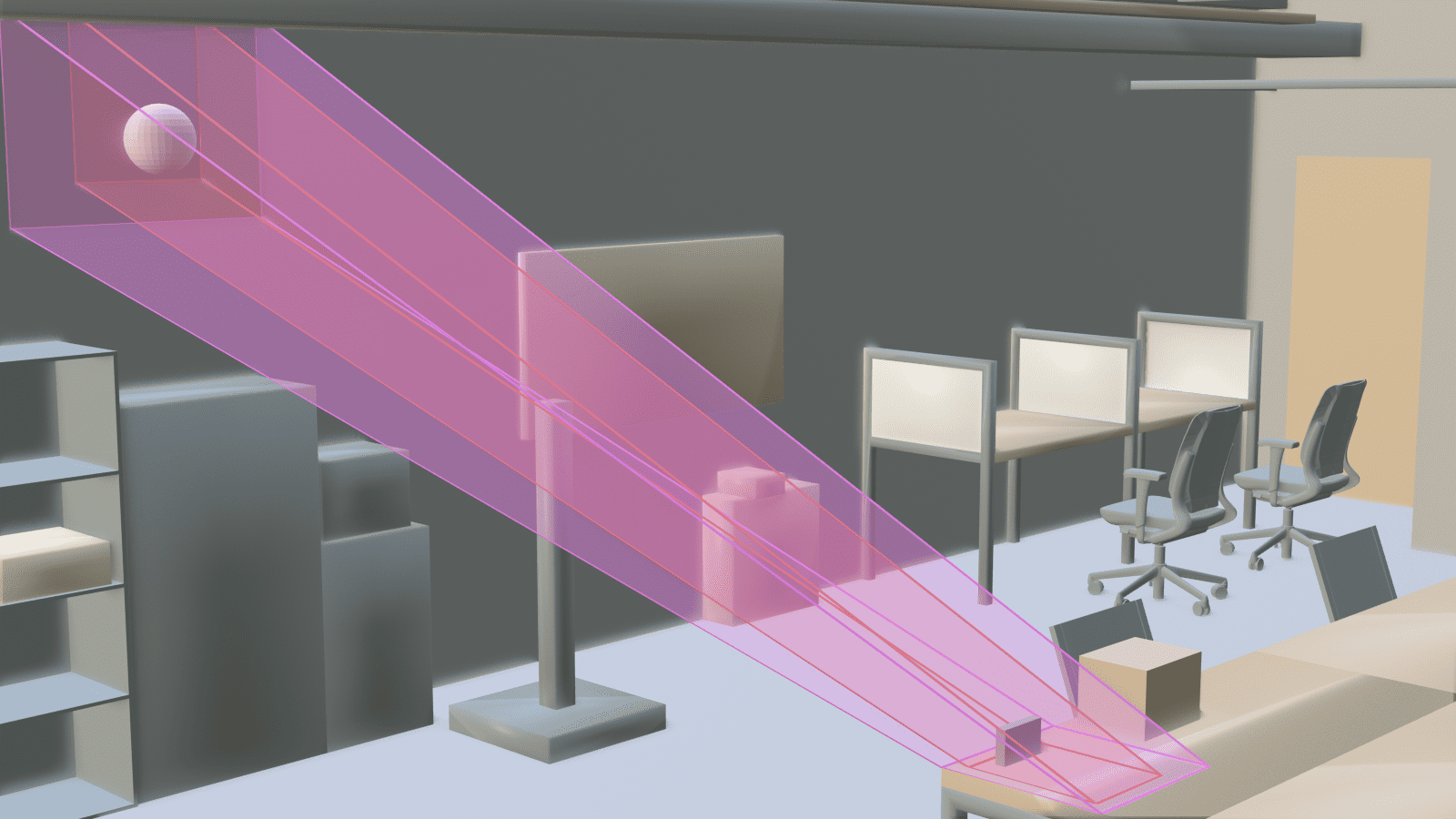}}
    \caption{Diffraction tunnels constructed in the lab environment.}
    \label{fig:diffra_tunnels}
    \vspace{-4mm}
\end{figure*}

\subsubsection{Diffraction Scenario}\label{sec:eval_diffra}

\begin{figure*}[t]
    \centering
    \makebox[0.49\linewidth]{\includegraphics[width=0.3\textwidth]{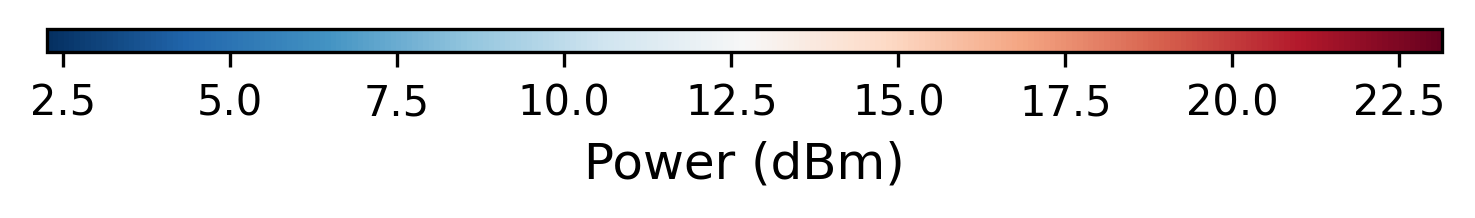}}%
    \makebox[0.49\linewidth]{\includegraphics[width=0.3\textwidth]{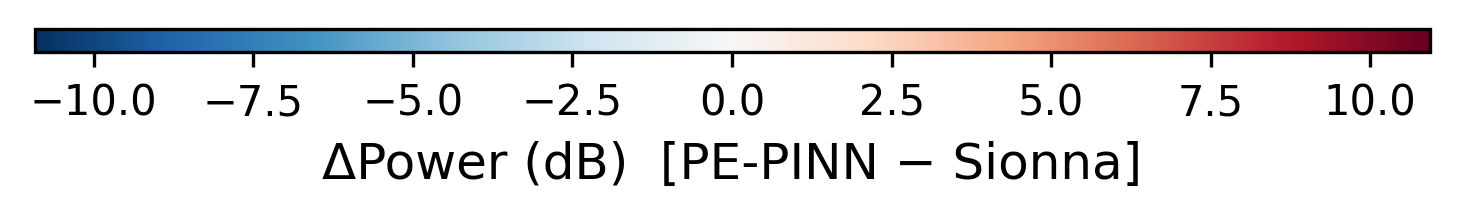}}\\[4pt]
    \subfigure[$P$, PE-PINN]{\includegraphics[width = 0.24\linewidth]{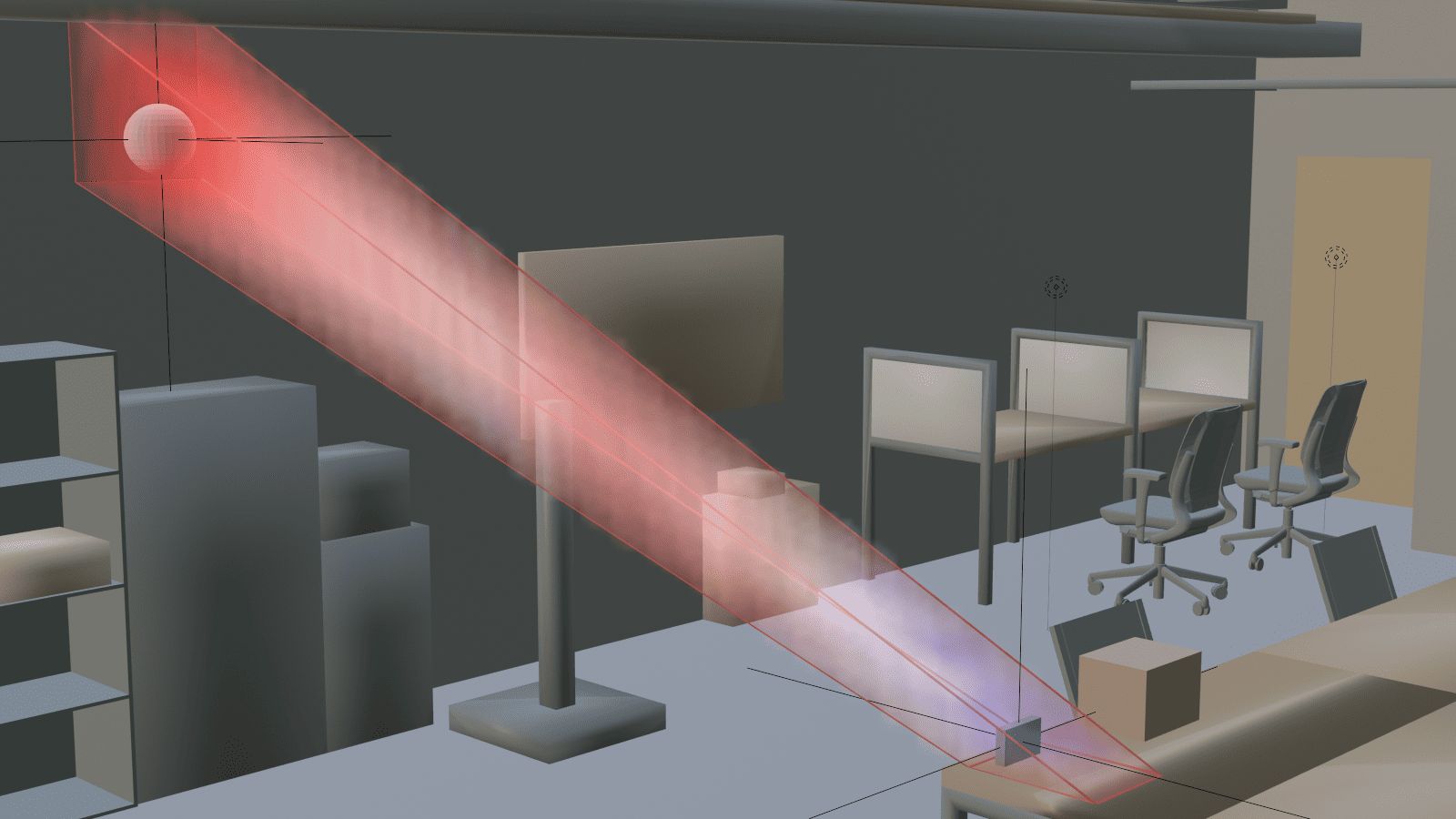}}
    \subfigure[$P$, SIONNA]{\includegraphics[width = 0.24\linewidth]{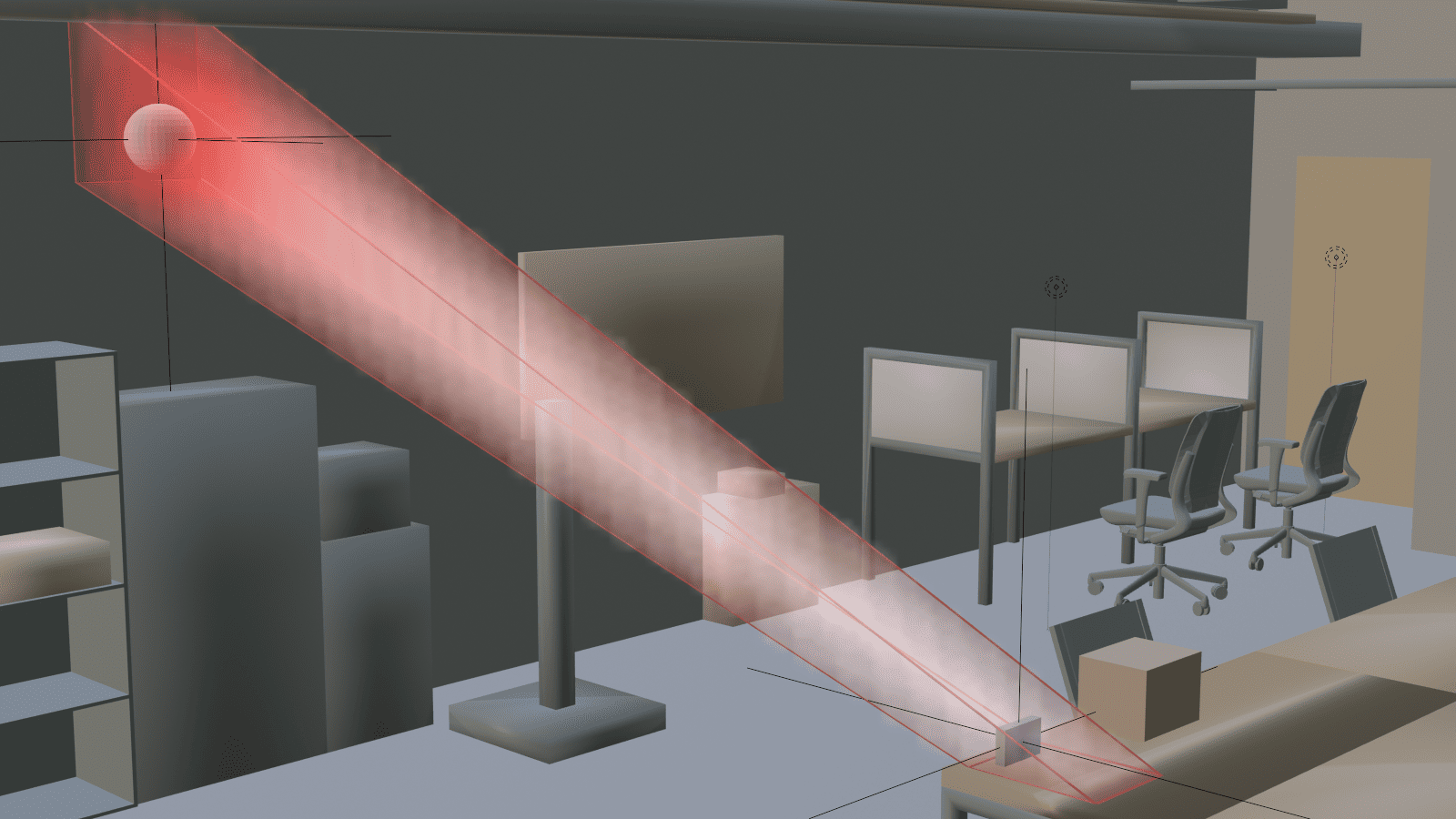}}
    \subfigure[$\Delta P$, side view]{\includegraphics[width = 0.24\linewidth]{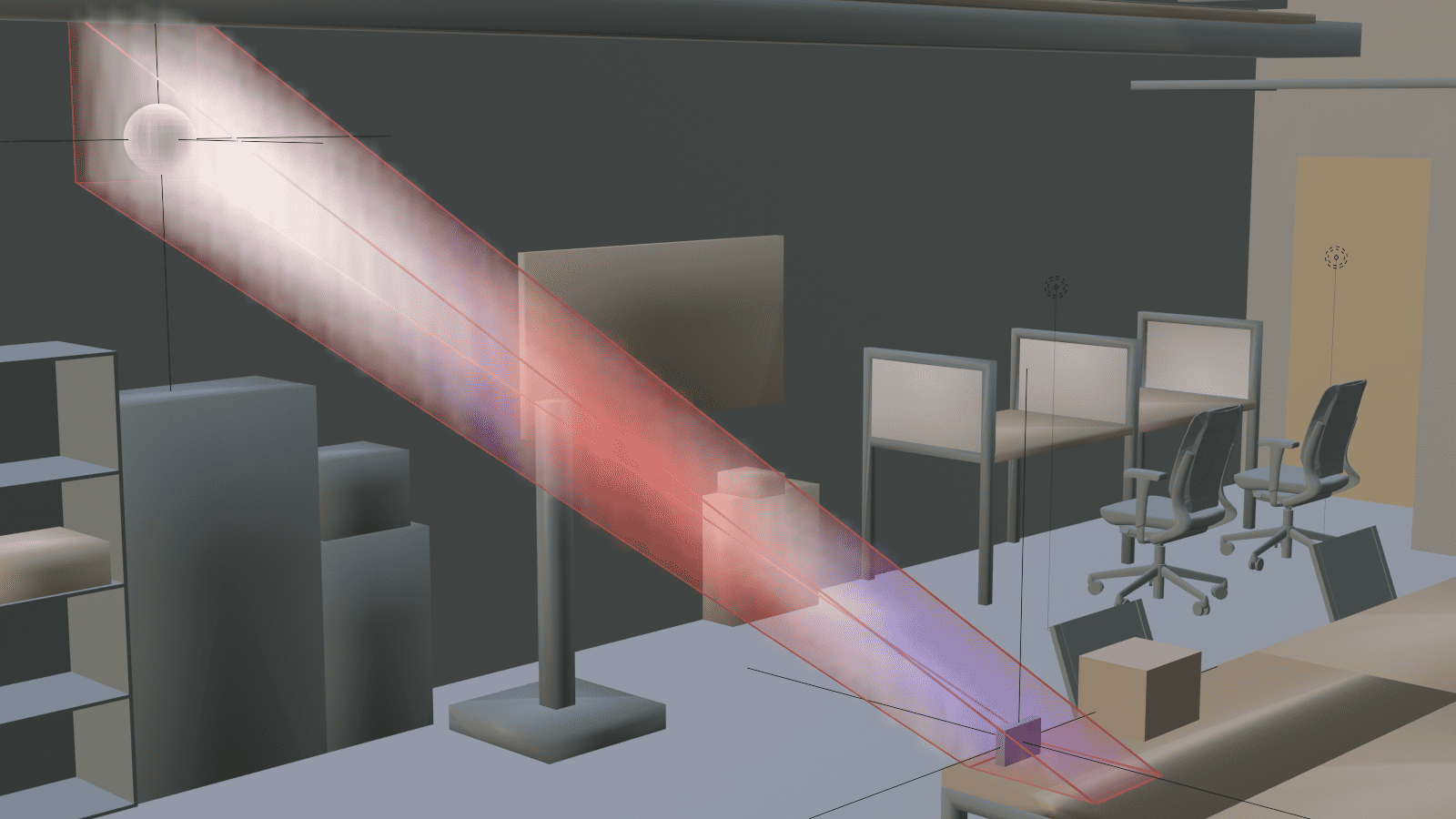}}
    \subfigure[$\Delta P$, top view]{\includegraphics[width = 0.24\linewidth]{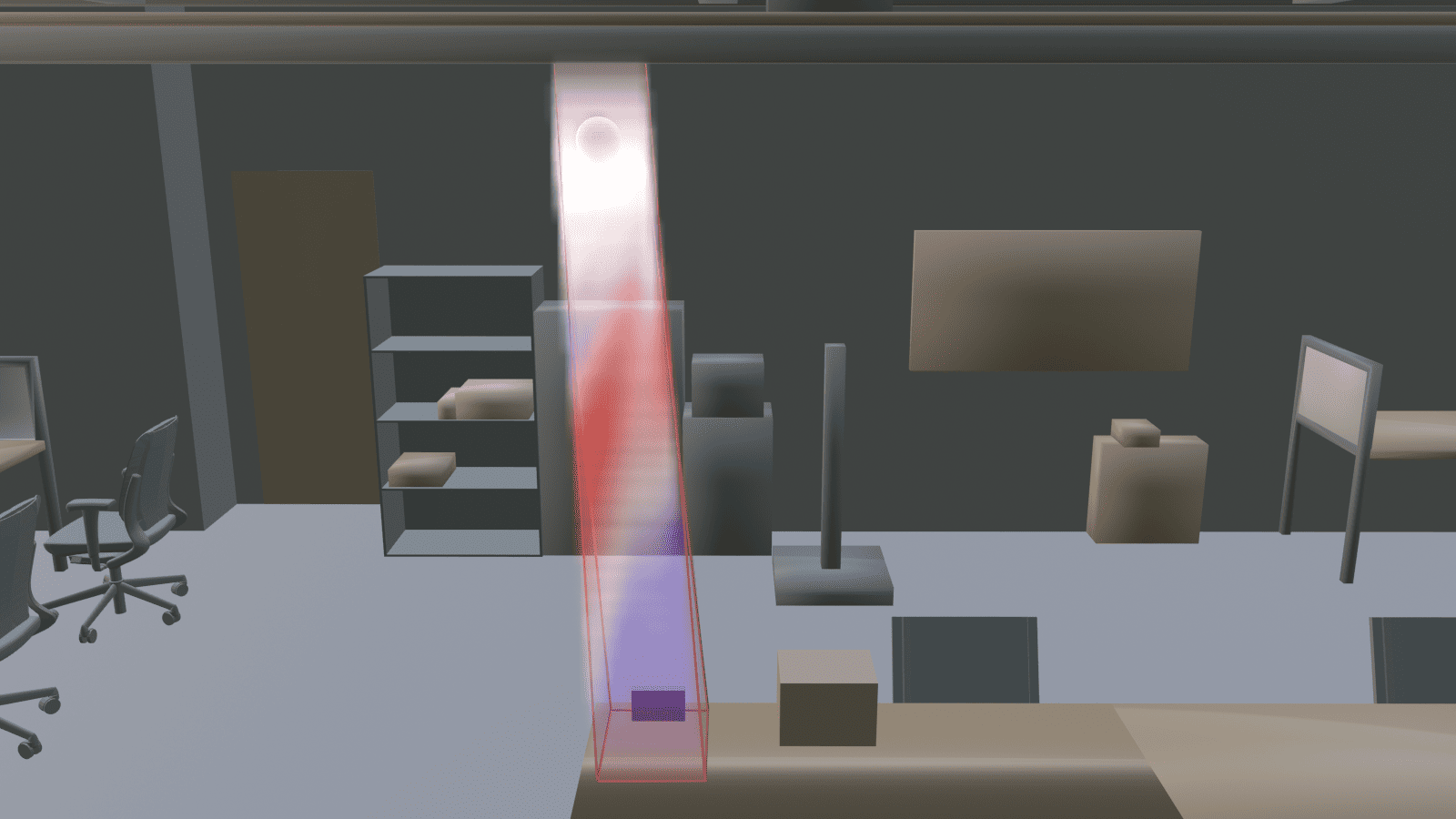}}\\
    \subfigure[$\phi$, PE-PINN]{\includegraphics[width = 0.24\linewidth]{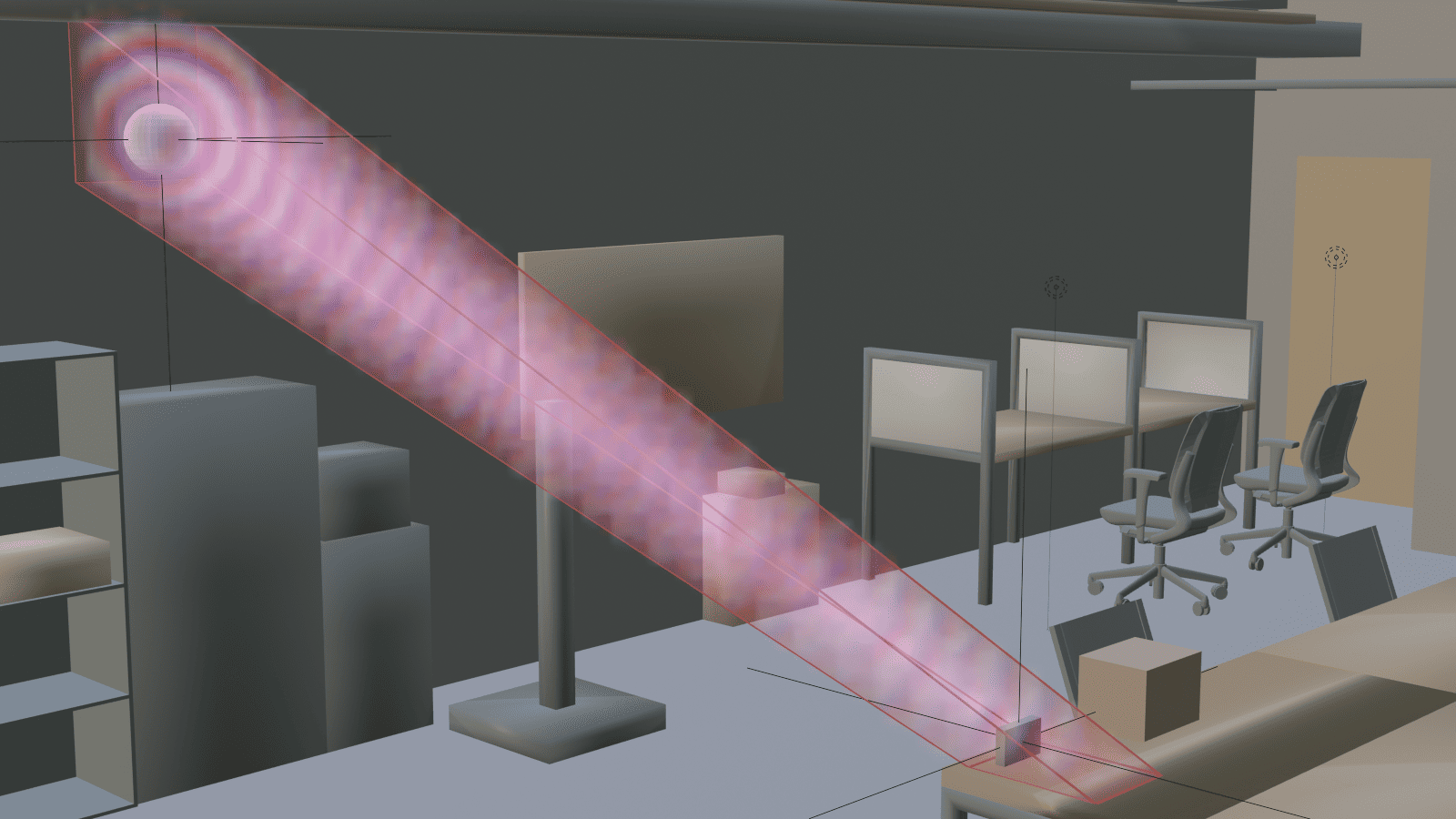}}
    \subfigure[$\phi$, SIONNA]{\includegraphics[width = 0.24\linewidth]{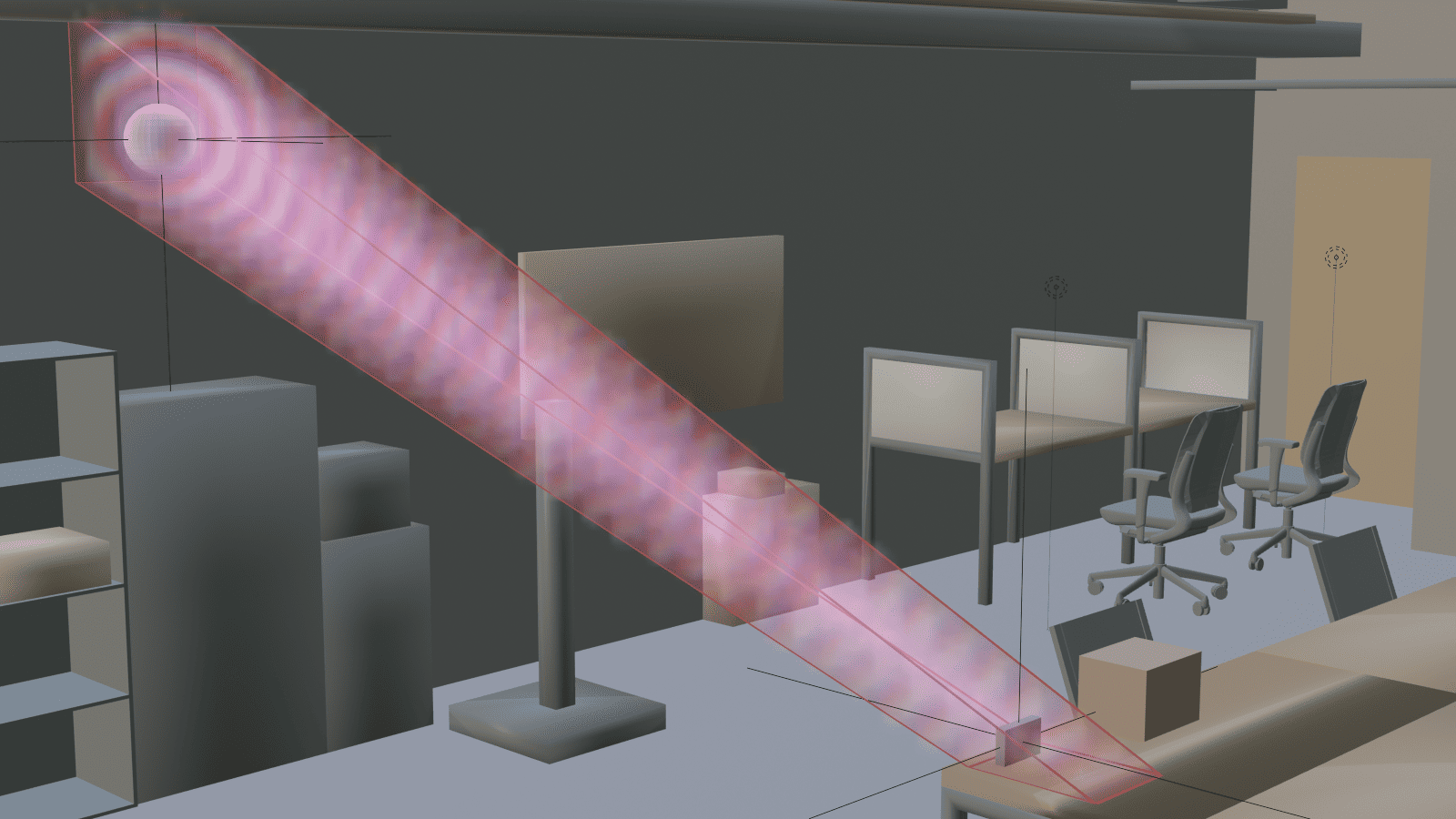}}
    \subfigure[$\Delta\phi$, side view]{\includegraphics[width = 0.24\linewidth]{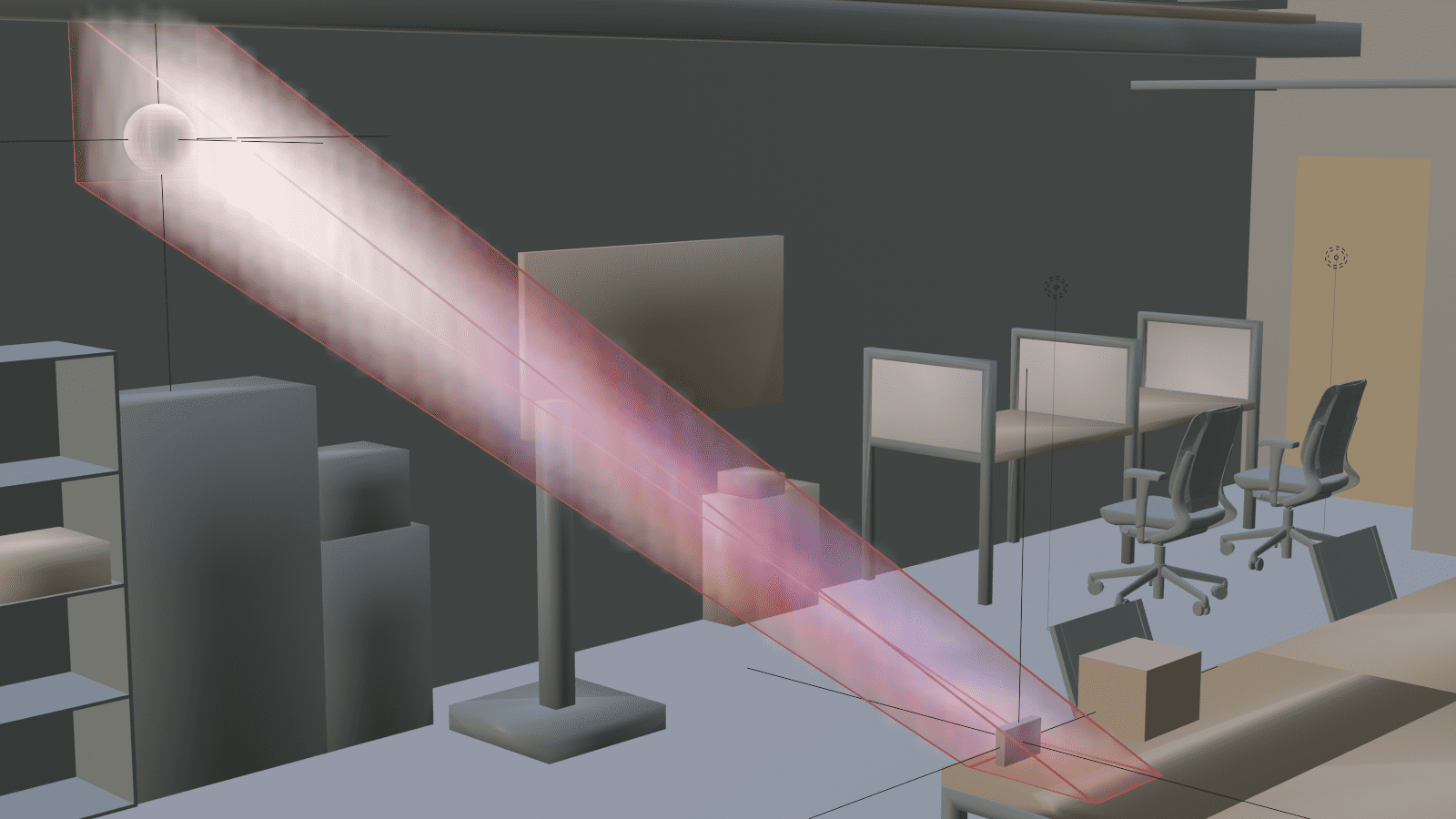}}
    \subfigure[$\Delta\phi$, top view]{\includegraphics[width = 0.24\linewidth]{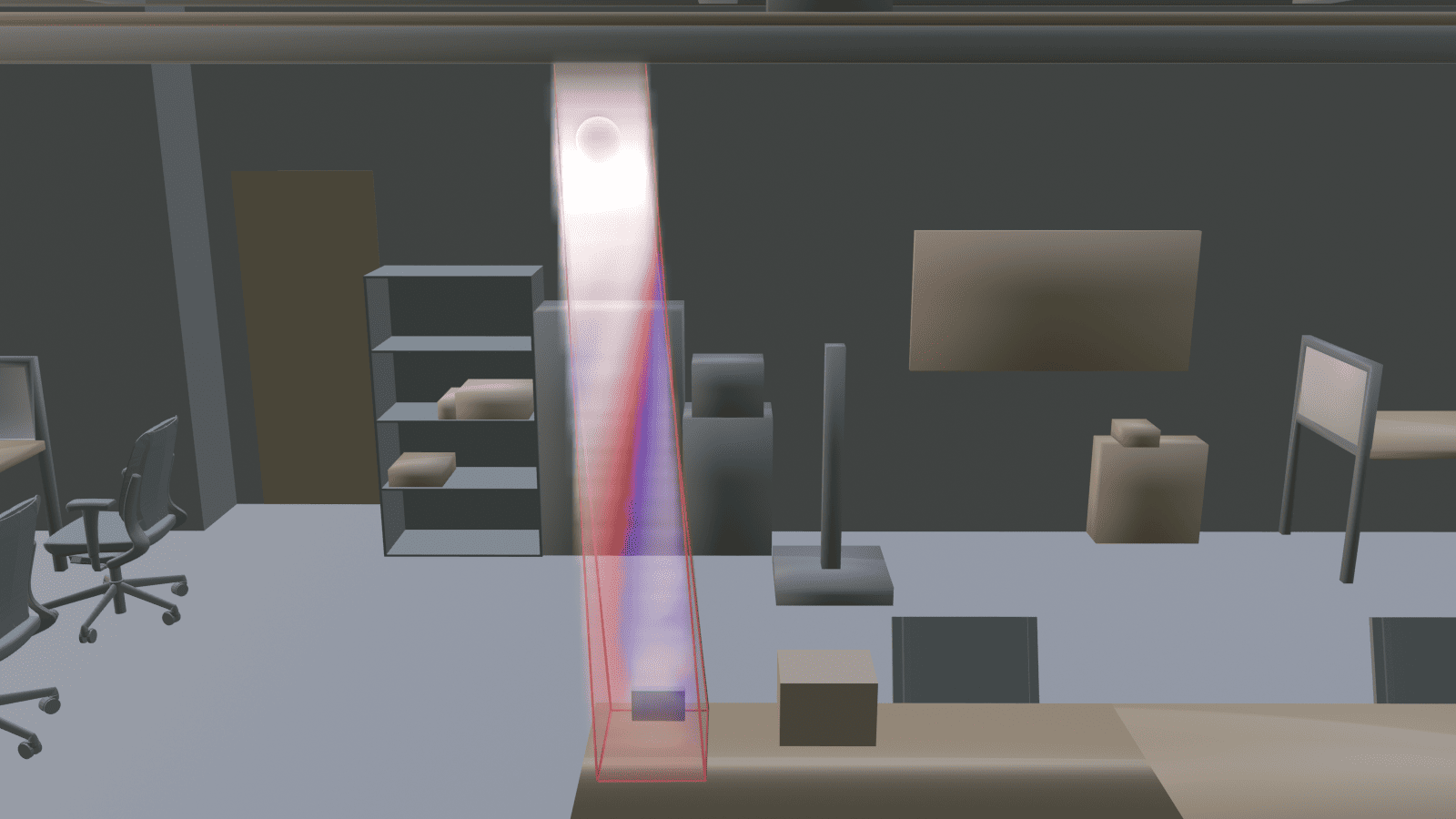}}\\
    \makebox[1\linewidth]{\includegraphics[width=0.3\textwidth]{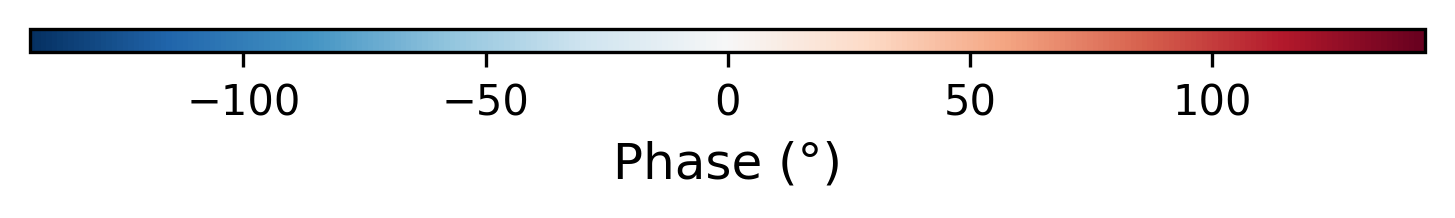}}
    \caption{Comparison of received power (dBm) and phase ($^\circ$) in the small diffraction tunnel: PE-PINN (a,e), Sionna (b,f), and their pointwise difference PE-PINN $-$ Sionna shown in side view (c,g) and top view (d,h).} \label{fig:tunnel_diffrasmall}
    \vspace{-4mm}
\end{figure*}
 
\begin{figure*}[t]
    \centering
    \makebox[0.49\linewidth]{\includegraphics[width=0.3\textwidth]{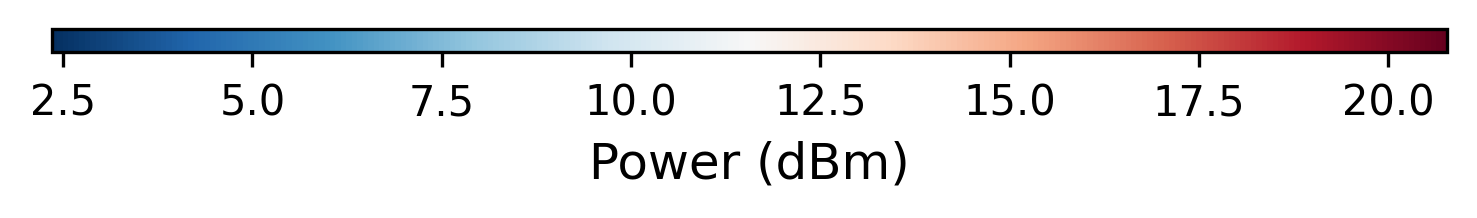}}%
    \makebox[0.49\linewidth]{\includegraphics[width=0.3\textwidth]{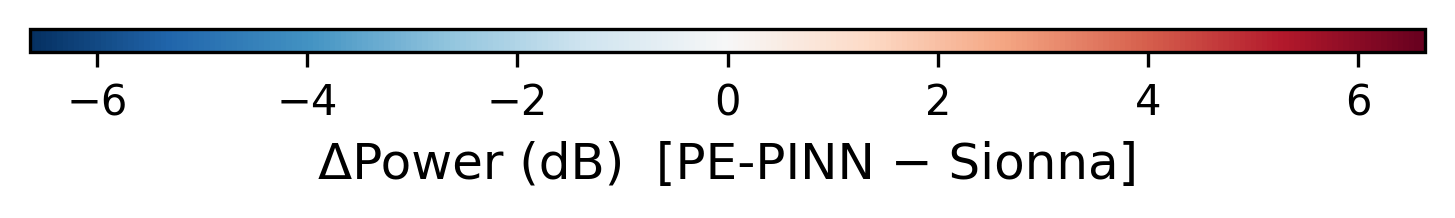}}\\[4pt]
    \subfigure[$P$, PE-PINN]{\includegraphics[width = 0.24\linewidth]{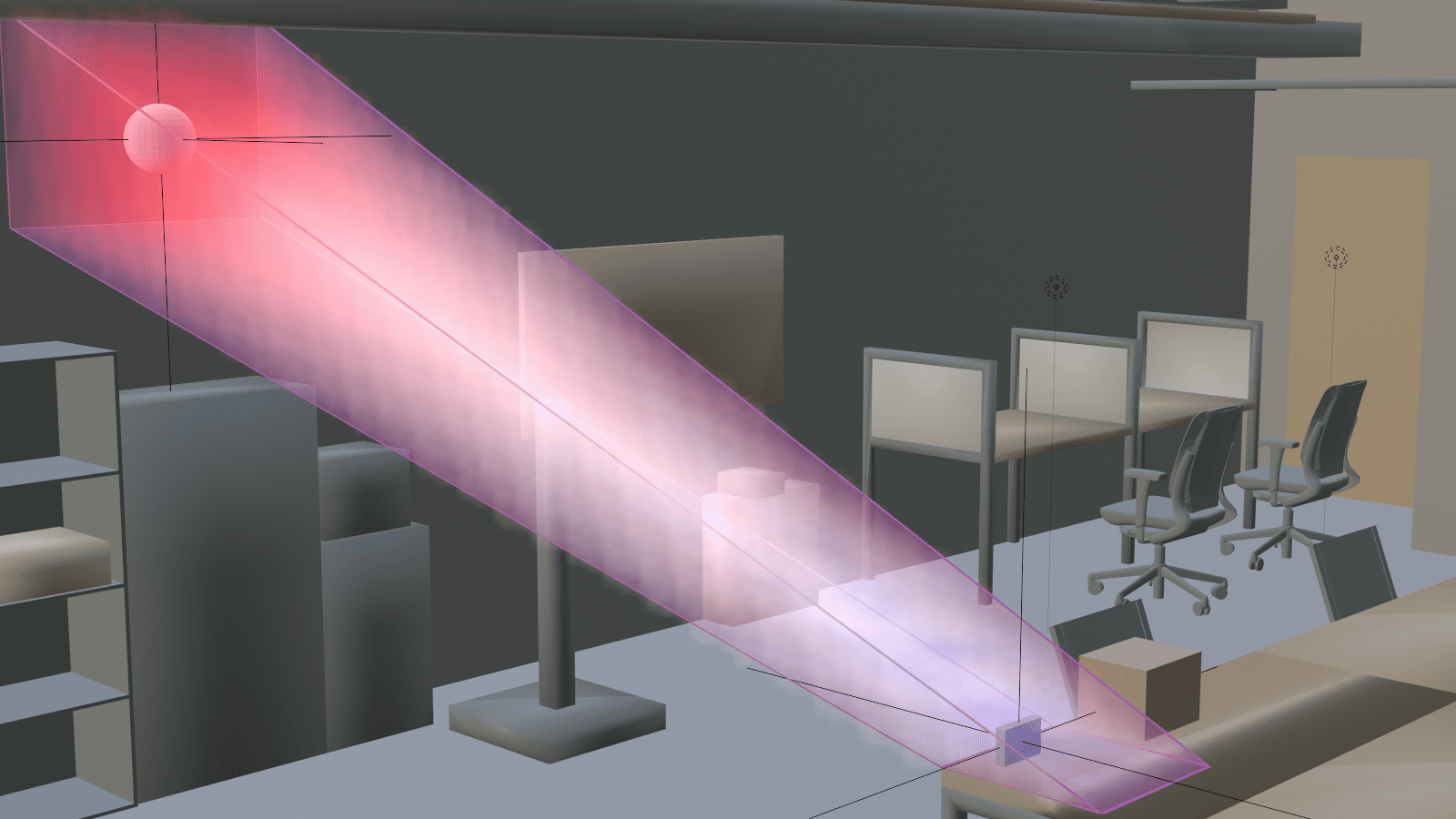}} 
    \subfigure[$P$, SIONNA]{\includegraphics[width = 0.24\linewidth]{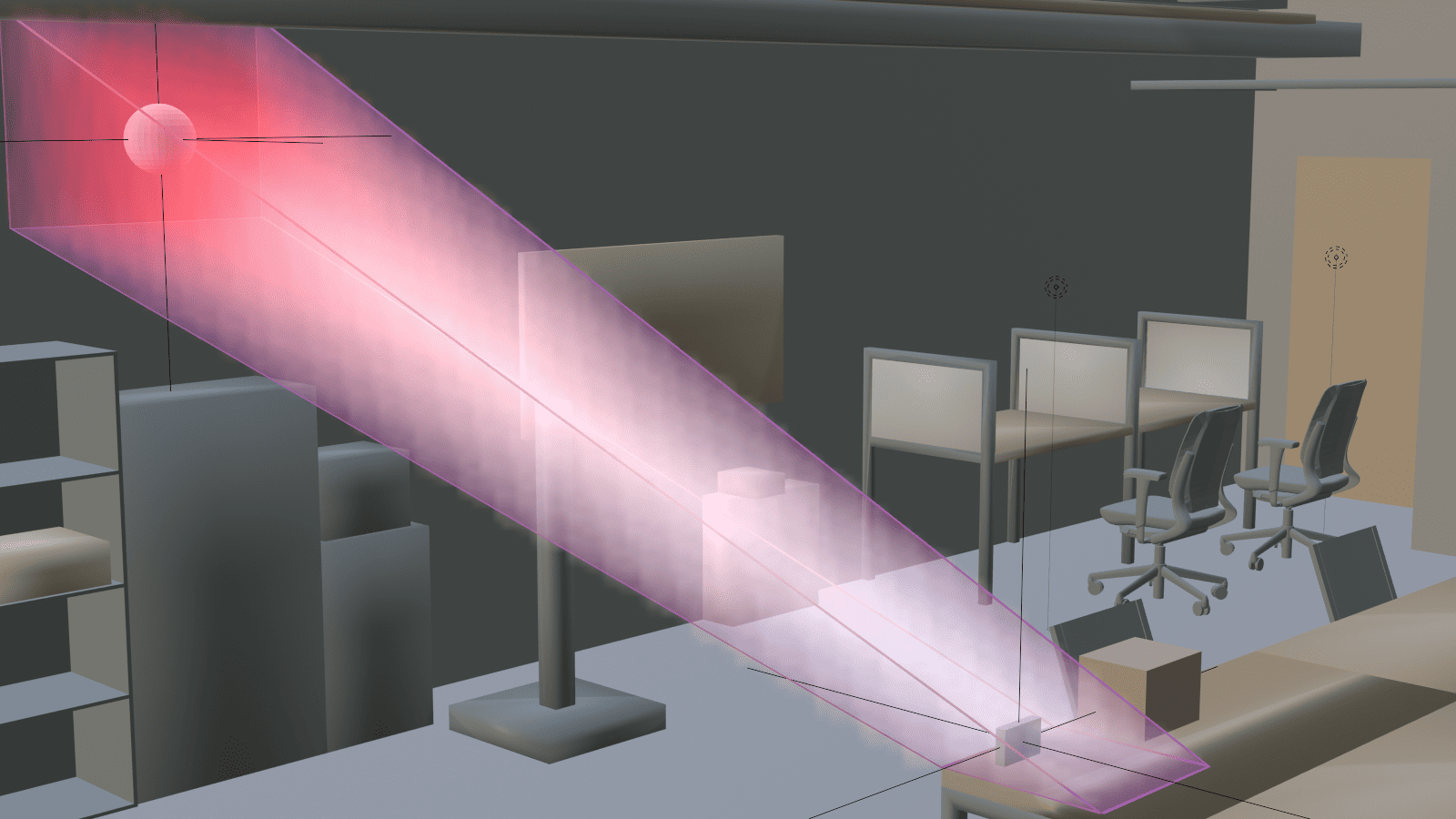}} 
    \subfigure[$\Delta P$, side view]{\includegraphics[width = 0.24\linewidth]{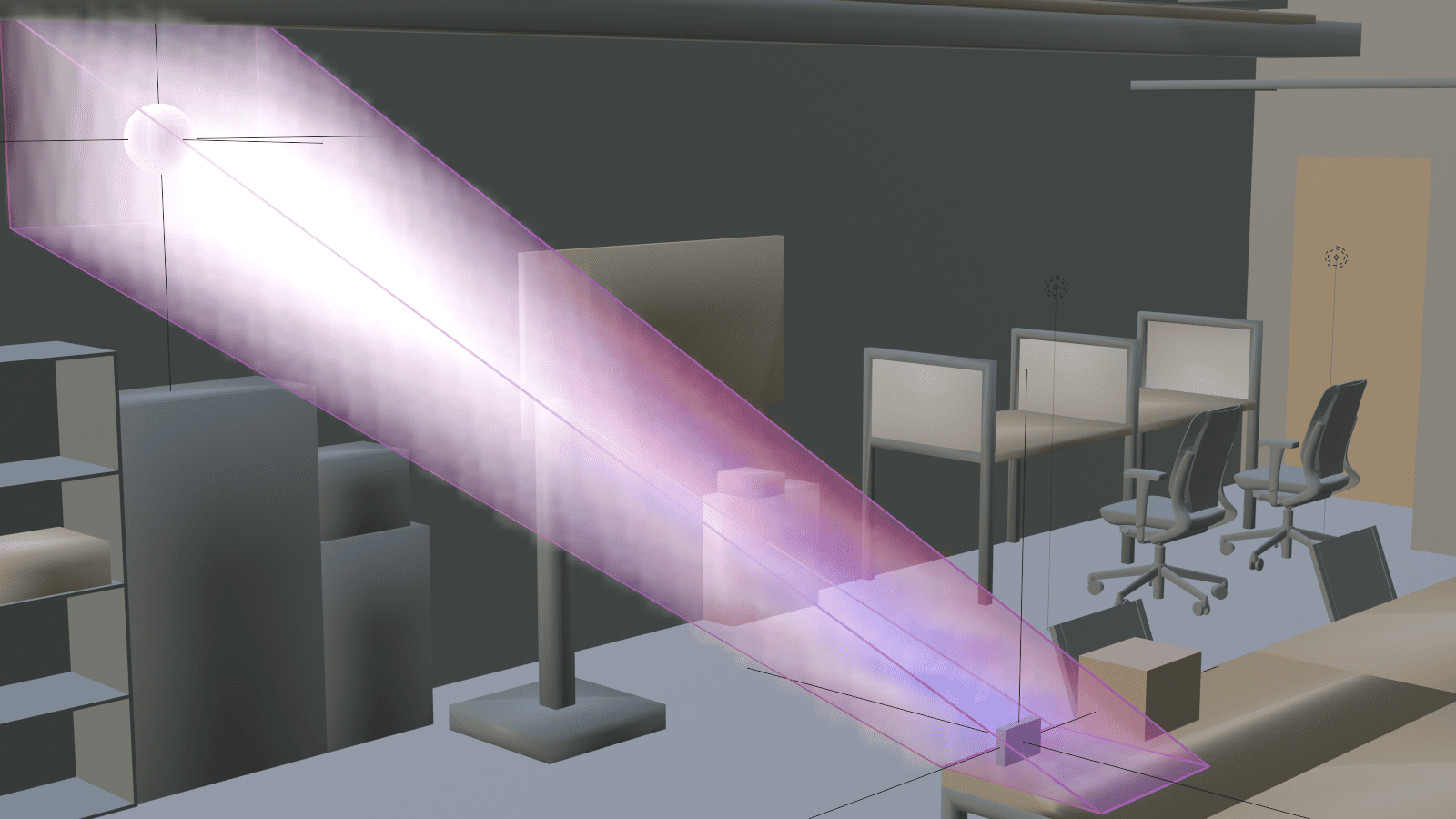}} 
    \subfigure[$\Delta P$, top view]{\includegraphics[width = 0.24\linewidth]{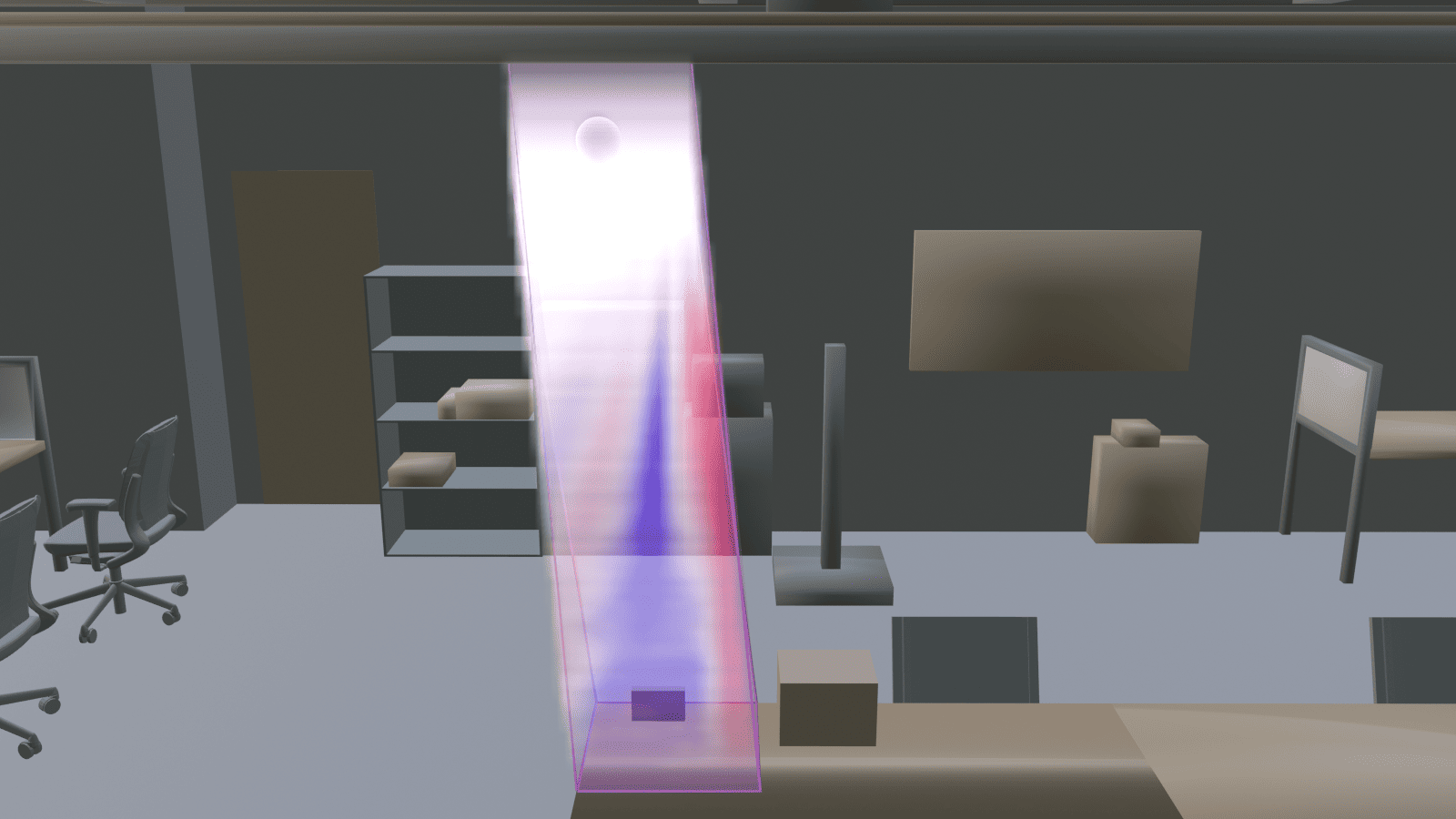}} \\
    \subfigure[$\phi$, PE-PINN]{\includegraphics[width = 0.24\linewidth]{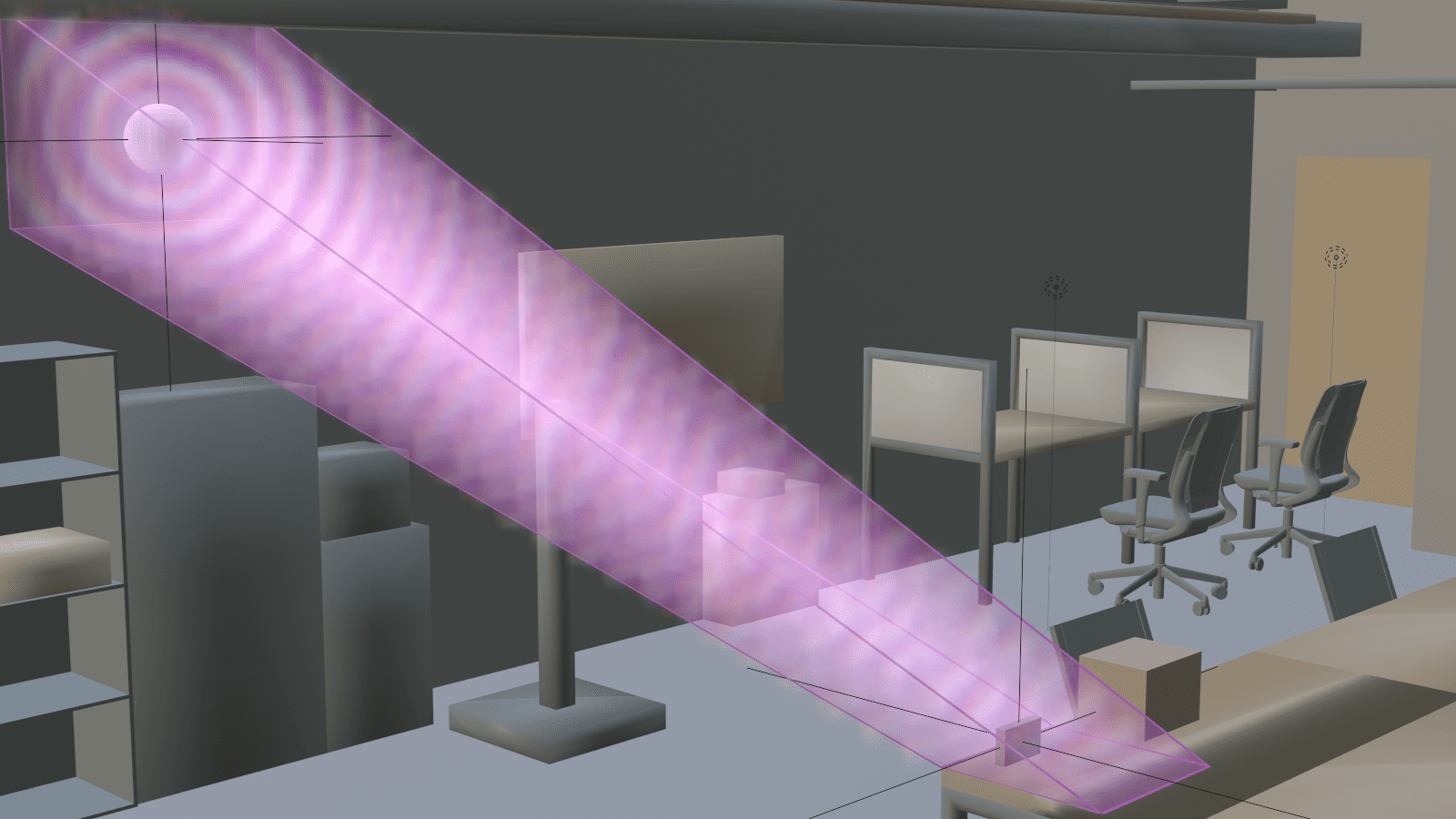}} 
    \subfigure[$\phi$, SIONNA]{\includegraphics[width = 0.24\linewidth]{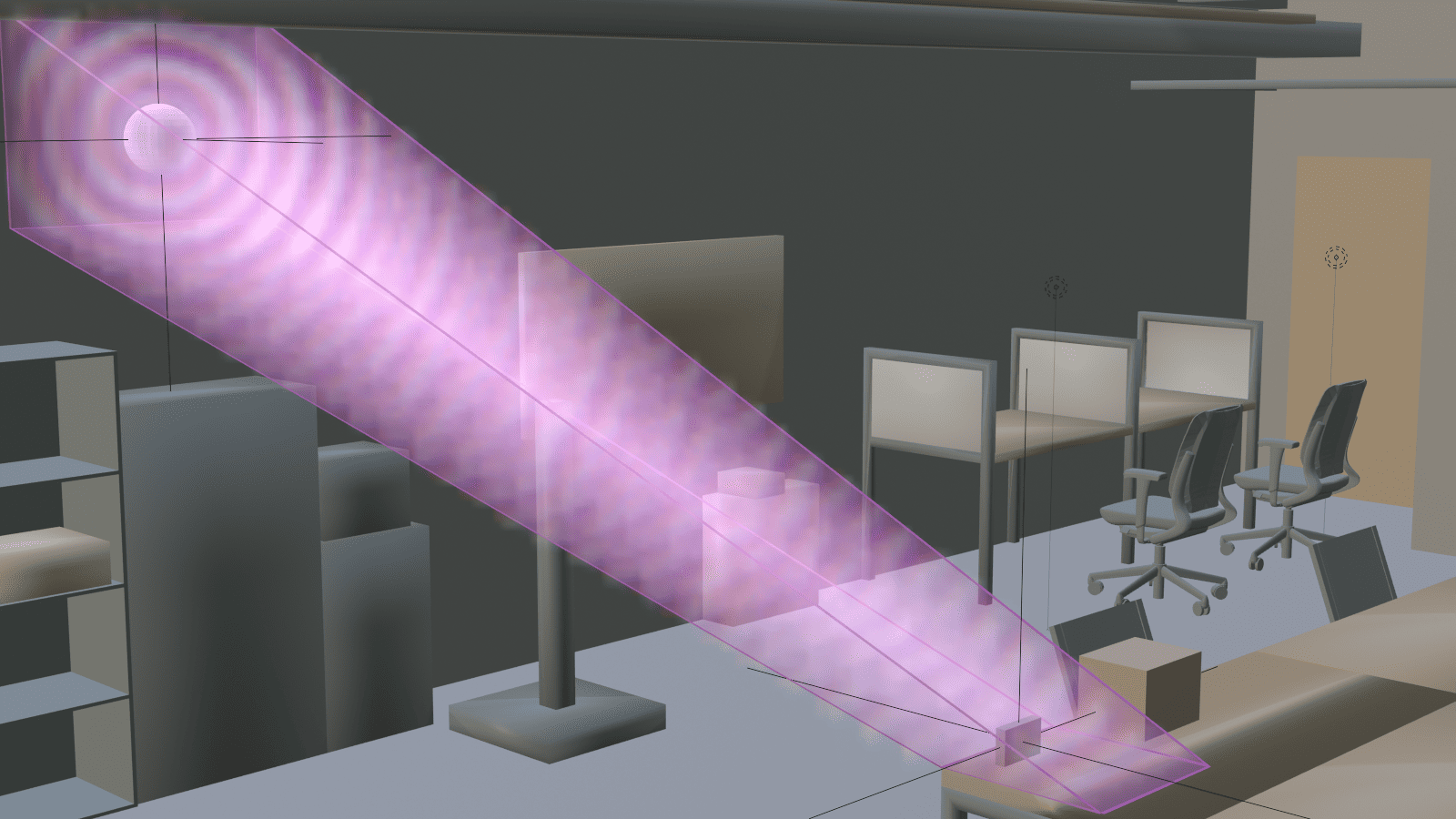}} 
    \subfigure[$\Delta\phi$, side view]{\includegraphics[width = 0.24\linewidth]{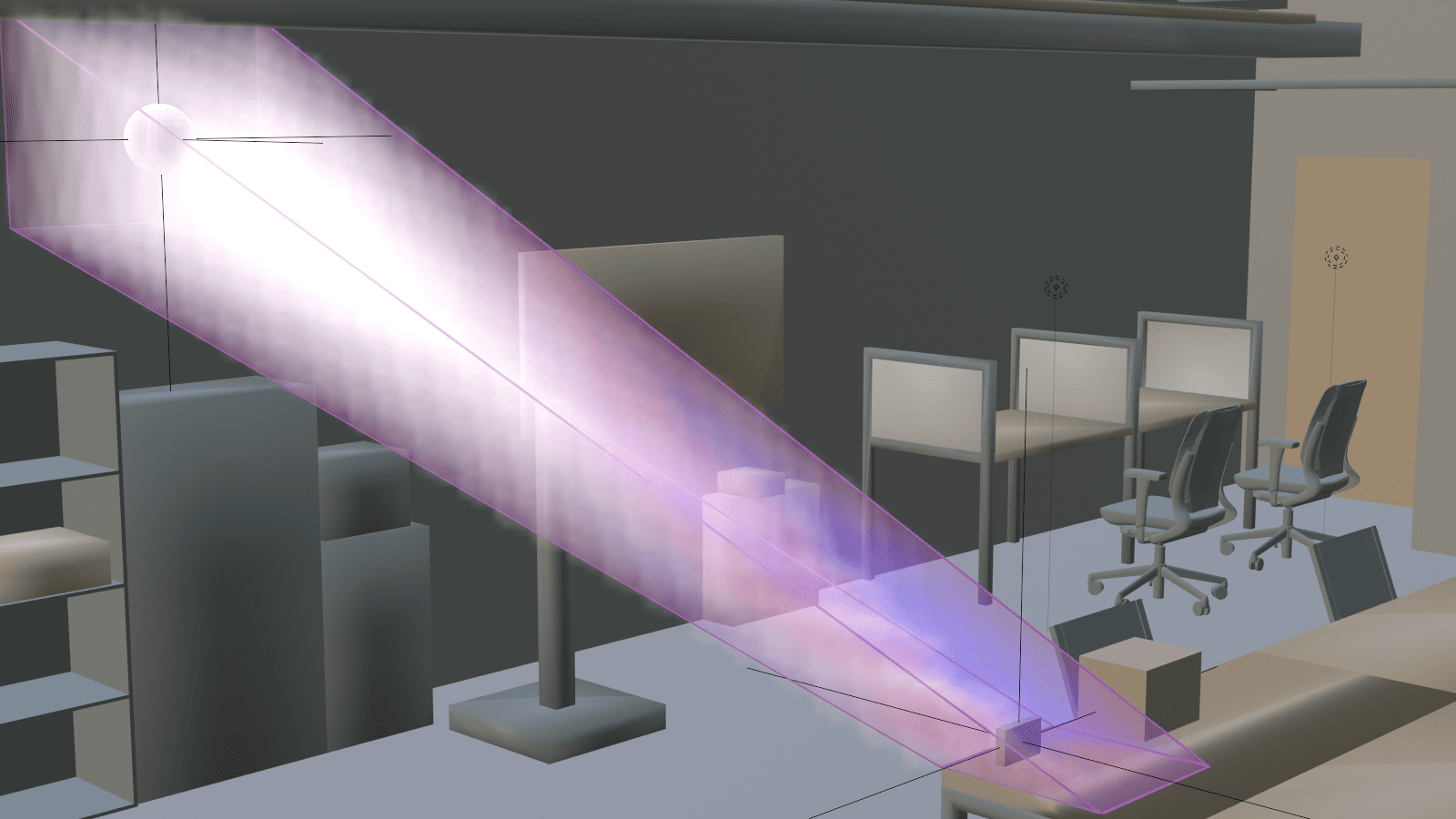}} 
    \subfigure[$\Delta\phi$, top view]{\includegraphics[width = 0.24\linewidth]{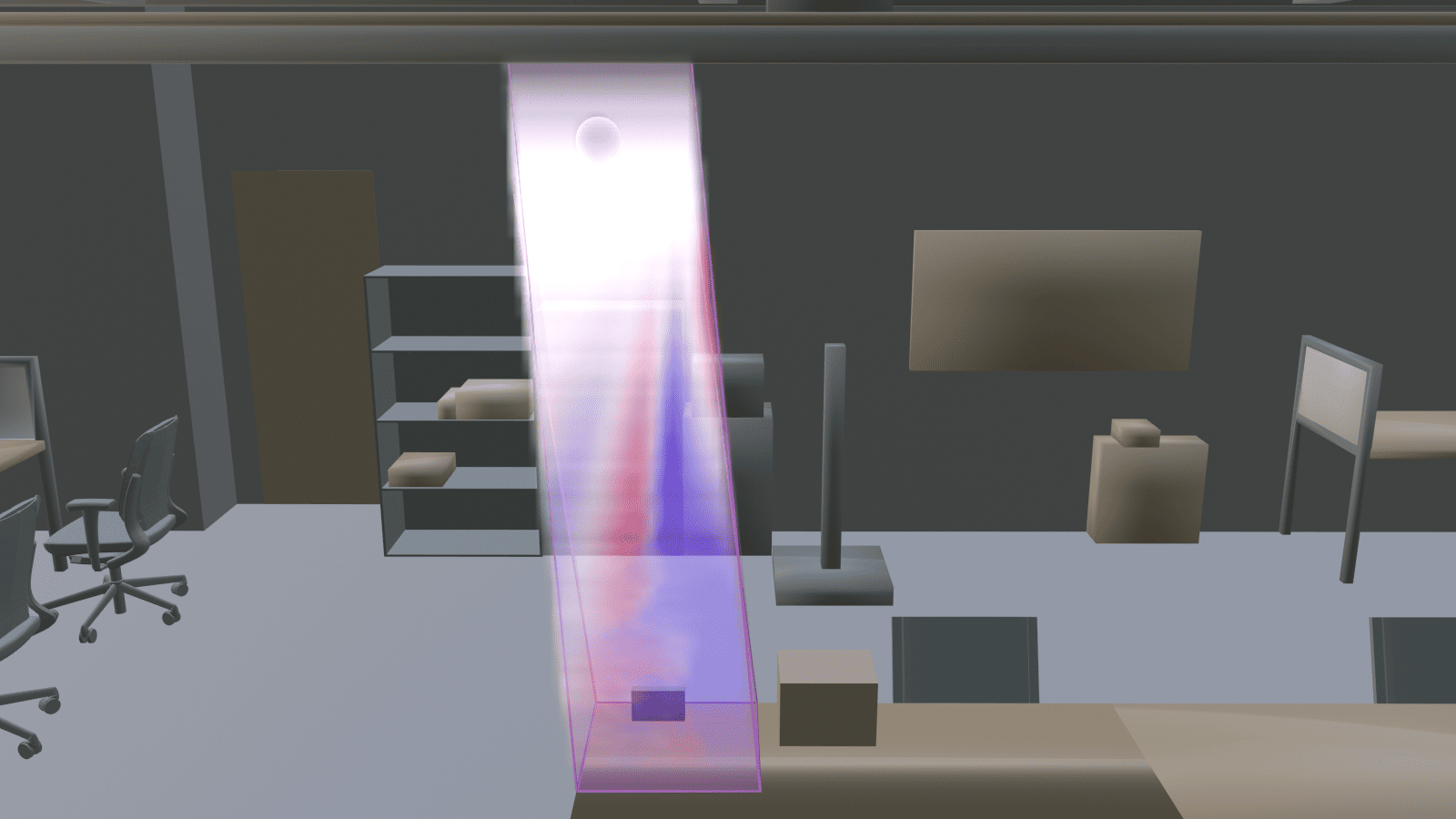}} \\
    \makebox[1\linewidth]{\includegraphics[width=0.3\textwidth]{FiguresResized/colorbar_phase.png}}
    \caption{Comparison of received power (dBm) and phase ($^\circ$) in the wide diffraction tunnel: PE-PINN (a,e), Sionna (b,f), and their pointwise difference PE-PINN $-$ Sionna shown in side view (c,g) and top view (d,h).}  
\label{fig:tunnel_diffrawide}
\vspace{-4mm}
\end{figure*}

As depicted in Fig.~\ref{fig:lab}, a metallic block (i.e., a USRP) placed on a desk serves as the diffracting obstacle. Two tunnels of different sizes are constructed around the metallic block to evaluate the PE-PINN solution under varying domain geometries, as illustrated in Fig.~\ref{fig:diffra_tunnels}, which also shows their spatial overlap.
Each tunnel is defined as a hexahedral domain that extends from the transmitter toward the obstacle.  The metallic block is modeled as a $0.16 \times 0.03 \times 0.10$~m rectangular block centered at $(1.5,\, 0.2,\, 0.75)$~m. The small diffraction tunnel has a near-end cross-section of $0.5 \times 0.6$~m at $y = -3.35$~m (source side) and tapers to a $0.3 \times 0.4$~m far-end cross-section near the metallic block at $z = 0.70$~m. The wide diffraction tunnel enlarges the near-end cross-section to $1.0 \times 0.75$~m and the far-end to $0.5 \times 0.5$~m, encompassing a broader region around the obstacle to capture additional diffraction contributions.

Following the PE-PINN architecture~\cite{zhang2026physicsinformedneuralnetworksarchitectural}, the source network $\mathcal{N}_{\text{src}}$ models the incident field using the spherical wave kernel centered at the transmitter location $\mathbf{x}_{\text{src}} = (2.0,\, -3.2,\, 2.3)$, with the electric field amplitude normalized to unity on a sphere of radius equal to one wavelength ($\lambda \approx 0.1249$~m at 2.4~GHz). The scatter network $\mathcal{N}_{\text{sct}}$ captures the scattering phenomena through a multi-kernel structure. Based on the transmitter position and the metallic block geometry, three illuminated faces are identified: the front face ($y^{-}$), the top face ($z^{+}$), and the right face ($x^{+}$). For each face, a reflection kernel is placed at the corresponding image source location: $\mathbf{x}_{\text{ref},1} = (2.0,\, 3.57,\, 2.3)$ for the front face, $\mathbf{x}_{\text{ref},2} = (2.0,\, -3.2,\, -0.70)$ for the top face, and $\mathbf{x}_{\text{ref},3} = (1.16,\, -3.2,\, 2.3)$ for the right face. A source kernel $\Psi_{\mathbf{x}_{\text{src}}}$ is retained in $\mathcal{N}_{\text{sct}}$ to account for the shadowing effect behind the obstacle. Additionally, seven edge diffraction kernels, as proposed in Sec.~\ref{sec:edge_kernel}, are constructed along the illuminated edges of the metallic block, excluding the two bottom edges on the illuminated faces that lie on the desk surface. Specifically, $\mathbf{e}_{1}$-$\mathbf{e}_{2}$ are placed along the two vertical edges of the front face ($y^{-}$), each discretized with $P=5$ secondary sources; $\mathbf{e}_{3}$ along the top horizontal edge of the front face with $P=8$; $\mathbf{e}_{4}$-$\mathbf{e}_{5}$ along the two short edges of the top face ($z^{+}$) with $P=3$ each; $\mathbf{e}_{6}$ along the back edge of the top face with $P=8$; and $\mathbf{e}_{7}$ along the back vertical edge of the right face ($x^{+}$) with $P=5$. The number of point sources per edge is chosen proportionally to the edge length. On all faces and edges of the metallic block, the PEC boundary condition $E^{\text{src}} + E^{\text{sct}} = 0$ is enforced. The total field is:
\begin{equation}
    E(\mathbf{x}) \approx \mathcal{N}_{\text{src}}(\mathbf{x}) 
    + \mathcal{N}_{\text{sct}}\!\left(\mathbf{x};\, 
    \Psi_{\mathbf{x}_{\text{src}}},\, 
    \{\Psi_{\mathbf{x}_{\text{ref},i}}\}_{i=1}^{3},\, 
    \{\Psi_{\mathbf{e}_{j}}\}_{j=1}^{7}\right).
\end{equation}
Figs.~\ref{fig:tunnel_diffrasmall} and~\ref{fig:tunnel_diffrawide} compare the received power (dBm) and phase ($^\circ$) between PE-PINN and Sionna for the small and wide diffraction tunnels, respectively, with their pointwise differences shown in side view and top view. The field distributions are visualized as semi-transparent volumetric renderings overlaid on the 3D lab environment, where the color at each pixel represents the accumulated contribution of all voxels along the viewing direction. The quantitative comparison is summarized in Table~\ref{tab:diffra_comparison}. Both tunnels exhibit visually similar power distributions along the main propagation path, but differences emerge around the metallic block, where the interference between the incident, reflected, and diffracted waves creates a complex field pattern that the two solvers resolve differently.


The quantitative comparison is summarized in Table~\ref{tab:diffra_comparison}. To isolate the region where PE-PINN and Sionna diverge most, owing to the complex interference of incident, reflected, and edge-diffracted waves, we define a \emph{Near-Block} evaluation zone for both tunnel sizes: a region extending 0.15\,m around the USRP bounding box in all directions plus 0.3\,m into the shadow region behind it. For a fair comparison, the wide tunnel's near-block metrics are computed using only those evaluation points that also fall within the small tunnel's hexahedral domain, so that both rows in the table correspond to the same spatial region. Over the full domain, the wide tunnel yields consistently lower errors (magnitude MAE $1.49$\,dB vs.\ $3.82$\,dB; phase MAE $8.94^\circ$ vs.\ $35.56^\circ$), largely because the surrounding free-space region, where both solvers naturally agree, dominates the statistics. In the near-block region, these advantages largely vanish: the magnitude MAE rises to $25.77$\,dB and $30.78$\,dB for the wide and small tunnels, respectively, and the phase RMSE to $83.51^\circ$ and $103.62^\circ$. The sharp increase from full-domain to near-block metrics in both configurations confirms that the discrepancy is highly localized around the USRP, a conclusion further corroborated by the difference heatmaps, which show the largest deviations at the same locations relative to the USRP regardless of tunnel size.

\begin{table}[t]
\centering
\caption{PE-PINN $-$ Sionna Comparison for Diffraction Scenario}\label{tab:diffra_comparison}
\begin{tabular}{l|l|cc|cc}
\toprule
 & & \multicolumn{2}{c|}{\textbf{Full Tunnel}} & \multicolumn{2}{c}{\textbf{Near-Block}} \\
 \textbf{Tunnel} & & \textbf{MAE} & \textbf{RMSE} & \textbf{MAE} & \textbf{RMSE} \\
\midrule
\multirow{2}{*}{\textbf{Small}}
 &$ \Delta$Power (dB)     & 3.82  & 17.14 & 30.78 & 90.92 \\
 & $ \Delta$Phase ($^\circ$)   & 35.56 & 61.54 & 91.38 & 103.62 \\
\midrule
\multirow{2}{*}{\textbf{Wide}}
 & $ \Delta$Power (dB)     & 1.49  & 10.86 & 25.77 & 90.04 \\
 & $ \Delta$Phase ($^\circ$)   & 8.94  & 20.55 & 70.10 & 83.51 \\
\bottomrule
\end{tabular}
\vspace{-2mm}
\end{table}

\subsubsection{Refraction Scenario}\label{sec:eval_refra}

\begin{figure}[b]
    \centering
    \includegraphics[width=.9\linewidth]{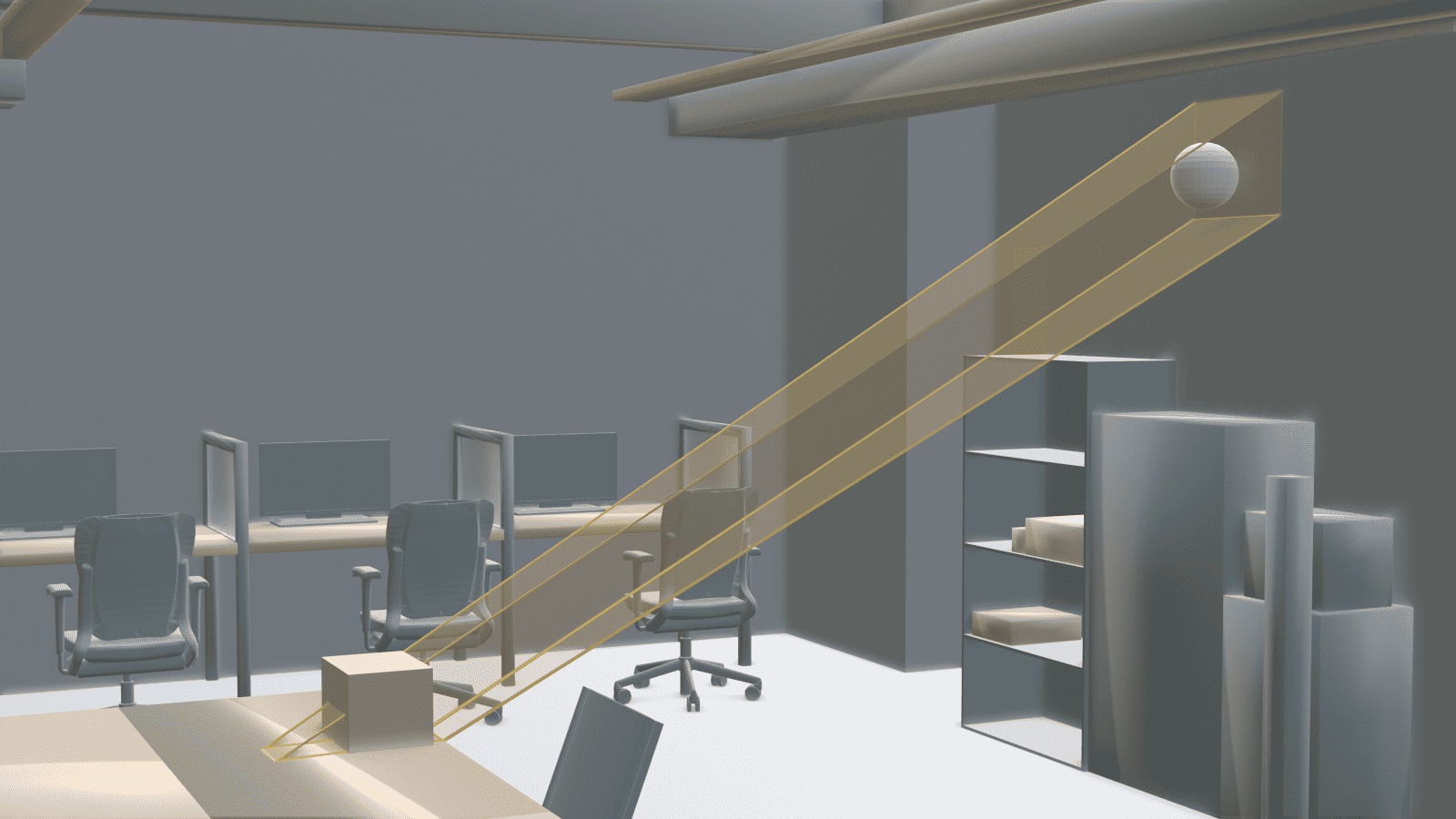}
    \caption{Refraction tunnel constructed in the lab environment.}
    \label{fig:refra_tunnel}
\end{figure}

\begin{figure*}[t]
    \centering
    \makebox[0.49\linewidth]{\includegraphics[width=.3\textwidth]{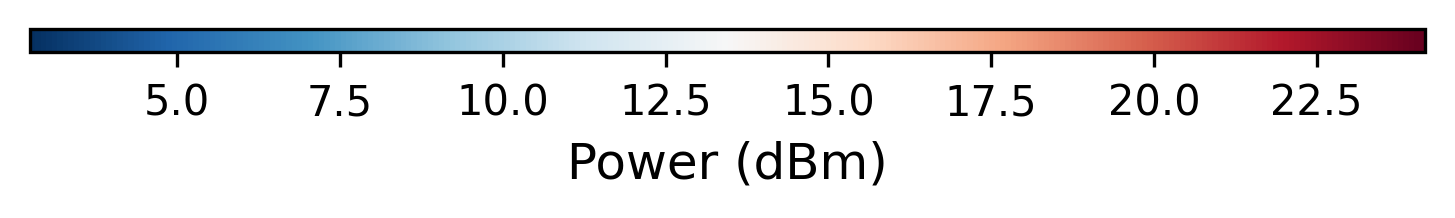}}%
    \makebox[0.49\linewidth]{\includegraphics[width=.3\textwidth]{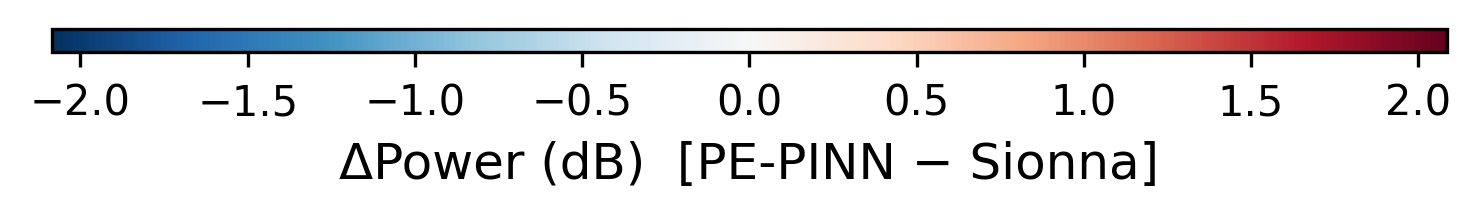}}\\[4pt]
    \subfigure[$P$, PE-PINN]{\includegraphics[width = 0.24\linewidth]{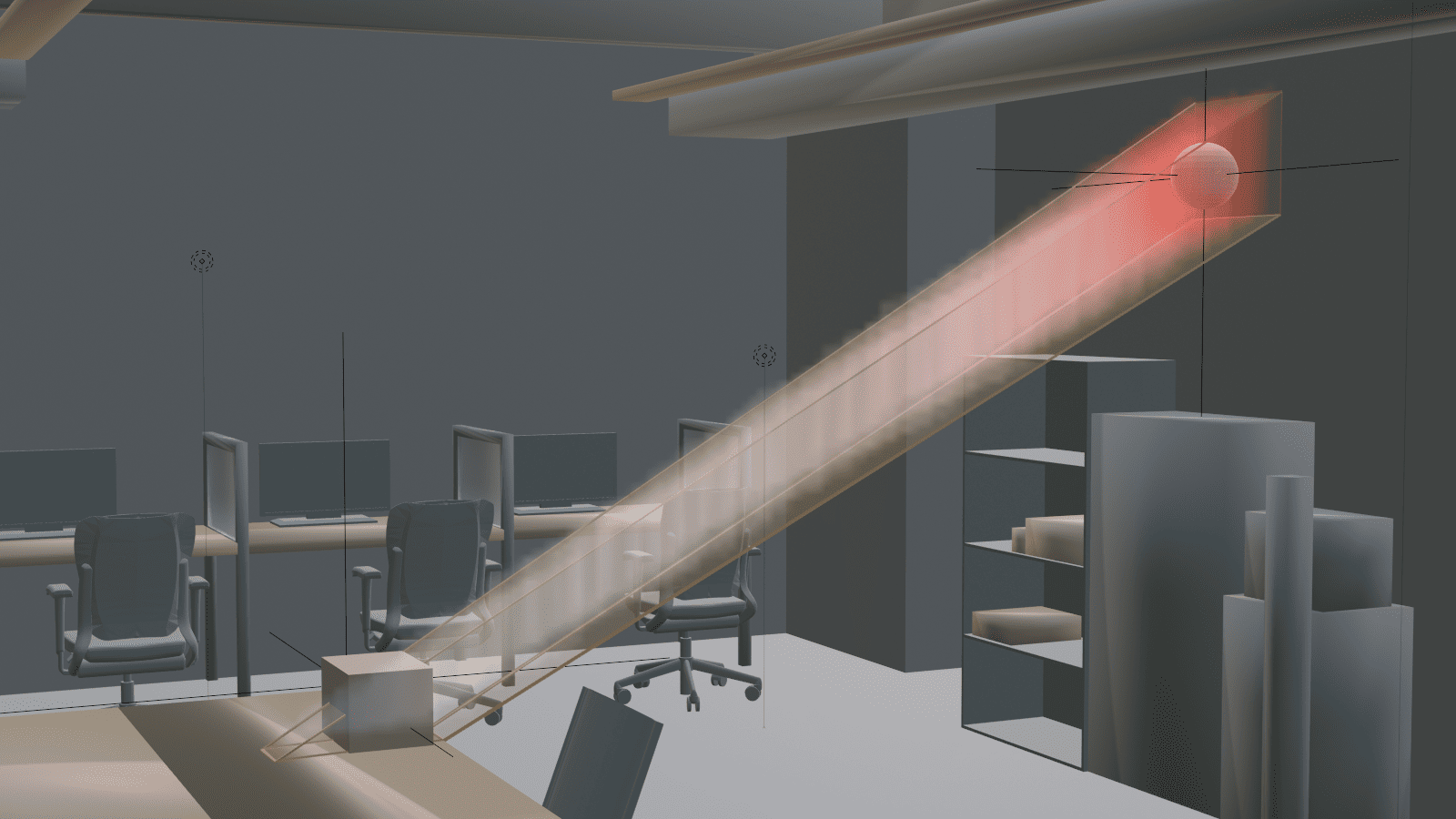}} 
    \subfigure[$P$, SIONNA]{\includegraphics[width = 0.24\linewidth]{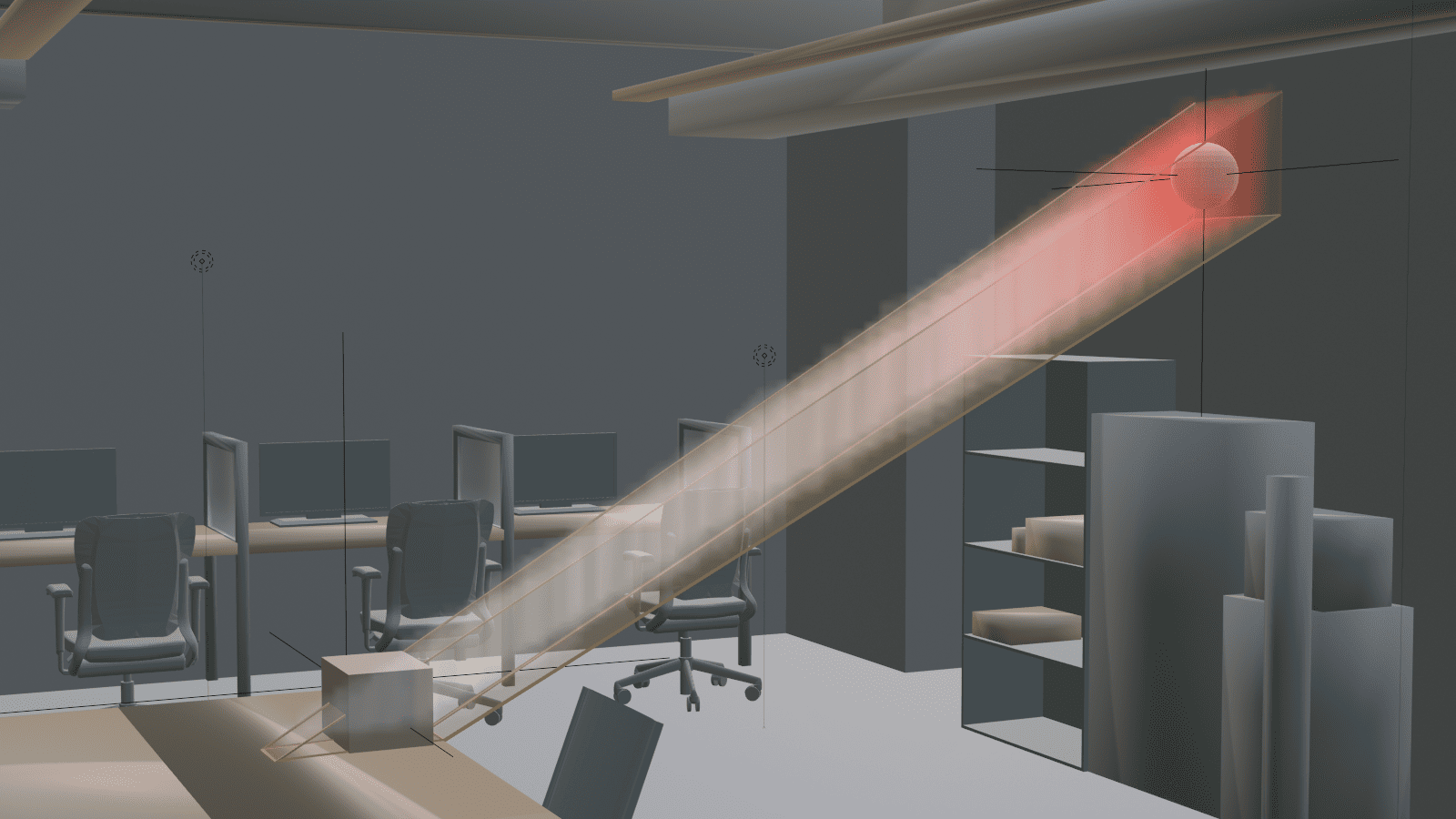}} 
    \subfigure[$\Delta P$, side view]{\includegraphics[width = 0.24\linewidth]{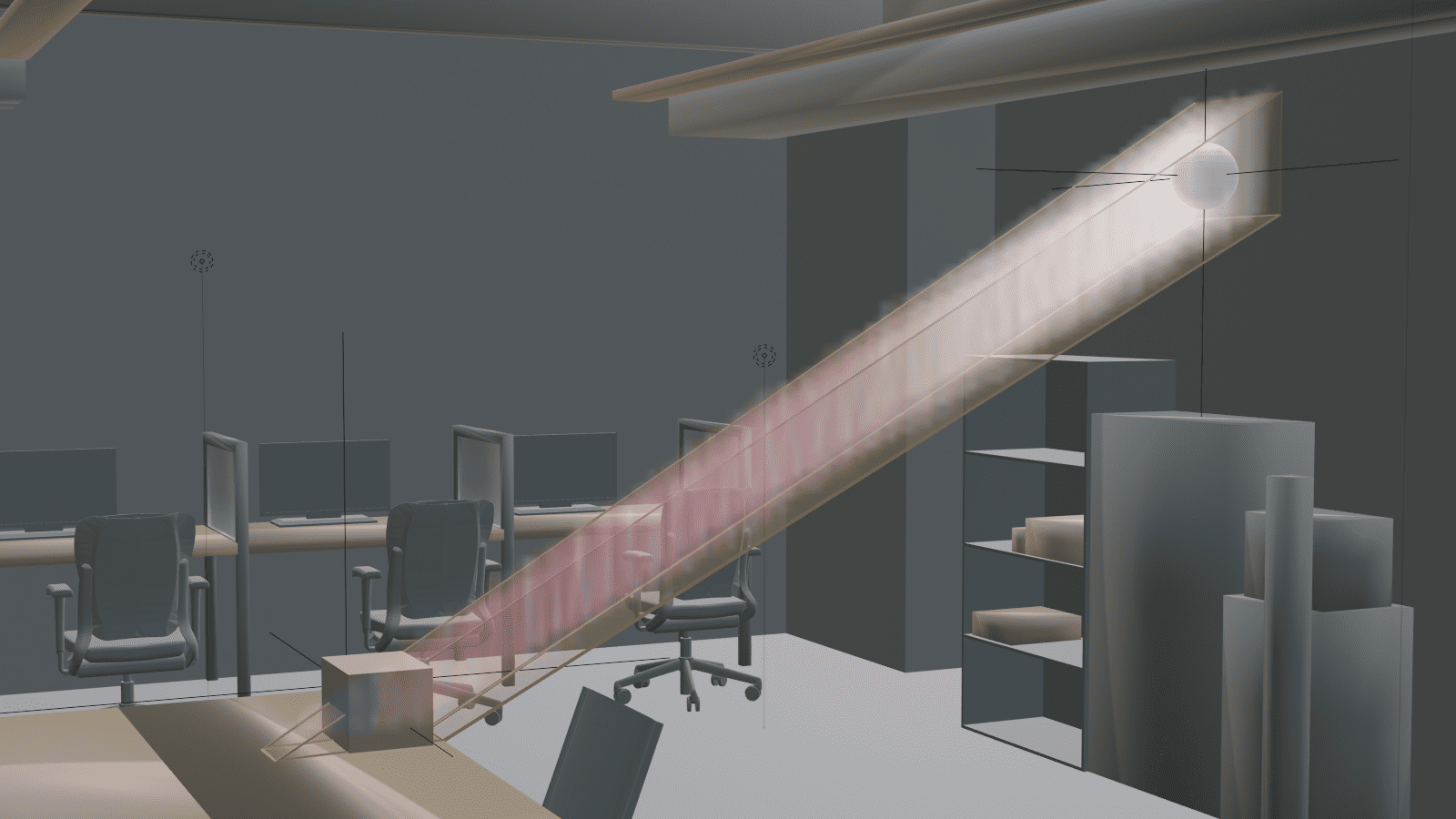}} 
    \subfigure[$\Delta P$, top view]{\includegraphics[width = 0.24\linewidth]{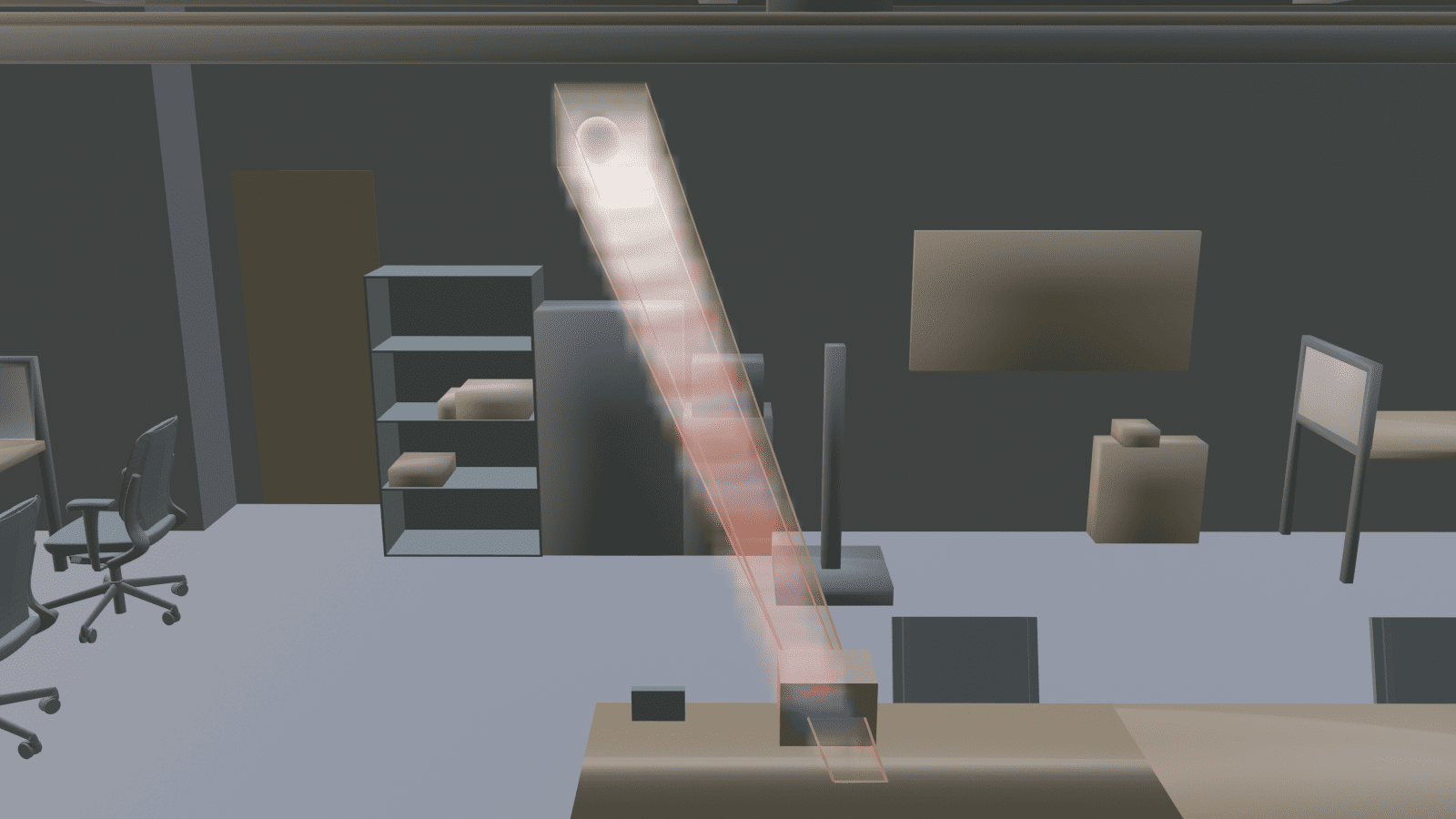}} \\
    \subfigure[$\phi$, PE-PINN]{\includegraphics[width = 0.24\linewidth]{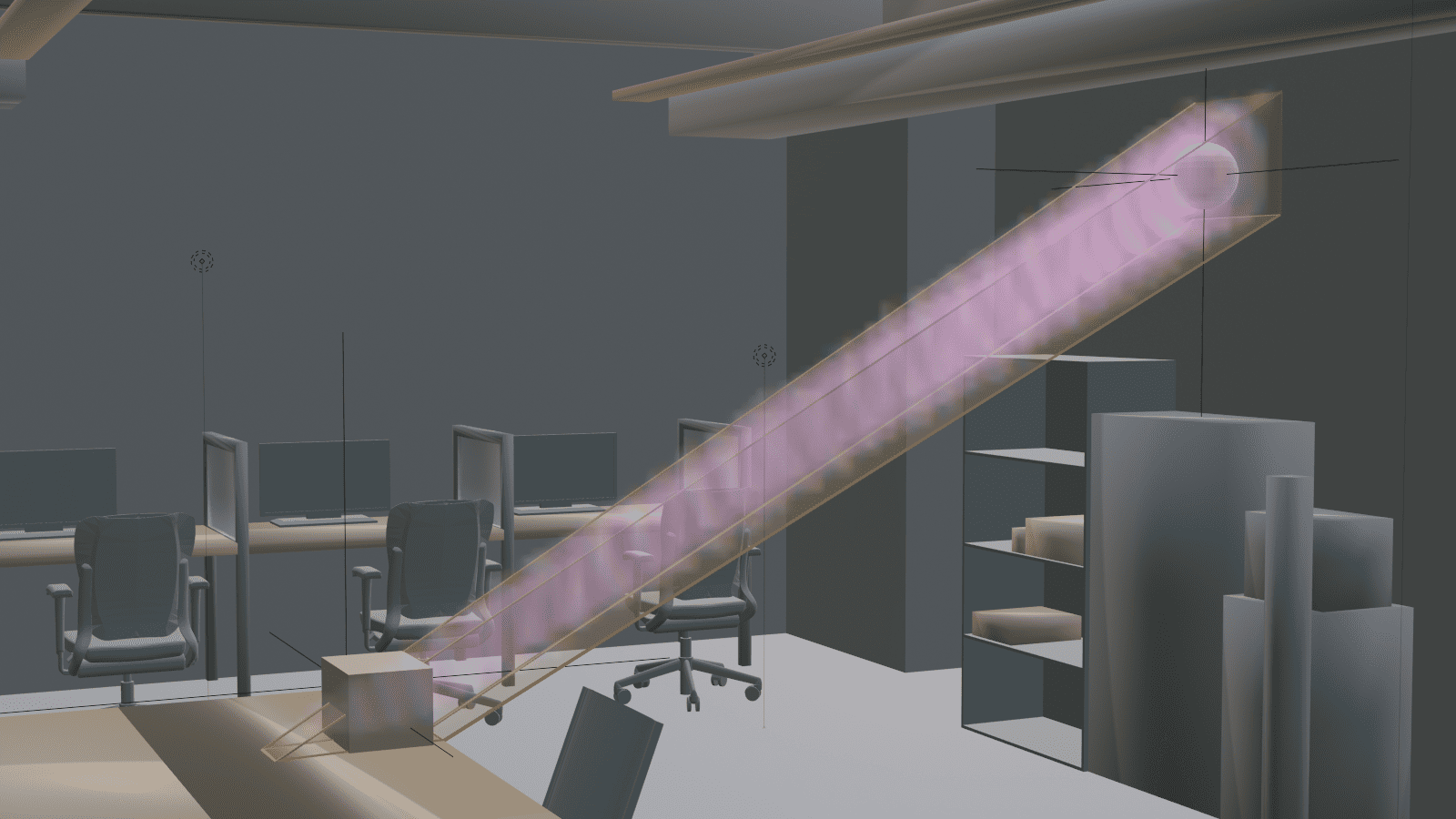}} 
    \subfigure[$\phi$, SIONNA]{\includegraphics[width = 0.24\linewidth]{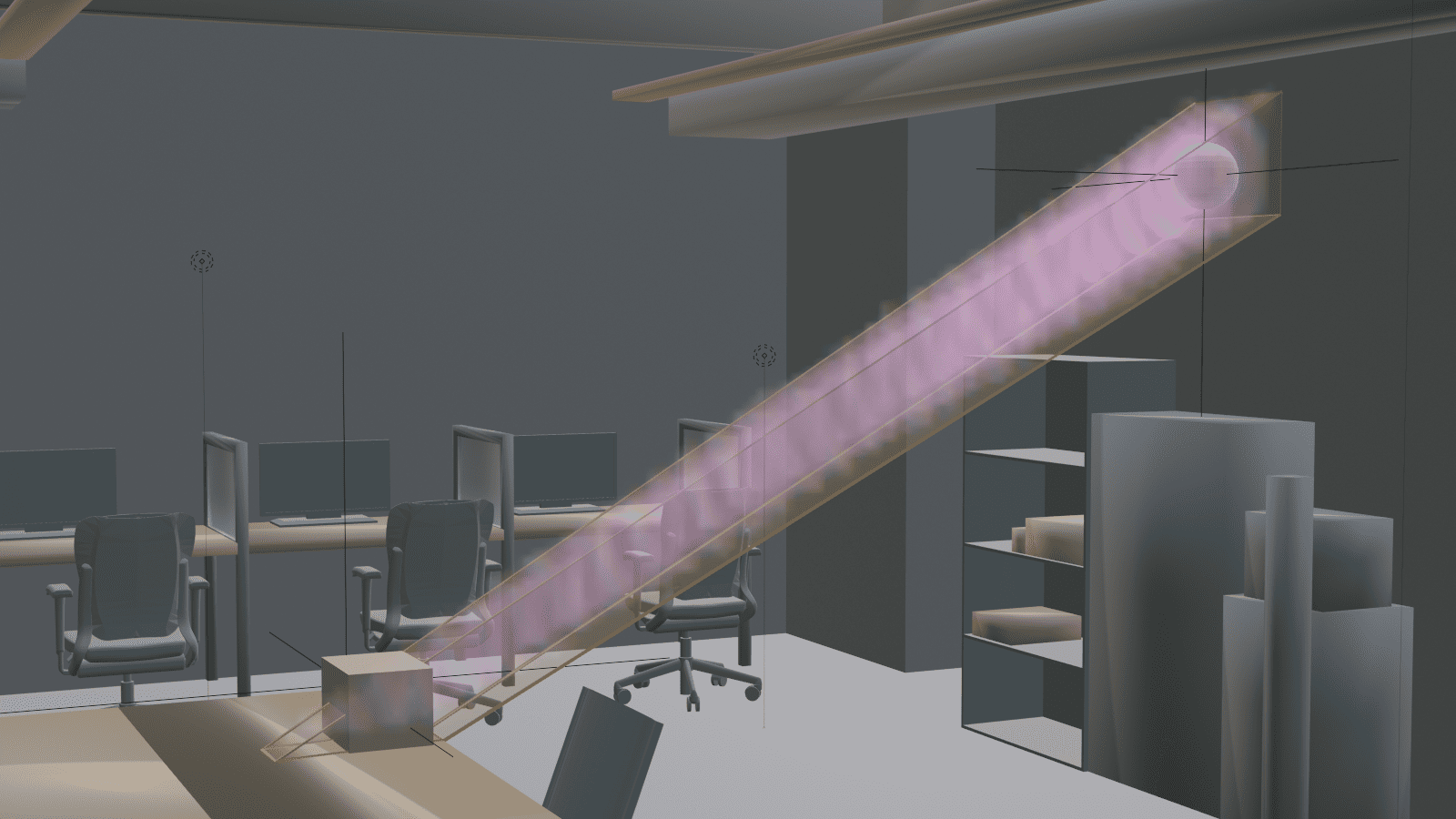}} 
    \subfigure[$\Delta\phi$, side view]{\includegraphics[width = 0.24\linewidth]{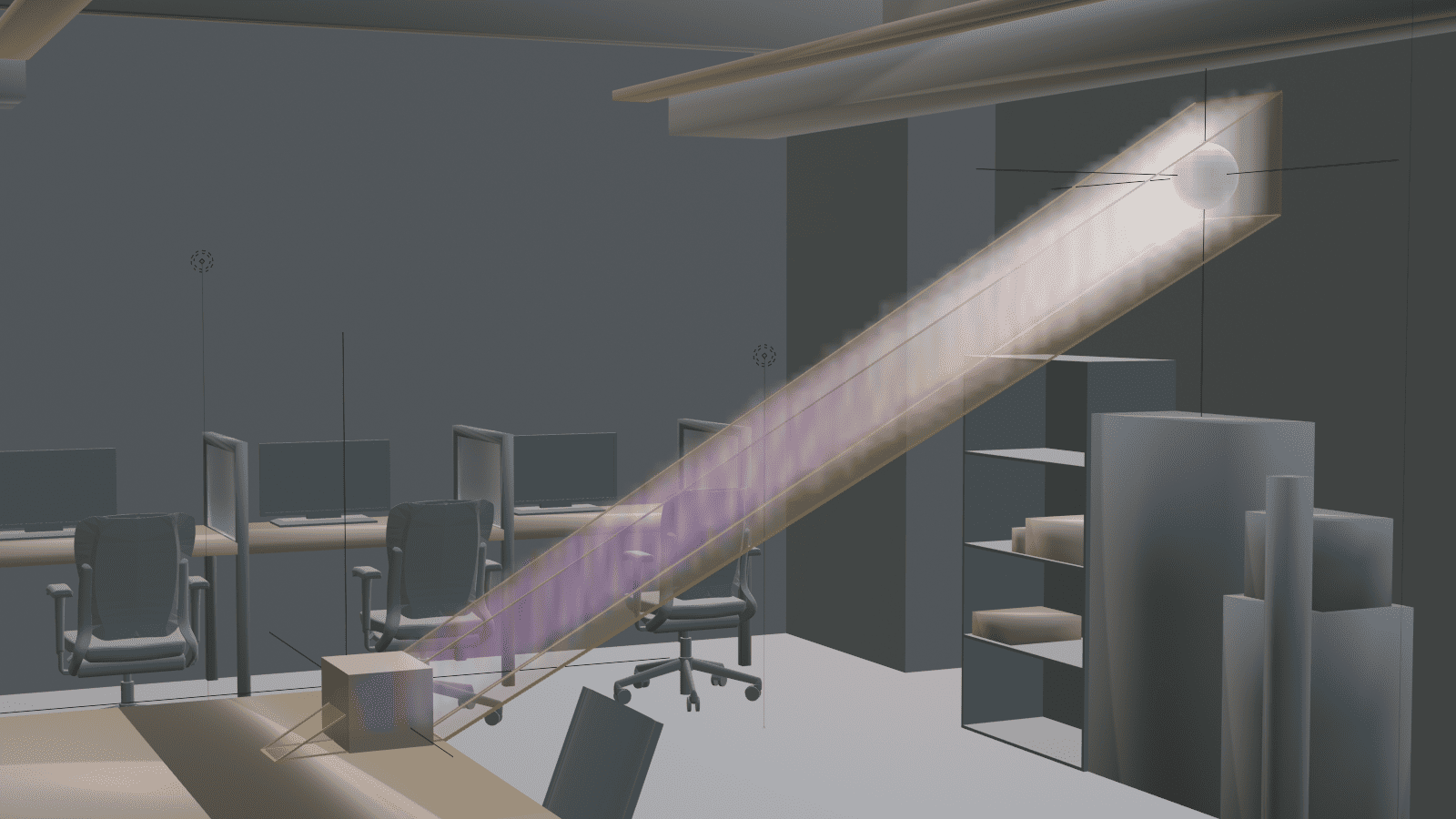}} 
    \subfigure[$\Delta\phi$, top view]{\includegraphics[width = 0.24\linewidth]{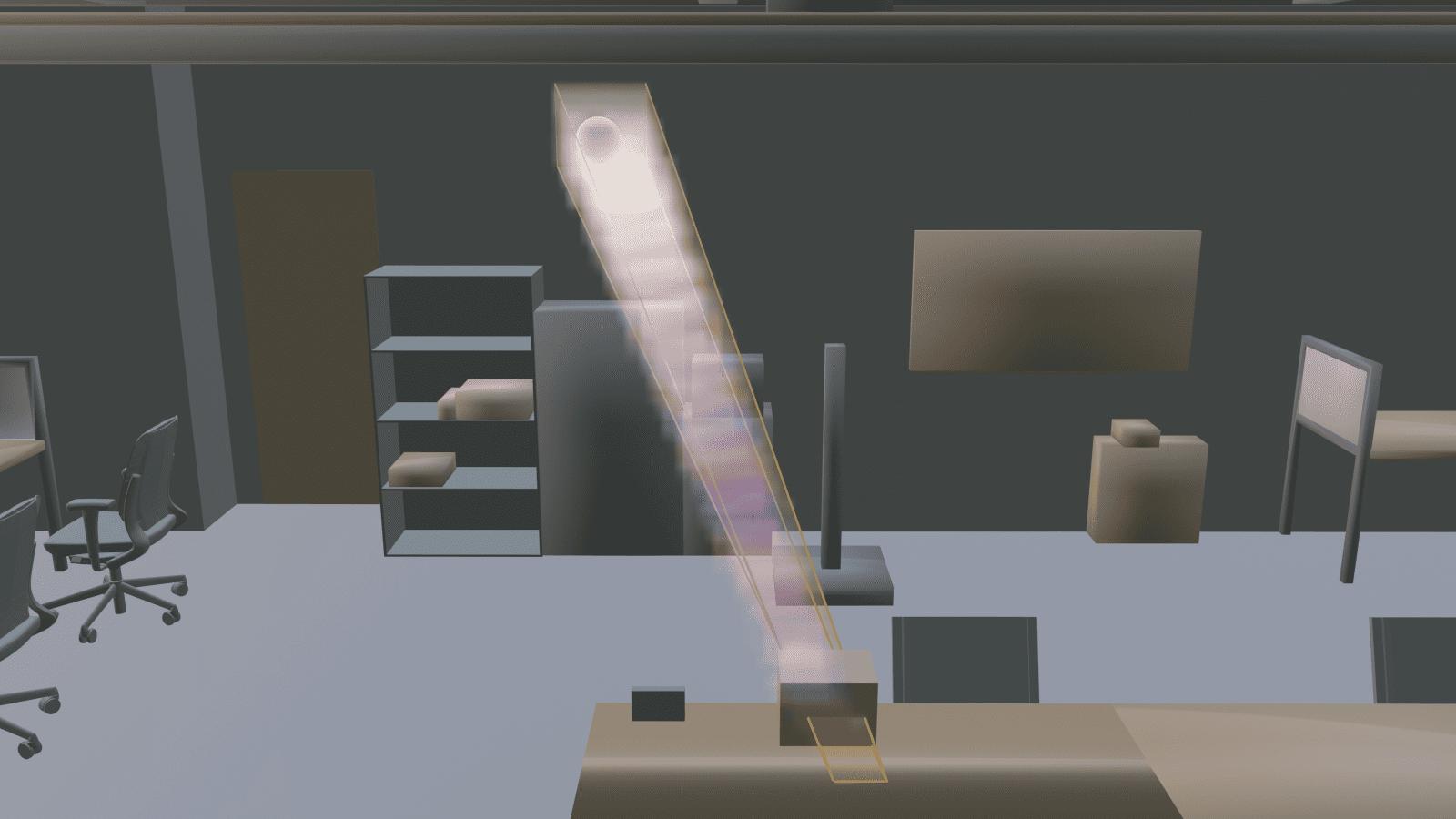}} \\
    \makebox[1\linewidth]{\includegraphics[width=0.3\textwidth]{FiguresResized/colorbar_phase.png}}
    \caption{Comparison of received power (dBm) and phase ($^\circ$) in the refraction tunnel: PE-PINN (a,e), Sionna (b,f), and their pointwise difference PE-PINN $-$ Sionna shown in side view (c,g) and top view (d,h).} \label{fig:tunnel_refra}
\end{figure*}

\begin{figure*}[t]
    \centering
    \subfigure[Sionna results.]{\includegraphics[width = 0.24\linewidth]{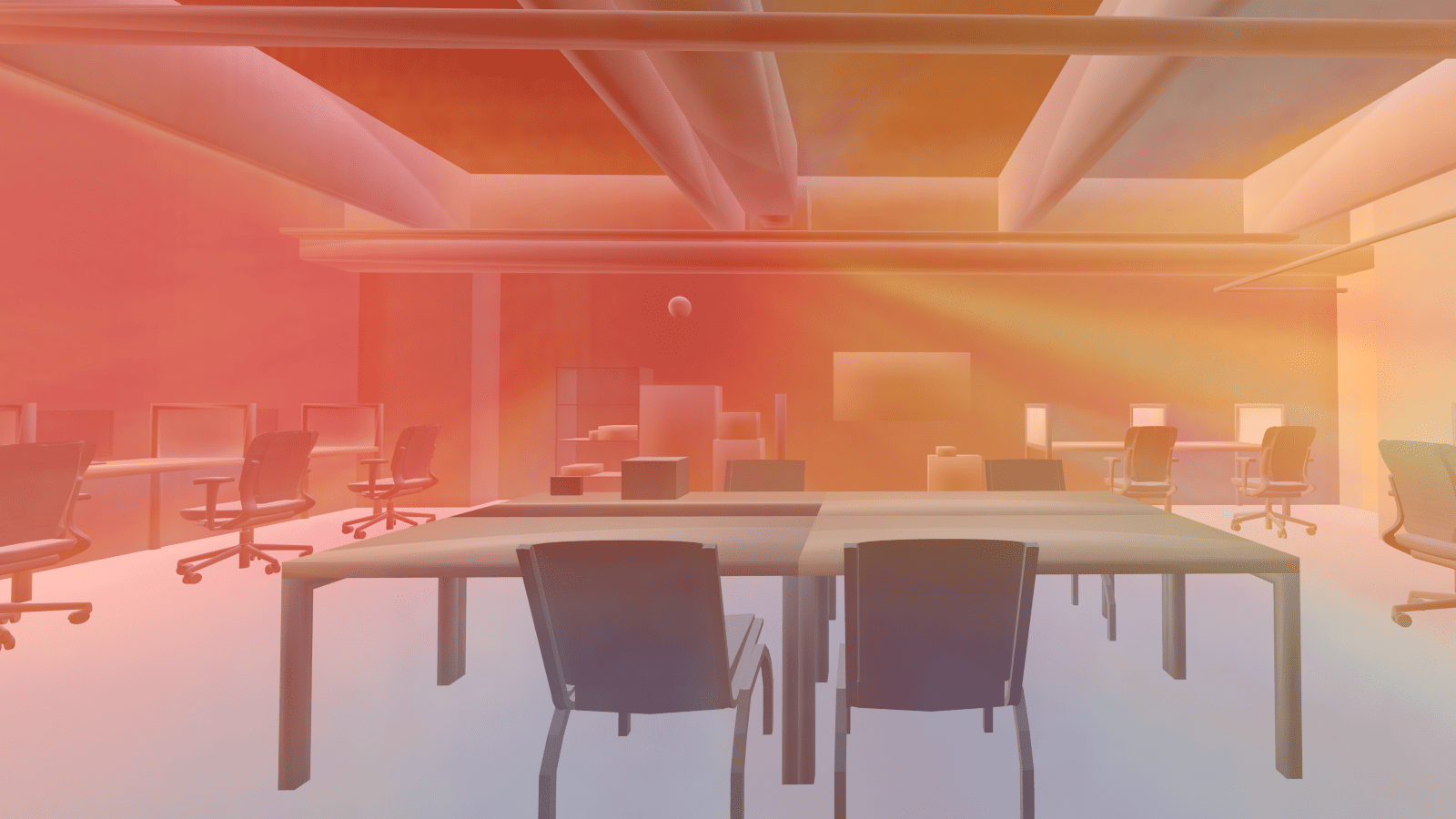}} 
    \subfigure[PE-PINN generated tunnels]{\includegraphics[width = 0.24\linewidth]{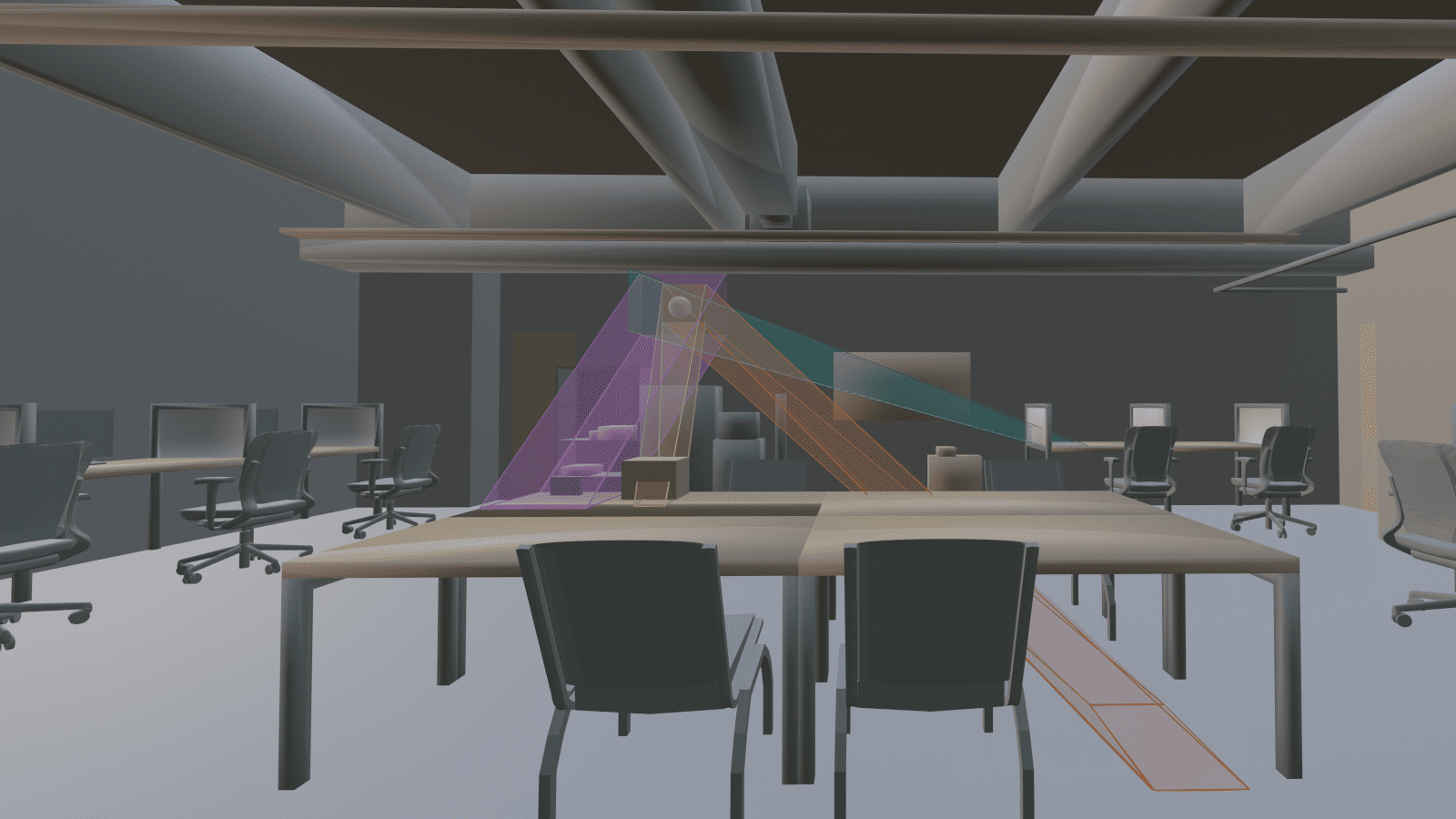}} 
    \subfigure[Corrected field.]{\includegraphics[width = 0.24\linewidth]{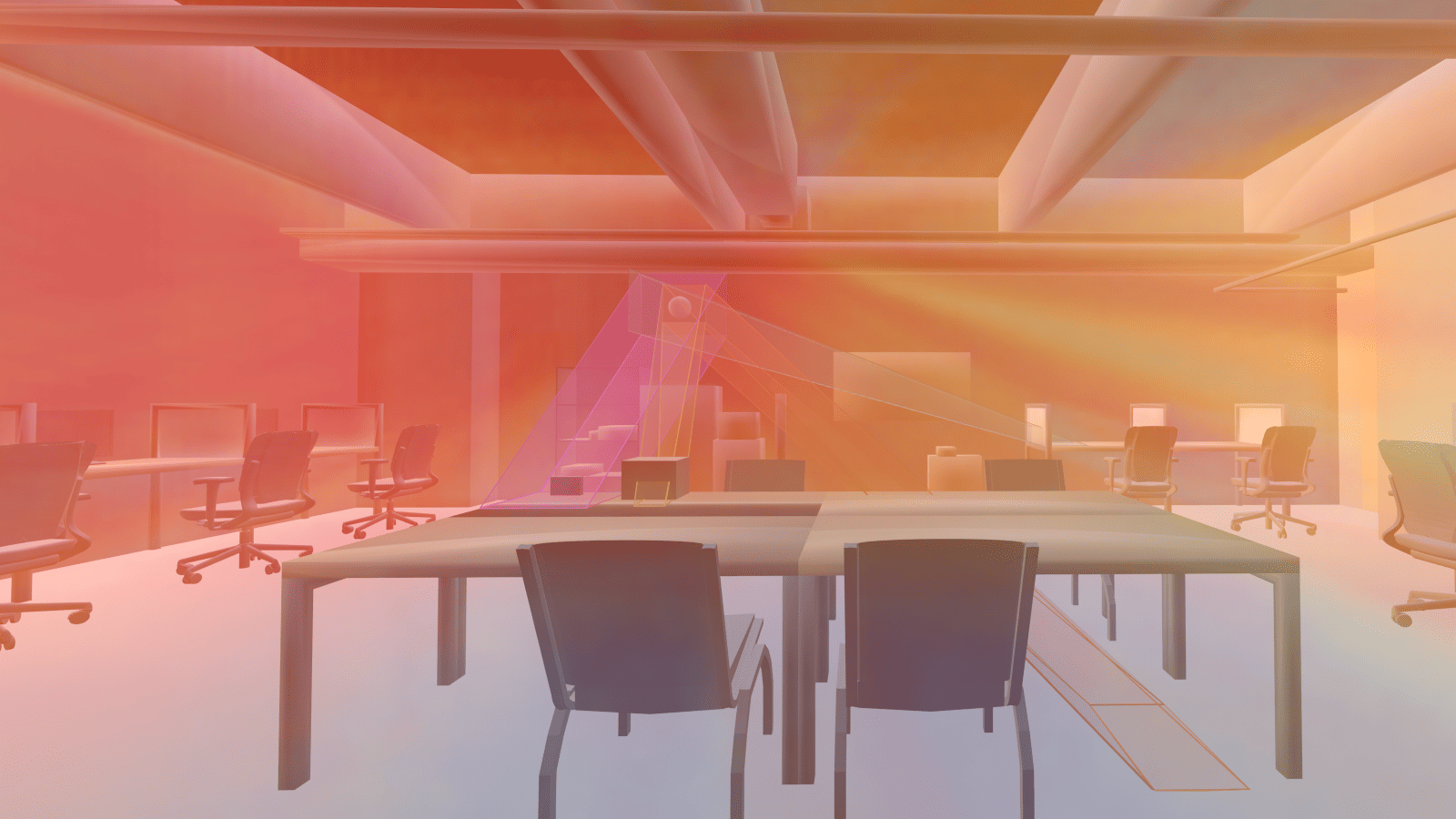}} 
    \subfigure[$\Delta P$ in the field.]{\includegraphics[width = 0.24\linewidth]{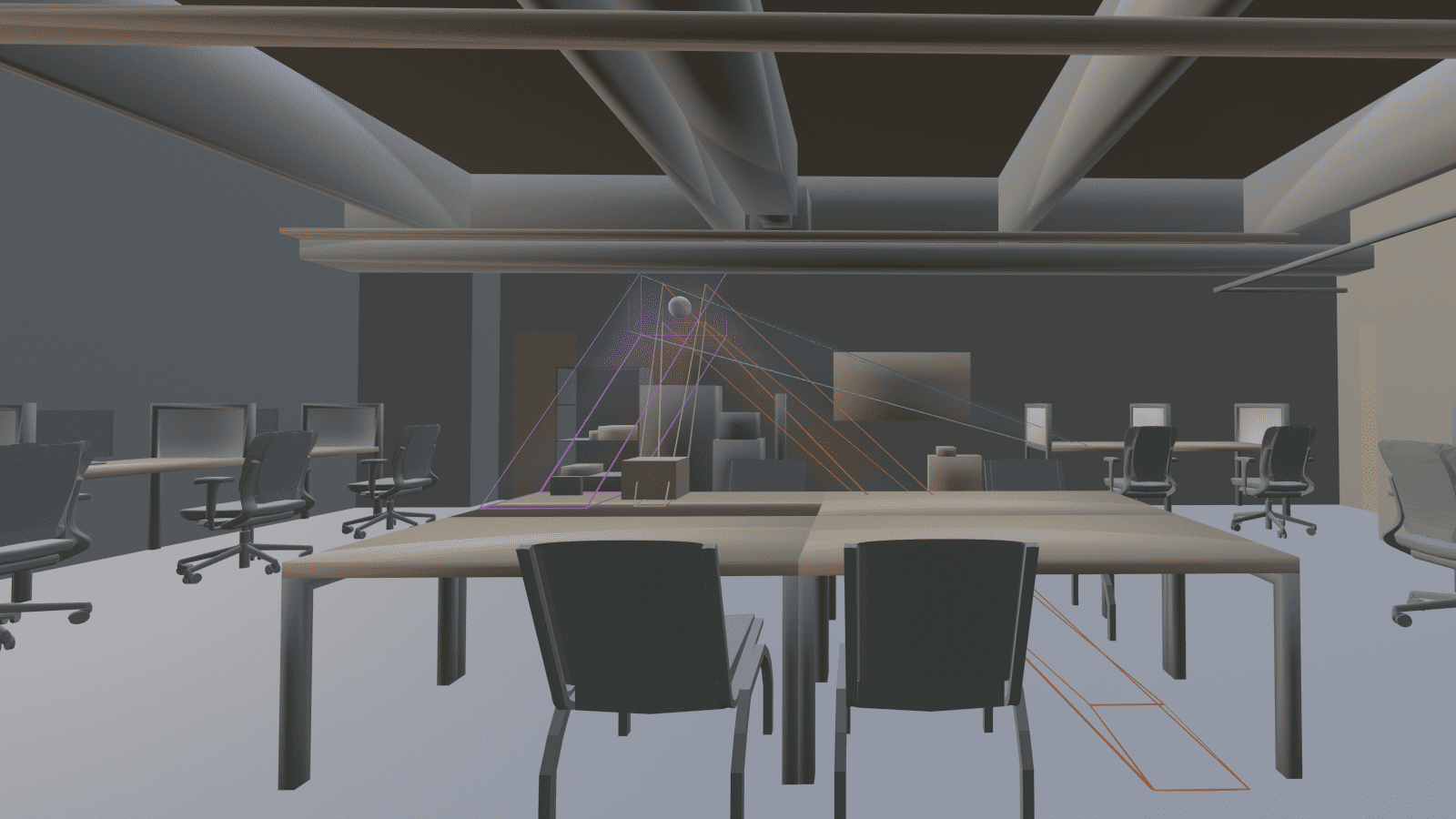}} \\
    \makebox[0.73\linewidth]{\includegraphics[width=0.3\linewidth]{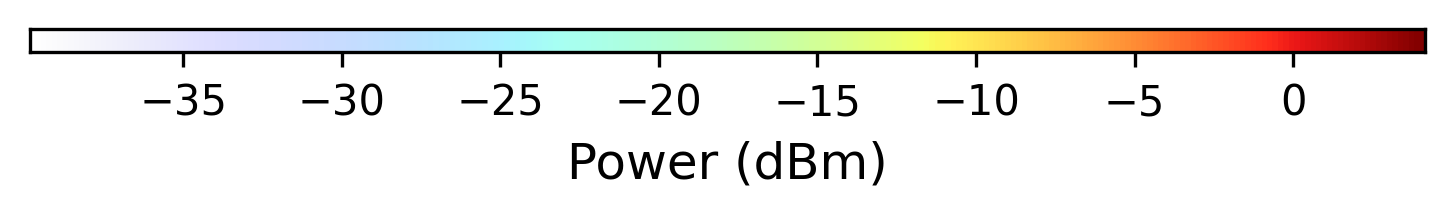}}
    \makebox[0.23\linewidth]{\includegraphics[width=0.2\linewidth]{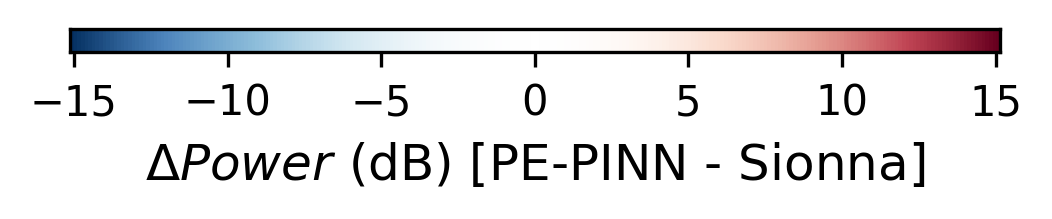}}
    \caption{Demonstration of a high-fidelity channel correction in the current radio DT at UW-Madison.} \label{fig:wholelabDT}
\end{figure*}

Assuming a stack of paper on the desk blocking the direct path between the transmitter and receivers. To correct the channel in the shadowing region, a refraction tunnel is constructed as shown in Fig.~\ref{fig:refra_tunnel}. Similar to the diffraction tunnel, the refraction tunnel is defined as a hexahedral domain extending from the transmitter toward the paper stack, with a near-end cross-section of $0.5 \times 0.45$~m at $y = -3.35$~m and a far-end cross-section of $0.2 \times 0.45$~m at $z = 0.70$~m. The paper stack is modeled as a dielectric strip using the ITU chipboard material~\cite{series2015effects} with a relative permittivity $\varepsilon_r = 2.58$ and a frequency dependent conductivity $\sigma = 0.0217 \times f_{\text{GHz}}^{0.78}$~S/m, occupying the region $y \in [0.14,\, 0.36]$~m. At the operating frequency of 2.4~GHz, this yields $\sigma \approx 0.042$~S/m, resulting in a complex wavenumber $k_c^2$ inside the strip as defined in~\eqref{eq:complex_k}. Unlike the diffraction case, the far end cross section is kept narrow so that the paper stack spans the entire width of the tunnel, forming a well-defined planar dielectric interface within the computational domain. The total field $E(\mathbf{x})$ is approximately:
\begin{equation}
    \begin{cases} 
    \mathcal{N}_{\text{src}}(\mathbf{x}) + \mathcal{N}_{\text{sct,L}}(\mathbf{x}; \Psi_{\mathbf{x}_{\text{ref},1}}, \Psi_{\mathbf{x}_{\text{ref},2}}), & y < 0.14, \\
    \mathcal{N}_{\text{sct,M}}(\mathbf{x}; \Psi_{\mathbf{x}_{\text{m},1}},\Psi_{\mathbf{x}_{\text{m},2}}), & 0.14 \le y \le 0.36, \\
    \mathcal{N}_{\text{sct,R}}(\mathbf{x}; \Psi_{\mathbf{x}_{\text{t},1}}), & y > 0.36.
    \end{cases}
\end{equation}
Following the PE-PINN architecture with material-aware domain decomposition~\cite{zhang2026physicsinformedneuralnetworksarchitectural}, the computational domain is partitioned into three regions along the $y$-axis: Region~1 ($y < 0.14$~m, free space on the source side), Region~2 ($0.14 < y < 0.36$~m, the lossy dielectric strip), and Region~3 ($y > 0.36$~m, free space on the transmission side). The Helmholtz Eq.~\eqref{eq:helmholtz_lossy} with complex $k_c^2$ is enforced within Region~2, while Regions~1 and~3 are governed by the Helmholtz equation with free-space wavenumber $k_0 = \omega\sqrt{\mu_0 \varepsilon_0}$, where $\mu_0$ and $\varepsilon_0$ denote the permeability and permittivity of free space, respectively.

The source network $\mathcal{N}_{\text{src}}$ remains identical to the diffraction case. A separate scatter network is assigned to each region. In Region~1, two reflection kernels are placed at $\mathbf{x}_{\text{ref},1} = (2.0,\, 3.48,\, 2.3)$ and $\mathbf{x}_{\text{ref},2} = (2.0,\, 3.92,\, 2.3)$, corresponding to the image sources of the transmitter mirrored across the two dielectric interfaces at $y = 0.14$~m and $y = 0.36$~m, respectively, to capture the reflected field from each surface of the strip. 
In Region~2, two kernels using the dielectric wavenumber $k_c$ are placed at $\mathbf{x}_{\text{m},1} = (2.0,\, -5.2,\, 2.3)$ and $\mathbf{x}_{\text{m},2} = (2.0,\, 6.0,\, 2.3)$ to model the forward and backward propagating waves within the lossy medium. In Region~3, a single transmitted-wave kernel is placed at $\mathbf{x}_{\text{t},1} = (2.0,\, -3.1,\, 2.3)$. The positions of these kernels are determined by combining the image principle with the optical path length correction: when a wave traverses a dielectric medium with relative permittivity $\varepsilon_r$, the effective optical path is scaled by $\sqrt{\varepsilon_r}$ compared to free space, so the image source must be placed at a correspondingly greater distance to produce the correct phase at the interface. Continuity of the tangential electric field $E$ and its normal derivative $\partial E / \partial y$ is enforced at both dielectric interfaces ($y = 0.14$~m and $y = 0.36$~m) through dedicated interface loss terms.

Fig.~\ref{fig:tunnel_refra} compares the received power and phase between PE-PINN and Sionna for the refraction tunnel, with quantitative results in Table~\ref{tab:refra_comparison}. Over the full domain the agreement is close (magnitude MAE $0.81$\,dB; phase MAE $4.63^\circ$). Within the \emph{Near-Strip} zone, defined as 0.25\,m before the dielectric strip plus the entire transmission side, the errors increase dramatically (magnitude MAE $7.44$\,dB; phase MAE $37.14^\circ$), confirming that the discrepancy is concentrated around the dielectric interface. The difference heatmaps reveal distinct physical mechanisms on each side of the strip: on the source side, an alternating $+/-$ pattern with a spatial period of roughly half a wavelength indicates a standing wave from incident and reflected interference at the interface, consistent with the preliminary results in Sec.~\ref{sec: prelimrefra}; on the transmission side, additional deviations reflect differences in how the two solvers model phase accumulation and amplitude attenuation through the lossy medium.

\begin{table}[ht!]
\centering
\caption{PE-PINN-Sionna Comparison for Refraction Scenario}\label{tab:refra_comparison}
\begin{tabular}{l|cc|cc}
\toprule
 & \multicolumn{2}{c|}{\textbf{Full Tunnel}} & \multicolumn{2}{c}{\textbf{Near-Strip}} \\
 & \textbf{MAE} & \textbf{RMSE} & \textbf{MAE} & \textbf{RMSE} \\
\midrule
$ \Delta$Power (dB)    & 0.81  & 11.14 & 7.44 & 53.49 \\
$ \Delta$Phase ($^\circ$)   & 4.63  & 13.65 & 37.14 & 61.74 \\
\bottomrule
\end{tabular}
\end{table}

\subsubsection{Tunnel replacement}
Fig.~\ref{fig:wholelabDT} illustrates the spatial power distribution across the indoor laboratory environment. The whole room field map in~(a) is obtained by Sionna with receivers placed on a uniform 5\,cm grid throughout the entire volume. The received power is rendered as a semi-transparent volumetric overlay on top of the 3D scene geometry, with color indicating power level following a jet colormap. To incorporate wave-level accuracy into selected regions of interest, we define hexahedral tunnel domains connecting the transmitter to key propagation paths, as highlighted in~(b). Within each tunnel, as shown in~(c) the Sionna values are replaced by PE-PINN predicted fields that satisfy the Helmholtz equation and account for reflection, refraction, diffraction, and lossy material interactions. (d) displays the power difference $\Delta P = P_{\mathrm{PINN}} - P_{\mathrm{Sionna}}$ within the tunnel domains. In total, four tunnel domains are replaced in this scene, two of which have been individually analyzed in Sections~\ref{sec:eval_diffra} and~\ref{sec:eval_refra}.

Since the tunnel is embedded within a complete indoor environment, multipath components reflected from other scatterers outside the tunnel may also arrive at receiver locations inside it. To quantify this effect, we decompose the full-scene Sionna solution within each tunnel into two groups: \emph{dominant paths}, whose interaction vertices all lie inside the tunnel volume, and \emph{other paths}, which involve at least one scattering event outside the tunnel. For the diffraction wide tunnel analyzed in Section~\ref{sec:eval_diffra}, the decomposition is evaluated over 11{,}297 receiver locations sampled on the 5\,cm uniform grid used for the full-scene simulation. The dominant paths exhibit a median power advantage of $7.85$\,dB and a mean advantage of $8.29$\,dB over the other paths. For the refraction tunnel analyzed in  Sec.~\ref{sec:eval_refra}, the same analysis over 3{,}185 receiver locations yields a mean advantage of $6.35$\,dB and a median power advantage of $8.24$\,dB.


\subsection{More Discussion}
The evaluation results consistently reveal a pronounced gap in channel fidelity between raw RT outputs from Sionna and full‑wave EM references, particularly in propagation regimes dominated by diffraction and refraction. In both controlled experiments and realistic indoor scenarios, Sionna exhibits significant magnitude and phase errors in shadowed regions and across dielectric objects. These discrepancies arise from the inherent reliance of RT on geometric‑optics approximations, which become unreliable when object dimensions are comparable to the wavelength or when near‑field and interference effects dominate.
By contrast, the PE‑PINN corrected channels closely track the full‑wave COMSOL ground truth in both magnitude and phase. Quantitative metrics and difference heatmaps show that PE‑PINN accurately recovers diffracted energy in shadow regions, models edge diffraction, and captures both phase accumulation and attenuation through dielectric media. The strong agreement between PE‑PINN and COMSOL, together with the distinct divergence from Sionna error patterns, confirms that the observed gains stem from physics‑consistent wave modeling rather than numerical smoothing or heuristic correction. These improvements are achieved precisely in the regimes where RT‑based methods are known to be least reliable.
Importantly, PE‑PINN is not intended to replace ray tracing or to construct wireless digital twins from scratch. Instead, it is designed as a complementary enhancement layer for DT frameworks that rely on RT engines such as Sionna for scalability and global multipath structure. Ray tracing remains effective for identifying dominant paths and system‑level propagation trends, while PE‑PINN selectively corrects RT predictions along critical propagation paths where wave effects beyond geometric optics dominate.
By confining PE‑PINN computations to localized tunnels around dominant paths, the proposed framework preserves the efficiency and compatibility of RT‑based DT pipelines while injecting full‑wave accuracy only where needed. In practice, PE‑PINN can either replace dominant path components or augment existing RT outputs, enabling substantially improved fidelity without the prohibitive cost of scene‑wide full‑wave simulation. These results suggest that the accuracy of wireless digital twins can benefit from both RT scalability, and physics‑aware correction in wave‑critical regions.


\section{Conclusion}\label{sec:conclusion}

This paper presented a physics‑constrained enhancement framework for wireless DTs that addresses the fidelity limitations of RT–based simulators in propagation regimes dominated by diffraction and refraction. Building on our prior PE‑PINN architecture, we introduced three targeted extensions: (i) a finite‑length edge diffraction kernel that discretizes each physical edge into $P$ coherent secondary point sources, (ii) a lossy dielectric formulation that embeds the complex wavenumber into in‑medium subnetworks, and (iii) a tunnel strategy that confines PE‑PINN computations to dominant propagation paths around obstacles of interest.
Through controlled three‑way comparisons against COMSOL references, we demonstrated that PE‑PINN closely matches the full‑wave ground truth in both diffraction and refraction scenarios, while Sionna exhibits pronounced errors in these shadowed and dielectric regions. We further applied the proposed framework to a DT of the UW-Madison wireless laboratory, showing that PE‑PINN can be flexibly integrated as either a localized replacement or augmentation of Sionna’s predictions along dominant propagation paths.
Future work will focus on improving scalability and automation.

\renewcommand\refname{Reference}
\bibliographystyle{IEEEtran}
\bibliography{Reference}



\end{document}